\documentclass[12pt]{article}
\usepackage{amssymb,amsmath}
\usepackage[pdftex]{graphicx}
\usepackage{subfig}
\usepackage{color}
\usepackage{float}
\usepackage{amsmath}
\usepackage{graphicx}
\usepackage{subfig}

\begin{document}
\newcommand{\room}{\;\;\,}
\newcommand{\Tr}{\mbox{Tr$\;$}}
\newcommand{\del}{\partial}
\newcommand{\mc}{\mathcal}
\newcommand{\p}[2]{\frac{\partial #1}{\partial #2}}
\newcommand{\pp}[3]{\frac{\partial^2 #1}{\partial #2\partial #3}}
\renewcommand{\[}{\left[}
\renewcommand{\]}{\right]}
\renewcommand{\(}{\left(}
\renewcommand{\)}{\right)}
\newcommand{\trix}[1]{\(\begin{array}{#1}}
\newcommand{\notrix}{\end{array}\)}
\setlength{\unitlength}{1.5cm}

\begin{titlepage}

%\vspace{-4cm}

\title{Wilson Lines and a Canonical Basis of $SU(4)$ Heterotic Standard Models {\LARGE \\[.5cm]  }}
                       
\author{{ Burt A.~Ovrut${}^{1}$, Austin Purves${}^{1}$ and Sogee Spinner${}^{2}$} \\[5mm]
    {\it  ${}^{1}$ Department of Physics, University of Pennsylvania} \\
   {\it Philadelphia, PA 19104--6396}\\[4mm]
   {\it ${}^{2}$ SISSA, Via Bonomea 265, I-34136 Trieste, Italy}
}

\date{}

\maketitle

\begin{abstract}
\noindent

\let\thefootnote\relax\footnotetext{ovrut@elcapitan.hep.upenn.edu, apurves@sas.upenn.edu, sspinner@sissa.it}The spontaneous breaking of $SU(4)$ heterotic standard models by ${\mathbb{Z}}_{3} \times {\mathbb{Z}}_{3}$ Wilson lines to the MSSM with three right-handed neutrino supermultiplets and gauge group $SU(3)_{C} \times SU(2)_{L} \times U(1) \times U(1)$ is explored. The two-dimensional subspace of the $Spin(10)$ Lie algebra that commutes with ${\mathfrak{su}}(3)_{C} \oplus {\mathfrak{su}}(2)_{L}$ is analyzed. It is shown that there is a unique basis for which 
%1) each quark/lepton and Higgs superfield arises from a different $Spin(10)$ multiplet, 2) the trace of the product of the Wilson line generators over complete $Spin(10)$ multiplets vanishes and 3) for which this trace is also zero over the extended MSSM spectrum. It follows that 
the initial soft supersymmetry breaking parameters are uncorrelated and for which  the $U(1) \times U(1)$ field strengths have no kinetic mixing at any scale. If the Wilson lines ``turn on'' at different scales, there is an intermediate regime with either a left-right or a Pati-Salam type model. We compute their  spectra directly from string theory, and adjust the associated mass parameter so that all gauge parameters exactly unify.
A detailed analysis of the running gauge couplings and soft gaugino masses is presented.

\vspace{.3in}
\noindent
\end{abstract}

\thispagestyle{empty}

\end{titlepage}

\section{Introduction}

The discovery of neutrino masses \cite{Nakamura:2010zzi} motivates the introduction of right-handed neutrinos into the standard model and, by extension, into the minimal $N=1$ supersymmetric standard model (MSSM) \cite{MSSM1, MSSM2, MSSM3}. Remarkably, this right-handed neutrino extended MSSM can arise from vacua of the $E_{8} \times E_{8}$ heterotic superstring \cite{e1, Lukas:1998yy, Lukas:1999kt, Donagi:1999jp, Donagi:1999ez, e2, e3}. Specifically, smooth compactifications on elliptically fibered ${\mathbb{Z}}_{3} \times {\mathbb{Z}}_{3}$ Schoen manifolds \cite{Donagi:2003tb, Braun:2004xv} with $SU(4)$ ``extension'' bundles \cite{Braun:2005ux, Braun:2005bw, Braun:2005zv, Braun:2005fk, Ambroso:2008kb, Buchbinder:2002ji, Buchbinder:2002pr} can lead to four-dimensional, $N=1$ supersymmetric theories with exactly the particle spectrum of the MSSM with three families of right-handed neutrino chiral supermultiplets, one per family, and no vector-like pairs or exotic states \cite{Braun:2005nv, lukas}. Furthermore, the theory is invariant under the $SU(3)_{C} \times SU(2)_{L} \times U(1)_{Y}$ gauge group of the standard model. 

However, the fact that this ``extended'' MSSM arises from heterotic string theory has important theoretical and phenomenological implications. First, and foremost, is the fact that the low energy gauge group ${\cal{G}}$ must contain one extra $U(1)$ factor. This is due to the fact that the last step in the symmetry breaking sequence $E_{8} \rightarrow Spin(10) \rightarrow {\cal{G}}$\footnote{We will refer to the $SO(10)$ group, as it is commonly called in the model building literature, as $Spin(10)$ to be more mathematically correct.} is accomplished by two ``Wilson lines,'' each corresponding to a generator of the ${\mathbb{Z}}_{3} \times {\mathbb{Z}}_{3}$ isometry group. Since these Wilson lines are Abelian, they preserve the rank of the gauge group. It follows that the rank 5 $Spin(10)$ group is spontaneously broken to the rank 4 standard model group plus an additional factor of $U(1)$, which can be associated with B-L (baryon minus lepton number). That is, ${\cal{G}}=SU(3)_{C} \times SU(2)_{L} \times U(1)_{Y} \times U(1)_{B-L}$. A renormalization group analysis of this 
B-L MSSM theory, including the radiative breaking of both B-L and electroweak symmetry, the associated B-L/EW hierarchy and predictions for the masses of the $Z^{\prime}$ boson, the Higgs bosons and all superpartners was presented in \cite{Ambroso:2009jd, Ambroso:2009sc}. The implications for proton decay, and dark matter, as well as a discussion of the associated cosmic strings, were given in \cite{Ambroso:2010pe} and \cite{Brelidze:2010hf} respectively, while consequences in the neutrino sector were discussed in \cite{Ambroso:2010pe, Mohapatra:1986aw, Ghosh:2010hy, Barger:2010iv}

Meanwhile, supersymmetric theories associated with B-L have had a long, rich history because of the intimate relationship between R-parity (first defined in~\cite{R1,R2}) and B-L: $R=(-1)^{3(B-L) + 2S}$, where $S$ is the spin. The fate of R-parity plays a crucial role in the phenomenology and cosmology of supersymmetry. This relation was originally explored in the context of a global symmetry~\cite{AM} and later expanded to gauged symmetries~\cite{Hayashi} and more minimally in~\cite{Mohapatra:1986aw}; all of which resulted in R-parity violation. This was followed by a medley of works exploring both facets of the fate of R-parity in a variety of contexts, including~\cite{Masiero:1990uj, Kuchimanchi:1993jg, Martin:1996kn, Aulakh:1999cd, Aulakh:2000sn, Feldman:2011ms}.

Recently, this B-L MSSM theory was advanced in a series of papers~\cite{FileviezPerez:2008sx, Barger:2008wn, Everett:2009vy} as the simplest B-L supersymmetric theory, since it only extends the MSSM by the right-handed neutrinos required by anomaly cancellation. An automatic prediction follows: R-parity violation is spontaneously broken in minimal B-L models because only the right-handed sneutrino can break the B-L symmetry in a realistic way. These papers explored many of the consequences of this theory and showed it to be consistent with present experimental data. Furthermore, these papers were recently supplemented by a discussion of the lepton number violating signals that could accompany the B-L MSSM at the LHC~\cite{FileviezPerez:2012mj}. Therefore, from both the top-down and bottom-up point of view, the B-L MSSM theory appears to be very compelling.

The analyses in \cite{Ambroso:2009jd, Ambroso:2009sc, Ambroso:2010pe}, in order to elucidate the mechanism for radiative B-L breaking and the B-L/EW hierarchy, necessarily was carried out over a relatively restricted set of initial parameters. In addition, the basis of generators of $Y$ and $B-L$ is not ``orthogonal'' in the Cartan subalgebra of ${\mathfrak{so}}(10)$. Hence, the associated $U(1)_{Y}$ and $U(1)_{B-L}$ field strengths exhibit ``kinetic mixing'', both initially and at all lower scales. This greatly complicates the RGEs of the gauge couplings and was analyzed only approximately in \cite{Ambroso:2009jd, Ambroso:2009sc, Ambroso:2010pe}. Finally, the running parameters of the B-L MSSM generically experience five mass ``thresholds'', that is, scales where the coefficients of the associated RG beta functions change. In \cite{Ambroso:2009jd, Ambroso:2009sc, Ambroso:2010pe}, this was approximated by only three such masses, the other two being sufficiently close to these that, in a restricted regime of parameter space, they could safely be ignored.

In this paper, we rectify the last two of these problems. The first issue, carrying out the RG analysis for a greatly expanded set of initial parameters, will be analyzed in forthcoming work 
\cite{Preparation}. We begin by presenting a detailed mathematical analysis of the Cartan subalgebra of ${\mathfrak{so}}(10)$, deriving the two-dimensional subalgebra that commutes with ${\mathfrak{su}}(3)_{C} \oplus {\mathfrak{su}}(2)_{L}$ and introducing the ``canonical'' basis which spans it. Wilson lines derived from any linearly independent elements $Y_{1}$ and $Y_{2}$ of this subalgebra  will spontaneously break $Spin(10) \rightarrow SU(3)_{C} \times SU(2)_{L} \times U(1)_{Y_{1}}\times U(1)_{Y_{2}}$. Many, but not all, such bases will lead to the right-handed neutrino extended MSSM spectrum at low energy. We will restrict our attention to those bases that do-- such as $Y_{Y}$,$Y_{B-L}$ and $Y_{T_{3R}}$,$Y_{B-L}$, as well as to the specific ``non-canonical'' basis introduced  and analyzed in Appendix A. We show that, as with the $Y$, B-L generators discussed in 
\cite{Ambroso:2009jd,Ambroso:2009sc,Ambroso:2010pe}, the quark/lepton and Higgs superfields in both the canonical and non-canonical bases each arise from different ${\bf 16}$ and ${\bf 10}$ representations of $Spin(10)$ respectively. It follows that the soft supersymmetry breaking terms, as well as the Yukawa couplings, are uncorrelated by their origin in $Spin(10)$ multiplets. This ``liberates'' the initial parameter space of these theories and has important implications for low energy.

Next, we introduce the Killing inner product on the Cartan subalgebra. It is shown that as long as the basis elements $Y_{1}$,$Y_{2}$ are orthogonal, that is, that their Killing bracket vanishes, then kinetic mixing of the associated gauge field strengths at the initial ``unification'' scale will also vanish. Both the canonical and specific non-canonical basis elements, but not the $Y_{Y}$,$Y_{B-L}$ generators of  \cite{Ambroso:2009jd, Ambroso:2009sc, Ambroso:2010pe}, satisfy this condition. Henceforth, we restrict our discussion to such orthogonal bases. Using the RG analyses presented in \cite{Babu:1996vt}, we then show that kinetic mixing will continue to vanish at all lower energy-momentum scales if and only if $Tr(Y_{1}Y_{2})$ over the entire matter and Higgs spectrum of the B-L MSSM is zero. The canonical basis is shown to satisfy this condition and, hence, never exhibits kinetic mixing. However, in Appendix A we show that the specific non-canonical basis does not. Furthermore, we prove a theorem that the only orthogonal basis satisfying this condition is precisely the canonical basis, and appropriate multiples thereof. We note in passing that the $Y_{Y}$,$Y_{B-L}$ generators of \cite{Ambroso:2009jd, Ambroso:2009sc, Ambroso:2010pe} also don't have vanishing trace over the B-L MSSM. Thus, in this paper, we identify a unique basis of generators $Y_{T_{3R}}$,$Y_{B-L}$ whose low energy gauge group, $U(1)_{Y_{T_{3R}}}\times U(1)_{Y_{B-L}}$, does not exhibit kinetic mixing at any scale.

For this reason, the bulk of this paper analyzes the mass thresholds, boundary conditions and RG running of the gauge parameters and soft gaugino masses associated with this canonical basis. An important aspect of the analysis is that the Wilson lines associated with each of $Y_{T_{3R}}$,$Y_{B-L}$ need not ``turn on'' at the same scale. Rather, because the inverse radius of the ``hole'' in the Calabi-Yau threefold that they wrap depends on moduli, one such mass scale can precede the other, perhaps by as much as an order of magnitude. Allowing for this, there are then two different ``breaking patterns''. In the first, $Spin(10)$ is broken by the  $Y_{B-L}$ Wilson line to an intermediate region containing a left-right type model. This has gauge group $SU(3)_{C} \times SU(2)_{L} \times SU(2)_{R} \times U(1)_{B-L}$ with a specific spectrum that we calculate directly from string theory. At a lower scale, $M_{I}$, the $Y_{T_{3R}}$ Wilson line turns on and the theory becomes the extended MSSM. Conversely, if the $Y_{T_{3R}}$ Wilson line turns on first, then $Spin(10)$ is broken to a Pati-Salam type theory with gauge group $SU(4)_{C} \times SU(2)_{L} \times U(1)_{T_{3R}}$ and an explicit spectrum. Again, we calculate this directly using string theory. At a lower scale, $M_{I}$, the $Y_{B-L}$ Wilson line turns on and the theory becomes the extended MSSM. It is proven that in either case, subject to imposing the experimental values of the gauge couplings at the $Z$-mass, $M_{I}$ can be chosen so as to enforce the unification of all gauge couplings, albeit for different values of $M_{I}$. Of course, if both Wilson lines turn on simultaneously, then $Spin(10)$ is broken immediately to the B-L MSSM. In this case, gauge unification cannot occur.

Using this technology, in the final section we present a detailed discussion of the five mass thresholds, the boundary conditions at these scales, explicit computer analysis' plotting the running of the gauge couplings and a discussion of the gaugino soft masses. This is done independently for both the left-right and Pati-Salam intermediate breaking patterns, imposing the experimental values of the gauge couplings at the $Z$-mass. These plots show the unification of these couplings, and the explicit mass threshold behavior. The unification scale, the unified gauge coupling and the value of $M_{I}$ associated with such unification are evaluated in each case. Finally, we present the results when both Wilson lines are simultaneous and explore the extent to which the gauge parameters ``miss'' unification.

%%%%%%%%%%%%%%%%%%%%%%%%%%%%%%%%%%%%%%%%
%
\section{The ${\mathbb{Z}}_{3} \times {\mathbb{Z}}_{3}$ Wilson Lines}
%
%%%%%%%%%%%%%%%%%%%%%%%%%%%%%%%%%%%%%%%%

We begin by searching for the most general $U(1) \times U(1)$ subgroup of $Spin(10)$ that commutes with $SU(3)_{C} \times SU(2)_{L}$. This is most easily carried out using the associated Lie algebra ${\mathfrak{so}}(10)$, whose relevant properties can be found, for example, in \cite{Georgi}.

We will identify the color subalgebra ${\mathfrak{su}(3)_C} \subset {\mathfrak{so}}(10)$  with the $\alpha^{1}$ and $\alpha^{2}$ nodes of the ${\mathfrak{so}}(10)$ Dynkin diagram shown in Figure 1.
\begin{figure}[h]
\begin{center}
\begin{picture}(5,2)(0,.2)
\put(.625,1){\line(1,0){.75}}
\put(1.625,1){\line(1,0){.75}}
\put(2.588,.912){\line(1,-1){.53}}
\put(2.588,1.088){\line(1,1){.53}}
\put(.5,1){\circle{.25}}
\put(1.5,1){\circle{.25}}
\put(2.5,1){\circle{.25}}
\put(3.207,.293){\circle{.25}}
\put(3.207,1.707){\circle{.25}}

\put(.4,.6){$\alpha_1$}
\put(1.4,.6){$\alpha_2$}
\put(2.4,.6){$\alpha_3$}
\put(3.457,1.6){$\alpha_4$}
\put(3.457,.2){$\alpha_5$}

\put(.2,1.3){\color{red} \line(1,0){1.6}}
\put(1.8,1.3){\color{red} \line(0,-1){.9}}
\put(.2,.4){\color{red} \line(1,0){1.6}}
\put(.2,1.3){\color{red} \line(0,-1){.9}}

\put(2.9,1.4){\color{blue} \line(1,0){1}}
\put(3.9,2){\color{blue} \line(0,-1){.6}}
\put(2.9,2){\color{blue} \line(1,0){1}}
\put(2.9,2){\color{blue} \line(0,-1){.6}}

\end{picture}
\end{center}
\label{Figure1}
\caption{\small The Dynkin diagram for $Spin(10)$ with the $\mathfrak{su}(3)_{C}$ and $\mathfrak{su}(2)_{L}$ generators highlighted in red and blue respectively.}
\end{figure}
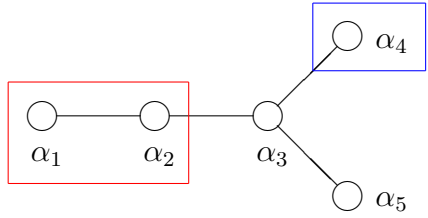
\noindent The complete set of ${\mathfrak{su}}(3)_C$ roots is given by 
\begin{equation}
\alpha^{1}=(1,-1,0,0,0), \quad \alpha^{2}=(0,1,-1,0,0), \quad \beta=(1,0,-1,0,0)
\label{1}
\end{equation}
as well as minus these roots. We denote the associated elements of ${\mathfrak{su}}(3)_C$ by $E_{\pm\alpha^{1}}, E_{\pm \alpha^{2}}$ and $E_{\pm \beta}$ respectively. When added to 
${\cal{H}}_{1}=\frac{1}{2}(H_{1}-H_{2})$ and ${\cal{H}}_{2}=\frac{1}{2}(H_{2}-H_{3})$ 
of the Cartan subalgebra ${\mathfrak{h}}$, these eight elements span ${\mathfrak{su}}(3)_C \subset {\mathfrak{so}}(10)$.
Similarly, we will identify the electroweak subalgebra ${\mathfrak{su}}(2)_{L} \subset {\mathfrak{so}}(10)$ with the $\alpha^{4}$ node of the ${\mathfrak{so}}(10)$ Dynkin diagram in Figure 1. The complete set of roots is given by
\begin{equation}
\alpha^{4}=(0,0,0,1,-1)
\label{2}
\end{equation}
and its minus. Denote the associated elements of ${\mathfrak{su}}(2)_L$ by $E_{\pm\alpha^{4}}$. Added to ${\cal{H}}_{4}=\frac{1}{2}(H_{4}-H_{5})$ of the Cartan subalgebra, these three elements span ${\mathfrak{su}}(2)_L \subset {\mathfrak{so}}(10)$.

To identify the most general $U(1) \times U(1)$ subgroup of $Spin(10)$ that commutes with $SU(3)_{C} \times SU(2)_{L}$, we simply search for the subspace of the five-dimensional Cartan subalgebra ${\mathfrak{h}}$ that commutes with all of the ${\mathfrak{su}}(3)_C \oplus {\mathfrak{su}}(2)_L$ generators listed above. Using the commutation relations
\begin{equation}
[H_{i},H_{j}]=0, \quad [H_{i},E_{\alpha}]=\alpha_{i}E_{\alpha}
\label{3}
\end{equation}
valid for {\it any} root $\alpha$, this is equivalent to solving for the most general element of 
${\mathfrak{h}}$ that annihilates $\alpha^{1},\alpha^{2},\beta,\alpha^{4}$. Writing this as
$H_{3\oplus2}={\sum}_{i=1}^{5}a^{i}H_{i}$,
it follows from \eqref{1} and \eqref{2} that 
\begin{equation}
H_{3\oplus2}=a(H_{1}+H_{2}+H_{3})+b(H_{4}+H_{5})
\label{4}
\end{equation}
for any real coefficients a and b. That is, the elements of ${\mathfrak{so}}(10)$ that commute with 
${\mathfrak{su}}(3)_C \oplus {\mathfrak{su}}(2)_L$ form the two-dimensional subspace ${\mathfrak{h}}_{3\oplus2}$ of the Cartan subalgebra spanned by 
\begin{equation}
H_{1}+H_{2}+H_{3}, \quad H_{4}+H_{5}~.
\label{5}
\end{equation}

Of course, any linearly independent basis of ${\mathfrak{h}}_{3\oplus2} \subset {\mathfrak{h}}$ is of potential physical interest. However, the basis \eqref{5} arises naturally in the above calculation. For this reason, and others to be specified below, we will refer to \eqref{5} as the ``canonical'' basis and discuss its properties first and in detail. In Appendix A, we briefly analyze a non-canonical basis and prove a general theorem about such bases.

%%%%%%%%%%%%%%%%%%%%%%
\subsection{The Canonical Basis \label{CB}}
%%%%%%%%%%%%%%%%%%%%%%

To identify the physical meaning of each of the canonical generators, it is useful to find their explicit form in the ${\bf 16}$ representation of  ${\mathfrak{so}}(10)$ since it contains a complete family of quarks/leptons including the right-handed neutrino. This can be accomplished using standard methods; see, for example, \cite{Georgi,Slansky:1981yr}. Written in a basis in which the ${\bf 16}$ decomposes under ${\mathfrak{su}}(3)_C \oplus {\mathfrak{su}}(2)_L$ as
\begin{equation}
{\bf 16}= (\overline{\bf 3}, {\bf 1})^{\oplus 2} \oplus ({\bf 3},{\bf 2}) \oplus ({\bf 1},{\bf 2})\oplus({\bf 1},{\bf 1})^{\oplus 2}~,
\label{6}
\end{equation}
that is, into the u, d, Q quarks and the L, $\nu$, e leptons respectively,
we find that
\begin{equation}
2(H_{1}+H_{2}+H_{3})= ((-1){\bf 1}_{3})^{\oplus 2}  \oplus (1){\bf 1}_{6} \oplus (-3){\bf 1}_{2} 
\oplus ((3){\bf 1}_{1})^{\oplus 2}~.
\label{7}
\end{equation}
We have multiplied $H_{1}+H_{2}+H_{3}$ by $2$ to ensure that all diagonal elements are integers.
It follows that one can identify
\begin{equation}
2(H_{1}+H_{2}+H_{3})=3(B-L)~.
\label{8}
\end{equation}
Similarly, written in the same basis we find 
\begin{equation}
H_{4}+H_{5}=  (-1){\bf 1}_{3} \oplus (1){\bf 1}_{3}  \oplus (0){\bf 1}_{6} \oplus (0){\bf 1}_{2} 
\oplus (-1){\bf 1}_{1} \oplus (1){\bf 1}_{1}~.
\label{9}
\end{equation}
If  hypercharge is defined using the relation $Q=T_{3L}+Y$, then one can identify
\begin{equation}
 H_{4}+H_{5}= 2(Y-\frac{1}{2}(B-L))~.
 \label{10}
 \end{equation}
 It is useful to note that
 \begin{equation}
Y-\frac{1}{2}(B-L)=T_{3R}~,
\label{11}
\end{equation}
where $T_{3R}$ is the diagonal generator of $SU(2)_{R}$. Noting that the complete set of roots associated with the $\alpha^{5}$ node of the ${\mathfrak{so}}(10)$ Dynkin diagram in Figure 1 is
\begin{equation}
\alpha^{5}=(0,0,0,1,1)
\label{12}
\end{equation}
and its minus, one can see immediately that the three elements ${\cal{H}}_{5}=\frac{1}{2}( H_{4}+H_{5})$ and $E_{\pm\alpha^{5}}$ span an $SU(2)$ Lie algebra. It follows from \eqref{11} that this can be identified with ${\mathfrak{su}}(2)_R \subset {\mathfrak{so}}(10)$. Having made these physical identifications, we henceforth denote these generators by
\begin{equation}
Y_{B-L}=2(H_{1}+H_{2}+H_{3}), \quad Y_{T_{3R}}=H_{4}+H_{5} ~.
\label{13}
\end{equation}
Note that their Killing brackets are given by
\begin{equation}
(Y_{B-L}|Y_{B-L})=12, \quad (Y_{T_{3R}}|Y_{T_{3R}})=2, \quad (Y_{B-L}|Y_{T_{3R}})=0
\label{14}
\end{equation}
where we have used the relation $(H_{i}|H_{j})=\delta_{ij}$.
To set our notation, we reiterate that 
\begin{equation}
[Y_{B-L}]_{\bf 16} = ((-1){\bf 1}_{3})^{\oplus 2}  \oplus (1){\bf 1}_{6} \oplus (-3){\bf 1}_{2} 
\oplus ((3){\bf 1}_{1})^{\oplus 2} 
\label{15}
\end{equation}
and
\begin{equation}
 [Y_{T_{3R}}]_{\bf 16} = (-1){\bf 1}_{3} \oplus (1){\bf 1}_{3}  \oplus (0){\bf 1}_{6} \oplus (0){\bf 1}_{2} \oplus (-1){\bf 1}_{1} \oplus (1){\bf 1}_{1}~. 
 \label{16}
\end{equation}
The Killing bracket is defined on ${\mathfrak{so}}(10)$ and is independent of the representation of the Lie algebra. That being said, the Killing bracket of any two elements $x,y \in {\mathfrak{so}}(10)$ can be evaluated in any representation $R$ using the formula
\begin{equation}
(x|y)=\frac{1}{I_{R}} Tr([x]_{R}[y]_{R}) ~,
\label{17}
\end{equation}
where 
\begin{equation}
I_{R}=\frac{d_{R}}{d_{\mathfrak{so}(10)}}C_{2}(R)
\label{18}
\end{equation}
is the Dynkin index, ${d_{\mathfrak{so}(10)}}=45$ and $C_{2}(R)$ is the quadratic Casimir invariant for the representation. Using this relation, one can check the validity of \eqref{15} and \eqref{16}. Note that $C_{2}({\bf 16})=\frac{45}{4}$ and, hence, $I_{{\bf 16}}=4$. Furthermore, from \eqref{15} and \eqref{16} one finds
\begin{eqnarray}
&&Tr([Y_{B-L}]_{\bf 16}[Y_{B-L}]_{\bf 16})=48, \quad Tr( [Y_{T_{3R}}]_{\bf 16} [Y_{T_{3R}}]_{\bf 16})=8, \nonumber \\
&&\qquad \qquad \qquad Tr( [Y_{B-L}]_{\bf 16}[Y_{T_{3R}}]_{\bf 16})=0 ~.
\label{19}
\end{eqnarray}
It then follows from \eqref{17} that $[Y_{B-L}]_{\bf 16}$ and $[Y_{T_{3R}}]_{\bf 16}$ satisfy the Killing relations \eqref{14}, as they must.

In the following, it will be important to know the explicit form of $Y_{B-L}$ and $Y_{T_{3R}}$ in the 
${\bf 10}$ representation of ${\mathfrak{so}}(10)$. Written in a basis in which ${\bf 10}$ decomposes under ${\mathfrak{su}}(3)_C \oplus {\mathfrak{su}}(2)_L$ as
\begin{equation}
{\bf 10}=  ({\bf 3},{\bf 1})\oplus(\overline{\bf 3}, {\bf 1}) \oplus ({\bf 1},{\bf 2})^{\oplus 2}~,
\label{20}
\end{equation}
that is, into the color triplet $H_{C}, {\bar{H}}_{C}$ and weak doublet $H, {\bar{H}}$ Higgs fields respectively, we find using \cite{Georgi} that
\begin{equation}
[Y_{B-L}]_{\bf 10}=(2){\bf 1}_{3} \oplus (-2){\bf 1}_{3} \oplus (0){\bf 1}_{2}\oplus (0){\bf 1}_{2}
\label{21}
\end{equation}
and
\begin{equation}
[Y_{T_{3R}}]_{\bf 10}=(0){\bf 1}_{3} \oplus (0){\bf 1}_{3} \oplus (1){\bf 1}_{2}\oplus (-1){\bf 1}_{2}~.
\label{22}
\end{equation}
As above, these expressions can be checked by computing their Killing brackets. We find 
\begin{eqnarray}
&&Tr([Y_{B-L}]_{\bf 10}[Y_{B-L}]_{\bf 10})=24, \quad Tr( [Y_{T_{3R}}]_{\bf 10} [Y_{T_{3R}}]_{\bf 10})=4, \nonumber \\
&&\qquad \qquad \qquad Tr( [Y_{B-L}]_{\bf 10}[Y_{T_{3R}}]_{\bf 10})=0 ~.
\label{23}
\end{eqnarray}
Noting that $C_{2}({\bf 10})=9$ and, hence, $I_{{\bf 10}}=2$, it follows from \eqref{17} that $[Y_{B-L}]_{\bf 10}$ and $[Y_{T_{3R}}]_{\bf 10}$ indeed satisfy the Killing relations \eqref{14}.

%%%%%%%%%%%%%%%%%%%%%%%%%%%%%
\subsection{Properties of the Canonical Basis \label{PCB}}
%%%%%%%%%%%%%%%%%%%%%%%%%%%%%

There are four fundamental properties possessed by the canonical basis that make it particularly interesting. These are derived in the following.

%%%%%%%%%%%%%%%%%%%%%%%%%%%%%%
\subsubsection*{Wilson Lines and the MSSM: \label{MSSM}}
%%%%%%%%%%%%%%%%%%%%%%%%%%%%%%

Let us consider the two Wilson lines associated with the canonical basis. As abstract $Spin(10)$ group elements, these are 
\begin{equation}
\chi_{B-L} = e^{iY_{B-L}\frac{2\pi}{3}}, \quad \chi_{T_{3R}}=e^{iY_{T_{3R}}\frac{2\pi}{3}} \ .
\label{24}
\end{equation}
Note that $\chi_{B-L}^{3}=\chi_{T_{3R}}^{3}=1$ and, hence, each generates a finite ${\mathbb{Z}}_{3}$ subroup of $Spin(10)$. We identify these with the ${\mathbb{Z}}_{3} \times {\mathbb{Z}}_{3}$ isometry of the Calabi-Yau threefold $X$. When turned on simultaneously, these Wilson lines spontaneously break 
\begin{equation}
Spin(10) \longrightarrow SU(3)_{C} \times SU(2)_{L} \times U(1)_{T_{R}} \times U(1)_{B-L} \ .
\label{24a}
\end{equation}
As discussed previously, the ${\mathbb{Z}}_{3} \times {\mathbb{Z}}_{3}$ isometry acts equivariantly on the chosen vector bundle $V$ and, hence, the associated sheaf cohomology groups of tensor products of $V$ carry a representation of ${\mathbb{Z}}_{3} \times {\mathbb{Z}}_{3}$. To determine the zero modes of the Dirac operator twisted by $V$ and, hence, the low energy spectrum, one takes each $H^{1}(X,U_{R}(V))$, tensors it with the associated representation $R$, and then chooses the invariant subspace 
$(H^{1}(X,U_{R}(V)) \otimes R)^{{\mathbb{Z}}_{3} \times {\mathbb{Z}}_{3}}$. Let us carry this out for each of the relevant representations of $Spin(10)$.

For $R={\bf 16}$, the associated sheaf cohomology is
\begin{equation}
H^{1}(X,V)=RG^{\oplus3}~,
\label{25}
\end{equation}
where $RG$ is the regular representation of ${\mathbb{Z}}_{3} \times {\mathbb{Z}}_{3}$ given by
\begin{equation}
RG=1 \oplus  \chi_{1} \oplus \chi_{2} \oplus \chi_{1}^{2} \oplus \chi_{2}^{2} \oplus \chi_{1}\chi_{2} \oplus \chi_{1}^{2}\chi_{2} \oplus \chi_{1}\chi_{2}^{2} \oplus \chi_{1}^{2}\chi_{2} ^{2}
\label{26}
\end{equation}
and $\chi_{1}, \chi_{2}$ are the third roots of unity which generate the first and second factors of 
${\mathbb{Z}}_{3} \times {\mathbb{Z}}_{3}$. Note that 
\begin{equation}
h^{1}(X,V)=27
\label{27}
\end{equation}
and, hence, there are $27$ ${\bf 16}$ representations of $Spin(10)$ in the spectrum prior to turning on the Wilson lines. Choosing the Wilson line generators in the canonical basis \eqref{13}, it follows from \eqref{15} and \eqref{16} that the action of the Wilson lines \eqref{24} on each ${\bf 16}$ is given by
\begin{eqnarray}
{\bf 16}&=& \chi_{T_{3R}}^{2} \cdot  \chi_{B-L}^{2} ({\bf{\bar{3}},\bf {1}},-1,-1)\oplus   \chi_{T_{3R}} \cdot \chi_{B-L}^{2} ({\bf{\bar{3}},\bf {1}},1,-1) \label{28} \\
&& \oplus 1 \cdot  \chi_{B-L} ({\bf{3}},{\bf {2}},0,1) \oplus 1 \cdot 1({\bf{1}},{\bf {2}},0,-3) \oplus  \chi_{T_{3R}}^{2} \cdot 1 ({\bf{1}},{\bf {1}},-1,3) \nonumber\\
&& \oplus  \chi_{T_{3R}} \cdot 1({\bf{1}},{\bf {1}},1,3)~. \nonumber
\end{eqnarray}
Choosing $\chi_{1}=\chi_{T_{3R}}$ and $\chi_{2}=\chi_{B-L}$ in \eqref{26}, we find that $(H^{1}(X,V) \otimes {\bf 16})^{{\mathbb{Z}}_{3} \times {\mathbb{Z}}_{3}}$ consists of {\it three families} of quarks and leptons, each family transforming as
\begin{equation} 
Q= (U,D)^T = ({\bf{3}},{\bf {2}},0,\frac{1}{3}), \quad u=({\bf{\bar{3}},\bf {1}},-\frac{1}{2},-\frac{1}{3}), \quad d=({\bf{\bar{3}},\bf {1}},\frac{1}{2},-\frac{1}{3})
\label{29}
\end{equation}
and 
\begin{equation}
L=(N,E)^T=({\bf{1}},{\bf {2}},0,-1), \quad \nu=({\bf{1}},{\bf {1}},-\frac{1}{2},1), \quad e=({\bf{1}},{\bf {1}},\frac{1}{2},1)
\label{30}
\end{equation}
under $SU(3)_{C} \times SU(2)_{L} \times U(1)_{T_{3R}} \times U(1)_{B-L}$.

For $R={\bf 10}$ the associated sheaf cohomology is 
\begin{equation}
H^{1}(X,\wedge^{2}V)=\chi_{1} \oplus \chi_{1}^{2} \oplus \chi_{1}\chi_{2}^{2} \oplus \chi_{1}^{2}\chi_{2} \ .
\label{31}
\end{equation}
Note that 
\begin{equation}
h^{1}(X,\wedge^{2}V)=4
\label{32}
\end{equation}
and, hence, there are 4 ${\bf 10}$ representations of $Spin(10)$ in the spectrum prior to turning on the Wilson lines. Choosing the Wilson line generators in the canonical basis \eqref{13}, it follows from \eqref{21} and \eqref{22} that the action of the Wilson lines \eqref{24} on each ${\bf 10}$ is given by
\begin{eqnarray}
{\bf 10}&= &1 \cdot \chi_{B-L}^{2} ({\bf 3},{\bf 1},0,2) \oplus 1 \cdot \chi_{B-L} ({\bar{\bf 3}},{\bf 1},0,-2) \nonumber \\
&&\oplus \chi_{T_{3R}} \cdot 1({\bf 1},{\bf 2},1,0) \oplus \chi_{T_{3R}}^{2} \cdot 1 ({\bf 1},{\bf 2},-1,0)~. 
\label{33}
\end{eqnarray}
Choosing $\chi_{1}=\chi_{T_{3R}}$ and $\chi_{2}=\chi_{B-L}$ in \eqref{31}, we find that $(H^{1}(X,\wedge^{2}V) \otimes {\bf 10})^{{\mathbb{Z}}_{3} \times {\mathbb{Z}}_{3}}$ consists of a {\it single pair} of Higgs doublets transforming as
\begin{equation}
H=({\bf 1},{\bf 2},\frac{1}{2},0), \quad \bar{H}= ({\bf 1},{\bf 2},-\frac{1}{2},0) 
\label{34}
\end{equation}
under $SU(3)_{C} \times SU(2)_{L} \times U(1)_{T_{R}} \times U(1)_{B-L}$. These results lead to the following important property of the canonical basis.\\

\noindent $\bullet$ {\it When the two Wilson lines corresponding to the canonical basis are turned on simultaneously, the resulting low energy spectrum is precisely that of the MSSM--that is, three families of quark/lepton chiral superfields, each family with a right-handed neutrino supermultiplet, and one pair of Higgs-Higgs conjugate chiral multiplets. There are no vector-like pairs or exotic particles.}\\

The canonical basis exhibits a second, related, property that has important consequences for the the low energy effective Lagrangian. Consider, once again, the $R={\bf 16}$ case and the ${\mathbb{Z}}_{3} \times {\mathbb{Z}}_{3}$ invariant tensor product of $H^{1}(X,V)$ in \eqref{25},\eqref{26} with the ${\bf 16}$ decomposition in \eqref{28}. Note that, with the exception of $\chi_{B-L}$, $\chi_{T_{3R}}\chi_{B-L}^{2}$ and $\chi_{T_{3R}}^{2}\chi_{B-L}^{2}$ in \eqref{26} which project out all terms in \eqref{28}, each of the remaining six entries in each RG form a ${\mathbb{Z}}_{3} \times {\mathbb{Z}}_{3}$ invariant with only one term in a ${\bf 16}$. That is, each quark and lepton chiral multiplet in the low energy theory arises from a different ${\bf 16}$ representation of $Spin(10)$. Now consider the $R={\bf 10}$ case. It is easily seen from \eqref{31} and \eqref{33} that, with the exception of $\chi_{T_{3R}}\chi_{B-L}^{2}$ and $\chi_{T_{3R}}^{2}\chi_{B-L}$ in $\eqref{31}$ which project out all terms in \eqref{33}, the remaining two entries $\chi_{T_{3R}}$ and $\chi_{T_{3R}}$ each form a ${\mathbb{Z}}_{3} \times {\mathbb{Z}}_{3}$ invariant  with only a single component of \eqref{33}, the Higgs and Higgs conjugate chiral multiplets respectively. Hence, each arises from a different {\bf 10} representation of $Spin(10)$. This leads to the second important property of the canonical basis. \\

\noindent ${\bullet}$ {\it Since each quark/lepton and Higgs superfield  of the low energy Lagrangian arises from a different ${\bf 16}$ and ${\bf 10}$ representation of $Spin(10)$ respectively, the parameters of the effective theory, and specifically the Yukawa couplings and the soft supersymmetry breaking parameters, are uncorrelated by the $Spin(10)$ unification. For example, the soft mass squared parameters of the right-handed sneutrinos need not be universal with the remaining slepton supersymmetry breaking parameters.}

%%%%%%%%%%%%%%%%%%%%%%%%
\subsubsection*{The Kinetic Mixing Parameter: }
%%%%%%%%%%%%%%%%%%%%%%%%

Prior to turning on the ${\mathbb{Z}}_{3} \times {\mathbb{Z}}_{3}$ Wilson lines, the conventionally normalized kinetic energy part of the gauge field Lagrangian is $Spin(10)$ invariant and given by
\begin{equation}
{\cal{L}}_{kinetic}=\frac{-1}{2I_{R}}Tr(F^{a}T^{a}_{R})^{2}~,
\label{35}
\end{equation}
where $\{T^{a}_{R}, a=1,\dots.45\}$ is an orthogonal basis of ${\mathfrak{so}}(10)$ in any representation $R$ , each basis element Killing normalized to $\frac{1}{\sqrt{2}}$. In particular, defining
\begin{equation}
T^{1}= Y-\frac{1}{2}(B-L)=\frac{1}{2} Y_{T_{3R}}, \quad T^{2}=\sqrt{\frac{3}{8}}(B-L)=\frac{1}{2\sqrt{6}}Y_{B-L}
\label{36}
\end{equation}
we see from \eqref{14} that 
\begin{equation}
(T^{1}|T^{1})=(T^{2}|T^{2})=\frac{1}{2}, \quad (T^{1}|T^{2})=0
\label{37}
\end{equation}
and, hence,
\begin{equation}
{\cal{L}}_{kinetic}=-\frac{1}{4}(F_{\mu\nu}^{1})^{2}-\frac{1}{4}(F_{\mu\nu}^{2})^{2} +\dots ~.
\label{38}
\end{equation}
That is, there is no kinetic mixing term of the form $F_{\mu\nu}^{1} F^{2\mu\nu}$. This is a consequence of the fact that the canonical basis elements $Y_{T_{3R}}$ and $Y_{B-L}$ are Killing orthogonal, and is of little importance while $Spin(10)$ remains unbroken. However, if both Wilson lines are turned on simultaneously, the gauge group is spontaneously broken to $SU(3)_{C} \times SU(2)_{L} \times U(1)_{T_{3R}} \times U(1)_{B-L}$. For general $U(1) \times U(1)$, the two Abelian field strengths can exhibit  kinetic mixing; that is,
\begin{equation}
{\cal{L}}_{kinetic}=-\frac{1}{4}((F_{\mu\nu}^{1})^{2}+2\alpha F_{\mu\nu}^{1} F^{2\mu\nu}+(F_{\mu\nu}^{2})^{2} +\dots) ~.
\label{39}
\end{equation}
for some real parameter $\alpha$. However, for $U(1)_{T_{3R}} \times U(1)_{B-L}$ the normalized canonical generators satisfy \eqref{37} and, specifically, are orthogonal in ${\mathfrak{so}}(10)$. It follows that the initial value of $\alpha$ at the unification scale, $M_{u}$, must vanish. This is the third important property of the canonical basis.\\

\noindent $\bullet$ {\it Since the generators of the canonical basis are Killing orthogonal in
${\mathfrak{so}}(10)$, the value of the kinetic field strength mixing parameter $\alpha$ must vanish at the unification scale. That is, $\alpha(M_{u})=0$.}\\

Once the $Spin(10)$ symmetry is broken by both Wilson lines, either by turning them on at the same scale or sequentially, as discussed below, one expects the mixing parameter $\alpha$ to regrow due to radiative corrections. In this case, the Abelian field strengths develop a non-vanishing mixing term which greatly complicates the renormalization group analysis of the low energy effective theory. Radiative kinetic mixing has been discussed by a number of authors, see, for example, \cite{Babu:1996vt,delAguila:1988jz,Holdom:1985ag,Dienes:1996zr,Foot:1991kb, Fonseca:2011vn}. Let us briefly review the analysis. Consider a theory with unspecified $U(1) \times U(1)$ gauge factors. Then, in general, at an arbitrary momentum scale 
\begin{equation}
{\cal{L}}_{kinetic}=-\frac{1}{4}((F_{\mu\nu}^{1})^{2}+2\alpha F_{\mu\nu}^{1} F^{2\mu\nu}+(F_{\mu\nu}^{2})^{2} +\dots) ~.
\label{40}
\end{equation}
The associated gauge covariant derivative is given by
\begin{equation}
D=\partial-iT^{1}g_{1}A^{1}-iT^{2}g_{2}A^{2} \ ,
\label{41}
\end{equation}
where we denote the coupling parameters and gauge fields associated with $T^{1}$ and $T^{2}$ by $g_{1},A^{1}$ and $g_{2},A^{2}$ respectively. Defining new gauge fields by $\vec{A}={\cal{O}}\vec{A}^{\prime}$ where
\begin{equation}
{\cal{O}}=\frac{1}{\sqrt{2}}
\begin{pmatrix}
 1 & 1 \\
-1 & 1 
\end{pmatrix} 
\label{42}
\end{equation}
diagonalizes the kinetic energy terms to
\begin{equation}
{\cal{L}}_{kinetic}=-\frac{1}{4}((1-\alpha)(F_{\mu\nu}^{\prime 1})^{2}+(1+\alpha)(F_{\mu\nu}^{\prime 2})^{2} +\dots) ~.
\label{43}
\end{equation}
Further rescaling of the gauge fields by $\vec{A}^{\prime}={\cal{D}}^{-\frac{1}{2}}\vec{A}^{\prime\prime}$ with 
\begin{equation}
{\cal{D}}^{-\frac{1}{2}}=
\begin{pmatrix}
 \frac{1}{\sqrt{1-\alpha}} & 0 \\
 0 &   \frac{1}{\sqrt{1+\alpha}}
\end{pmatrix}
\label{44}
\end{equation}
leads to a canonically normalized kinetic term
\begin{equation}
{\cal{L}}_{kinetic}=-\frac{1}{4}((F_{\mu\nu}^{\prime \prime1})^{2}+(F_{\mu\nu}^{\prime \prime 2})^{2} +\dots) ~.
\label{45}
\end{equation}
However, the covariant derivative now has off-diagonal gauge couplings
\begin{equation}
D=\partial -i(T^{1},T^{2}) 
\begin{pmatrix}
\frac{g_{1}}{\sqrt{1-\alpha}} & \frac{g_{1}}{\sqrt{1+\alpha}} \\
\frac{-g_{2}}{\sqrt{1-\alpha}} & \frac{g_{2}}{\sqrt{1+\alpha}}
\end{pmatrix}
\begin{pmatrix}
A^{\prime \prime 1} \\
A^{\prime \prime 2} 
\end{pmatrix} \ .
\label{46}
\end{equation}
Note that the four gauge couplings are not independent, being functions of the three parameters $\alpha, g_{1}$ and $g_{2}$ in the original Lagrangian. It is not surprising, therefore, that a further field redefinition will eliminate one of them. The transformation should be orthogonal so as to leave the field strength kinetic term diagonal and canonically normalized. This can be achieved by setting $\vec{A}^{\prime\prime}
={\cal{P}}\vec{\cal{A}}$ where
\begin{equation}
{\cal{P}}=\frac{1}{\sqrt{2}}
\begin{pmatrix}
 \sqrt{1-\alpha} & - \sqrt{1+\alpha}  \\
 \sqrt{1+\alpha}  &  \sqrt{1-\alpha}  
\end{pmatrix} \ . 
\label{47}
\end{equation}
We find that the covariant derivative now becomes
\begin{equation}
D=\partial -i(T^{1},T^{2}) 
\begin{pmatrix}
{\cal{G}}_{1} & {\cal{G}}_{M} \\
0 & {\cal{G}}_{2}
\end{pmatrix}
\begin{pmatrix}
{\cal{A}}^{1} \\
{\cal{A}}^{2} 
\end{pmatrix} \ ,
\label{48}
\end{equation}
with
\begin{equation}
{\cal{G}}_{1}=g_{1}, \quad {\cal{G}}_{2}=\frac{g_{2}}{\sqrt{1-\alpha^{2}}}, \quad {\cal{G}}_{M}=\frac{-g_{1}\alpha}{\sqrt{1-\alpha^{2}}} \ .
\label{49}
\end{equation}
Note that in the limit that $\alpha \rightarrow 0$, ${\cal{G}}_{2}=g_{2}$ and ${\cal{G}}_{M}=0$.

The renormalization group equations for the gauge couplings in this ``upper triangular'' realization were given in \cite{Babu:1996vt}. Here, however, it suffices to present the RGE for the off-diagonal coupling ${\cal{G}}_{M}$. It is found to be
\begin{equation}
\frac{d{\cal{G}}_{M}}{dt}=\frac{1}{16\pi^{2}} \beta_{M}
\label{50}
\end{equation}
where
\begin{equation}
\beta_{M}={\cal{G}}_{2}^{2}{\cal{G}}_{M}B_{22}+{\cal{G}}_{M}^{3}B_{11}+2{\cal{G}}_{1}^{2}{\cal{G}}_{M}B_{11}+2{\cal{G}}_{2}{\cal{G}}_{M}^{2}B_{12}+{\cal{G}}_{1}^{2}{\cal{G}}_{2}B_{12}
\label{51}
\end{equation}
and
\begin{equation}
B_{ij}=Tr(T^{i}T^{j}) \ .
\label{52}
\end{equation}
The trace in \eqref{52} is over the entire matter and Higgs spectrum of the MSSM. Note that all of the terms in the $\beta$ function \eqref{51}, with the exception the last term, contain at least one power of ${\cal{G}}_{M}$. If the mixing parameter $\alpha$ and, hence, the off-diagonal coupling ${\cal{G}}_{M}$ vanish at some initial scale, as they will for our canonical basis, then the terms containing ${\cal{G}}_{M}$ will not, by themselves, generate a non-zero mixing parameter at any lower scale. However, a non-vanishing ${\cal{G}}_{M}$ will be generated by the last term. The only exception to this is if the charges $T^{1}$ and $T^{2}$ are such that
\begin{equation}
B_{12}=Tr(T^{1}T^{2})=0 \ .
\label{53}
\end{equation}
Generically, this will not be the case for arbitrary charges of $U(1) \times U(1)$; see Appendix A.3. However, let us break $Spin(10)$ to $U(1)_{T_{3R}}\times U(1)_{B-L}$ with both Wilson lines of the canonical basis. The associated normalized charges $T^{1}$ and $T^{2}$ were presented in \eqref{36} and satisfy
\begin{equation}
(T^{1}|T^{2})=0 \ .
\label{54}
\end{equation}
It then follows from \eqref{17} that 
\begin{equation}
Tr([T^{1}]_{R}[T^{2}]_{R})=0
\label{55}
\end{equation}
for any complete ${\mathfrak{so}}(10)$ representation $R$. Recalling that each quark/lepton family with a right-handed neutrino fills out a complete ${\bf 16}$ multiplet, one can conclude that
\begin{equation}
Tr([T^{1}]_{quarks/leptons}[T^{2}]_{quarks/leptons})=0 \ .
\label{56}
\end{equation}
However, in the reduction to the zero-mode spectrum the color triplet Higgs $H_{C}$ and ${\bar{H}}_{C}$ are projected out. Hence, the electroweak Higgs doublets $H$ and ${\bar{H}}$ do not make up a complete ${\bf 10}$ of ${\mathfrak{so}}(10)$. Therefore, the trace of $T^{1}T^{2}$ over the Higgs fields of the MSSM is not guaranteed to vanish. It is straightforward to compute this trace using \eqref{21} and \eqref{22}. If we ignore the color triplet components, then
\begin{equation}
[Y_{T_{3R}}]_{H,{\bar{H}}}= (1){\bf 1}_{2}\oplus (-1){\bf 1}_{2}
\label{57}
\end{equation}
and
\begin{equation}
[Y_{B-L}]_{H,{\bar{H}}}=(0){\bf 1}_{2}\oplus (0){\bf 1}_{2} \ .
\label{58}
\end{equation}
It then follows from \eqref{36} and the $0$-entries in \eqref{58} that
\begin{equation}
Tr([T^{1}]_{H,{\bar{H}}}[T^{2}]_{H,{\bar{H}}})= \frac{1}{4\sqrt{6}}Tr([Y_{T_{3R}}]_{H,{\bar{H}}}[Y_{B-L}]_{H,{\bar{H}}})=0 \ .
\label{59}
\end{equation}
We conclude from \eqref{56} and \eqref{59} that
\begin{equation}
B_{12}=0 \ .
\label{60}
\end{equation}
Therefore, for the canonical basis if the initial value of $\alpha$ and, hence, ${\cal{G}}_{M}$ vanish, then both will remain zero at any lower scale. This is the fourth important property possessed by the canonical basis.\\

\noindent $\bullet$ {\it The generators of the canonical basis are such that $Tr(T^{1}T^{2})=0$ when the trace is performed over the matter and Higgs spectrum of the MSSM. This guarantees that if the original kinetic mixing parameter vanishes, then $\alpha$ and, hence, ${\cal{G}}_{M}$ will remain zero under the RG at any scale. This property of not having kinetic mixing greatly simplifies the renormalization group analysis of the $SU(3)_{C} \times SU(2)_{L} \times U(1)_{T_{3R}}\times U(1)_{B-L}$ low energy theory}.\\

\subsection{Sequential Wilson Line Breaking \label{swb}}

In Subsection \ref{MSSM} we introduced the two Wilson lines $\chi_{B-L}$ and $\chi_{T_{3R}}$ associated with the canonical basis. As abstract $Spin(10)$ group elements, these were given in \eqref{24}. Each generates a finite ${\mathbb{Z}}_{3}$ subroup of $Spin(10)$, together representing the ${\mathbb{Z}}_{3} \times {\mathbb{Z}}_{3}$ isometry of the Calabi-Yau threefold $X$. When turned on {\it simultaneously}, they spontaneously break $Spin (10)$ to the gauge group $SU(3)_{C} \times SU(2)_{L} \times U(1)_{T_{3R}} \times U(1)_{B-L}$. The associated low energy spectrum was computed in Subsection \ref{MSSM} and found to be exactly that of the MSSM; that is, three families of quarks/leptons, each with a right-handed neutrino supermultiplet, as well as one pair of Higgs-Higgs conjugated superfields.

Since $\pi_{1}(X/({{\mathbb{Z}}_{3} \times {\mathbb{Z}}_{3}}))={\mathbb{Z}}_{3} \times {\mathbb{Z}}_{3}$, there are two independent classes of non-contractible curves in the quotient of $X$. Each Wilson line corresponds to the ${\mathbb{Z}}_{3}$ holonomy group of a flat gauge bundle wrapped around a curve in one of these classes.  A ``mass scale''  can be assigned to each Wilson line; namely, the inverse radius of the associated non-contractible curve. This radius will depend on the geometric moduli of the Calabi-Yau threefold, leading to a larger (smaller) mass scale for a smaller (larger) radius curve. It follows that the two Wilson lines can ``turn on'' at different scales, depending on the moduli of the geometry. There are three possibilities-- a) Both Wilson lines have approximately the same scale corresponding to the unification mass. This was the situation discussed in Subsection \ref{MSSM}. b) The $\chi_{B-L}$ Wilson line turns on at the unification mass, followed sequentially at a smaller scale by $\chi_{T_{3R}}$. c) The converse situation, that is, the  $\chi_{T_{3R}}$ Wilson line turns on at the unification mass followed by $\chi_{B-L}$ at a lower scale. If the mass scales of the Wilson lines are sufficiently separated, these scenarios can have different low energy precision predictions. In this subsection, we will explore the gauge groups and spectra associated with scenarios b) and c).

\subsubsection*{$ M_{\chi_{B-L}} > M_{\chi_{T_{3R}}}:$}

Recall from \eqref{13} that $Y_{B-L}=2(H_{1}+H_{2}+H_{3})$. By construction, $Y_{B-L}$ commutes with the generators of ${\mathfrak{su}}(3)_C \oplus {\mathfrak{su}}(2)_L$. Furthermore, it is clear that it annihilates the $\alpha^{5}$ root given in \eqref{12} and, hence, also commutes with ${\mathfrak{su}}(2)_R$.  It is straightforward to check that these are the only subalgebras of ${\mathfrak{so}}(10)$ that it commutes with. For example, the root associated with the remaining 
$\alpha^{3}$ node of the Dynkin diagram in Figure 1 is given by
\begin{equation}
\alpha^{3}=(0,0,1,-1,0) \ .
\label{61}
\end{equation}
This is clearly not annihilated by $Y_{B-L}$. That is, the commutant of $Y_{B-L}$ in ${\mathfrak{so}}(10)$ is ${\mathfrak{su}}(3)_C \oplus {\mathfrak{su}}(2)_L \oplus {\mathfrak{su}}(2)_R$.
It follows that when the $Y_{B-L}$ Wilson line is turned on, it spontaneously breaks 
\begin{equation}
Spin(10) \longrightarrow SU(3)_{C} \times SU(2)_{L} \times SU(2)_{R} \times U(1)_{B-L} \ ,
\label{62}
\end{equation}
the so-called left-right model~\cite{Mohapatra:1974gc, Senjanovic:1975rk, Senjanovic:1978ev}.
Let us now determine the zero-mode spectrum associated with this breaking. 

First consider the $R={\bf 16}$ representations of $Spin(10)$. The associated sheaf cohomology is given in \eqref{25} where, now, prior to turning on $\chi_{T_{3R}}$
\begin{equation}
RG=1 \oplus  1 \oplus  \chi_{B-L} \oplus 1 \oplus   \chi_{B-L}^{2} \oplus  \chi_{B-L} \oplus  \chi_{B-L} \oplus  \chi_{B-L}^{2} \oplus  \chi_{B-L}^{2} \ .
\label{63}
\end{equation}
Similarly, it follows from \eqref{15} that, prior to turning on $\chi_{T_{3R}}$, the action of $\chi_{B-L}$ on each ${\bf 16}$ representation is given by
\begin{eqnarray}
{\bf 16}&=& \chi_{B-L}^{2} ({\bf{\bar{3}},\bf {1}},{\bf 2},-1) \oplus \chi_{B-L} ({\bf{3}},{\bf {2}},{\bf 1},1)\label{64} \\
&&  \oplus 1({\bf{1}},{\bf {2}},{\bf 1},-3) \oplus  1({\bf{1}},{\bf {1}},{\bf 2},3) \ . \nonumber
\end{eqnarray}
We find from \eqref{63} and \eqref{64} that $(H^{1}(X,V) \otimes {\bf 16})^{{\mathbb{Z}}_{3}^{B-L}}$ consists of {\it nine families} of matter multiplets, each family transforming as
\begin{equation}
Q=({\bf 3},{\bf 2},{\bf 1},\frac{1}{3}), \quad Q_{R}=
\begin{pmatrix}
 d \\
 u 
\end{pmatrix} 
=({\bar{\bf 3}},{\bf 1}, {\bf 2}, -\frac{1}{3})
\label{65}
\end{equation}
and
\begin{equation}
L=({\bf 1}, {\bf 2},{\bf 1}, -1), \quad L_{R}=
\begin{pmatrix}
e \\
\nu 
\end{pmatrix} 
=({\bf 1},{\bf 1}, {\bf 2}, 1)
\label{66}
\end{equation}
under $SU(3)_{C} \times SU(2)_{L} \times SU(2)_{R} \times U(1)_{B-L} $.

For $R={\bf 10}$ the associated sheaf cohomology is given in \eqref{31} where, now, prior to turning on $\chi_{T_{3R}}$ 
\begin{equation}
H^{1}(X,\wedge^{2}V)=1 \oplus 1 \oplus \chi_{B-L}^{2} \oplus  \chi_{B-L} \ .
\label{67}
\end{equation}
Similarly, it follows from \eqref{21} that, prior to turning on $\chi_{T_{3R}}$, the action of 
$\chi_{B-L}$ on each {\bf 10} representation is given by
\begin{equation}
{\bf 10}= \chi_{B-L}^{2} ({\bf 3}, {\bf 1}, {\bf 1}, 2) \oplus \chi_{B-L} ({\bar{\bf 3}}, {\bf 1}, {\bf 1}, -2) \oplus 1({\bf 1}, {\bf 2}, {\bf 2}, 0) \ .
\label{68}
\end{equation}
Tensoring \eqref{67} and \eqref{68} together and taking the invariant subspace, we find that $(H^{1}(X,\wedge^{2}V) \otimes {\bf 10})^{{\mathbb{Z}}_{3}^{B-L}}$ consists of {\it two pairs} of Higgs-Higgs conjugate multiplets, each transforming as
\begin{equation}
\mathcal{H} =
\begin{pmatrix}
 H \\
 {\bar{H}}
\end{pmatrix} 
=({\bf 1}, {\bf 2}, {\bf 2}, 0) 
\label{69}
\end{equation}
and a {\it single pair} of colored Higgs-Higgs conjugate  multiplets transforming as 
\begin{equation}
H_{C}=({\bf 3}, {\bf 1}, {\bf 1}, \frac{2}{3}), \quad {\bar{H}}_{C}=({\bar{\bf 3}}, {\bf 1}, {\bf 1}, -\frac{2}{3})
\label{70}
\end{equation}
under  $SU(3)_{C} \times SU(2)_{L} \times SU(2)_{R} \times U(1)_{B-L} $. 

We conclude that in the ``intermediate'' energy region between the unification scale, $M_{u}=M_{\chi_{B-L}}$ , where $\chi_{B-L}$ is turned on and the intermediate scale, which we denote by $M_I = M_{\chi_{T_{3R}}}$, when $\chi_{T_{3R}}$ becomes significant, the effective theory will consist of the zero-modes given in \eqref{65},\eqref{66} and \eqref{69},\eqref{70}. The parameters of their Lagrangian are subject to several constraints. First, note that each $1$ in RG given in \eqref{63} forms a ${\mathbb{Z}}_{3}^{B-L}$ invariant with a pair $L \oplus L_{R}$ in the associated ${\bf 16}$. It follows that the components of each pair, there are nine pairs in total, will have correlated Yukawa and soft supersymmetry breaking parameters in the intermediate region due to the $Spin(10)$ unification symmetry. Second, even though each $Q_{R}$,$L_{R}$ and $\mathcal{H}$ arises from a unique ${\bf 16}$ and ${\bf 10}$  of $Spin(10)$ respectively, their components superfields will have identical Yukawa and soft supersymmetry breaking parameters due to the $SU(2)_{R}$ symmetry.

What happens when $\chi_{T_{3R}}$ is eventually switched on? By construction, this Wilson line commutes with $SU(3)_{C} \times SU(2)_{L} \times U(1)_{B-L} $, but will spontaneously break
\begin{equation}
SU(2)_{R} \longrightarrow U(1)_{T_{3R}} \ .
\label{71}
\end{equation}
Decomposing the $SU(2)_{R}$ component of the ${\bf 16}$ representations with respect to $U(1)_{T_{3R}}$ gives
\begin{eqnarray}
&& Q=({\bf 3},{\bf 2},{\bf 1},1) \rightarrow 1({\bf 3},{\bf 2}, 0,1) \nonumber \\
&& Q_{R}=({\bar{\bf 3}},{\bf 1}, {\bf 2}, -1) \rightarrow \chi_{T_{3R}}^{2}({\bar{\bf 3}},{\bf 1}, -1, -1) \oplus  \chi_{T_{3R}}({\bar{\bf 3}},{\bf 1}, 1, -1) \nonumber \\
&& L=({\bf 1}, {\bf 2},{\bf 1}, -3) \rightarrow 1({\bf 1}, {\bf 2},0, -3) \label{72} \\
&& L_{R}=({\bf 1},{\bf 1}, {\bf 2}, 3) \rightarrow \chi_{T_{3R}}^{2}({\bf 1},{\bf 1}, -1, 3) \oplus  \chi_{T_{3R}}({\bf 1},{\bf 1}, 1, 3) \nonumber
\end{eqnarray}
where we have explicitly displayed the action of $\chi_{T_{3R}}$. Inserting these expressions into \eqref{64} exactly reproduces the decomposition of the {\bf 16} given in \eqref{28}.
Similarly, the same decomposition of the $SU(2)_{R}$ component of the ${\bf 10}$ representations yields
\begin{eqnarray}
&& \mathcal{H}=({\bf 1}, {\bf 2}, {\bf 2}, 0) \rightarrow  \chi_{T_{3R}}({\bf 1}, {\bf 2}, 1, 0) \oplus  \chi_{T_{3R}}^{2}({\bf 1}, {\bf 2}, -1, 0) \nonumber \\
&& H_{C}=({\bf 3}, {\bf 1}, {\bf 1}, 2) \rightarrow 1({\bf 3}, {\bf 1}, 0, 2) \label{73} \\
&&  {\bar{H}}_{C}=({\bar{\bf 3}}, {\bf 1}, {\bf 1}, -2) \rightarrow  1({\bar{\bf 3}}, {\bf 1}, 0, -2) \nonumber
\end{eqnarray}
Inserting these expressions into \eqref{68} reproduces the decomposition of the {\bf 10} given in \eqref{33}. It follows that turning on the $\chi_{T_{3R}}$ Wilson line spontaneously breaks $SU(3)_{C} \times SU(2)_{L} \times SU(2)_{R} \times U(1)_{B-L} \rightarrow SU(3)_{C} \times SU(2)_{L} \times U(1)_{T_{3R}} \times U(1)_{B-L}$ with exactly the spectrum of the MSSM, as it must. Furthermore, the requirement of ${\mathbb{Z}}_{3}^{R}$ invariance chooses at most one of $L$ or $L_{R}$ from each $L\oplus L_{R}$ pair. Thus, their  correlation is broken since at most one of each pair will descend to lower energy. Additionally,
each MSSM field will arise as a ${\mathbb{Z}}_{3}^{R}$ invariant from a different intermediate region multiplet.  Hence, the correlation of the component fields of the $SU(2)_{R}$ doublets $Q_{R}$,$L_{R}$ and $\mathcal{H}$ will also be broken, since each doublet contributes at most one of its components to the MSSM.
Finally, we note that since $T^{1}=\frac{1}{2}Y_{T_{3R}}$ is embedded in the non-Abelian subalgebra ${\mathfrak{su}}(2)_{R}$, the associated gauge field strength cannot mix with the Abelian $T^{2}=\frac{1}{2\sqrt{6}}Y_{B-L}$ field strength in the intermediate regime. It follows that at the scale $M_{T_{3R}}$ the mixing parameter $\alpha$ and, hence, ${\cal{G}}_{M}$ must vanish.

Let us now consider the converse situation where $\chi_{T_{3R}}$ is turned on at the unification mass followed sequentially by $\chi_{B-L}$ at a lower scale.

\subsubsection*{$M_{\chi_{T_{3R}}} > M_{\chi_{B-L}}:$}

Recall from \eqref{13} that $Y_{T_{3R}}=H_{4}+H_{5}$. By construction, $Y_{T_{3R}}$ commutes with the generators of ${\mathfrak{su}}(3)_C \oplus {\mathfrak{su}}(2)_L$. However, it does not annihilate the remaining node  $\alpha^{3}$ of the Dynkin diagram in Figure 1. Hence, one might conclude that ${\mathfrak{su}}(3)_C \oplus {\mathfrak{su}}(2)_L$ is its largest non-Abelian commutant. However, examining all the remaining roots of ${\mathfrak{so}}(10)$ we find that $Y_{T_{3R}}$ annihilates the non-simple root $\alpha^{0}$ given by
\begin{equation}
\alpha^{0}=(-1,-1,0,0,0) \ ,
\label{74}
\end{equation}
as well as the 2 Weyl reflections of $\alpha^{0}$ around $\alpha^{1}, \alpha^{2}$ and their minuses. These 6 roots, along with the Abelian generator ${\cal{H}}_{0}=\frac{1}{2}(H_{1}+H_{2})$,  extend ${\mathfrak{su}}(3)_C $ to the fifteen dimensional subalgebra ${\mathfrak{su}}(4)_{C}$. This is shown in the ``extended'' Dynkin diagram presented in Figure 2.
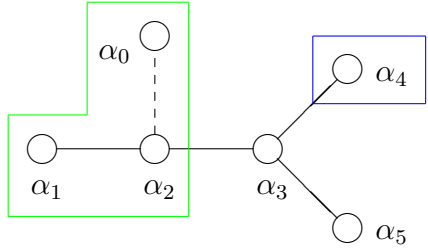
\begin{figure}[h]
\begin{center}
\begin{picture}(5,2.25)(0,.2)

\put(.625,1){\line(1,0){.75}}
\put(1.625,1){\line(1,0){.75}}
\put(2.588,.912){\line(1,-1){.53}}
\put(2.588,1.088){\line(1,1){.53}}
\put(.5,1){\circle{.25}}
\put(1.5,1){\circle{.25}}
\put(2.5,1){\circle{.25}}
\put(3.207,.293){\circle{.25}}
\put(3.207,1.707){\circle{.25}}

\put(.4,.6){$\alpha_1$}
\put(1.4,.6){$\alpha_2$}
\put(2.4,.6){$\alpha_3$}
\put(3.457,1.6){$\alpha_4$}
\put(3.457,.2){$\alpha_5$}

\put(.2,1.3){\color{green} \line(1,0){.7}}
\put(1.8,2.3){\color{green} \line(0,-1){1.9}}
\put(.9,2.3){\color{green} \line(1,0){.9}}
\put(.9,1.3){\color{green} \line(0,1){1}}
\put(.2,.4){\color{green} \line(1,0){1.6}}
\put(.2,1.3){\color{green} \line(0,-1){.9}}

\put(2.9,1.4){\color{blue} \line(1,0){1}}
\put(3.9,2){\color{blue} \line(0,-1){.6}}
\put(2.9,2){\color{blue} \line(1,0){1}}
\put(2.9,2){\color{blue} \line(0,-1){.6}}

\multiput(1.5,1.125)(0,.15){5}{\line(0,1){.08}}
\put(1.5,2){\circle{.25}}
\put(1,1.8){$\alpha_0$}

\end{picture}
\end{center}
\label{fig:2}
\caption{\small The extended Dynkin diagram of $Spin(10)$ with the ${\mathfrak{su}}(4)_{C}\supset \mathfrak{su}(3)_{C}$ and $\mathfrak{su}(2)_{L}$ subgroups highlighted in green and blue respectively. }
\end{figure}\\
Therefore, the commutant of $Y_{T_{3R}}$ is  actually the enlarged subalgebra ${\mathfrak{su}}(4) _{C}\oplus {\mathfrak{su}}(2)_L$. 
It follows that when the $Y_{T_{3R}}$ Wilson line is turned on, it spontaneously breaks 
\begin{equation}
Spin(10) \longrightarrow SU(4)_{C} \times SU(2)_{L} \times U(1)_{T_{3R}} \ ,
\label{75}
\end{equation}
a gauge group closely related to the so-called Pati-Salam model\footnote{Pati-Salam contains the full $SU(2)_R$ group instead of $U(1)_{T_{3R}}$.}~\cite{Pati:1974yy}.
We note that $Y_{B-L}$ does not annihilate $\alpha^{0}$ and, hence, $U(1)_{B-L} \subset SU(4)_{C}$.
Let us now determine the zero-mode spectrum associated with this breaking. 

First consider the $R={\bf 16}$ representations of $Spin(10)$. The associated sheaf cohomology is given in \eqref{26} where, now, prior to turning on $\chi_{B-L}$
\begin{equation}
RG=1 \oplus  \chi_{T_{3R}} \oplus  1 \oplus  \chi_{T_{3R}}^{2} \oplus  1 \oplus  \chi_{T_{3R}} \oplus  \chi_{T_{3R}}^{2} \oplus  \chi_{T_{3R}} \oplus  \chi_{T_{3R}}^{2} \ .
\label{76}
\end{equation}
Similarly, it follows from \eqref{16} that, prior to turning on $\chi_{B-L}$, the action of $\chi_{T_{3R}}$ on each ${\bf 16}$ representation is given by
\begin{equation}
{\bf 16}=1({\bf 4},{\bf 2},0) \oplus \chi_{T_{3R}}^{2}({\bar{\bf{4}}},{\bf 1}, -1) \oplus  \chi_{T_{3R}}({\bar{\bf{4}}},{\bf 1}, 1) \ .
\label{77}
\end{equation}
We find from \eqref{76} and \eqref{77} that $(H^{1}(X,V) \otimes {\bf 16})^{{\mathbb{Z}}_{3}^{R}}$ consists of {\it nine families} of matter multiplets, each family transforming as
\begin{equation}
\begin{pmatrix}
 Q \\
 L 
\end{pmatrix} 
=({\bf 4},{\bf 2},0), \quad
\begin{pmatrix}
 u \\
 \nu 
\end{pmatrix} 
=({\bar{\bf{4}}},{\bf 1}, -\frac{1}{2}), \quad
\begin{pmatrix}
 d \\
 e 
\end{pmatrix} 
=({\bar{\bf{4}}},{\bf 1}, \frac{1}{2})  
\label{78}
\end{equation}
under  $SU(4)_{C} \times SU(2)_{L} \times U(1)_{T_{3R}}$.

For $R={\bf 10}$ the associated sheaf cohomology is given in \eqref{31} where, now, prior to turning on $\chi_{B-L}$ 
\begin{equation}
H^{1}(X,\wedge^{2}V)=  \chi_{T_{3R}} \oplus  \chi_{T_{3R}}^{2} \oplus \chi_{T_{3R}}  \oplus  \chi_{T_{3R}}^{2}  \ .
\label{79}
\end{equation}
Similarly, it follows from \eqref{22} that, prior to turning on $\chi_{B-L}$, the action of 
$\chi_{T_{3R}}$ on each {\bf 10} representation is given by
\begin{equation}
{\bf 10}=1({\bf 6},{\bf 1},0) \oplus \chi_{T_{3R}}({\bf 1},{\bf 2}, 1) \oplus \chi_{T_{3R}}^{2}({\bf 1},{\bf 2}, -1) \ .
\label{80}
\end{equation}
Tensoring \eqref{79} and \eqref{80} together and taking the invariant subspace, we find that $(H^{1}(X,\wedge^{2}V) \otimes {\bf 10})^{{\mathbb{Z}}_{3}^{R}}$ consists of {\it two pairs} of electroweak Higgs-Higgs conjugate multiplets, each transforming as
\begin{equation}
H=({\bf 1},{\bf 2}, \frac{1}{2}), \quad {\bar{H}}=({\bf 1},{\bf 2}, -\frac{1}{2}) 
\label{81}
\end{equation}
under $SU(4)_{C} \times SU(2)_{L} \times U(1)_{T_{3R}}$.

We conclude that in the ``intermediate'' energy region between the unification scale, $M_{u}=M_{\chi_{T_{3R}}}$, where $\chi_{T_{3R}}$ is turned on and the intermediate scale, $M_I = M_{\chi_{B-L}}$, when $\chi_{B-L}$ becomes significant, the effective theory will consist of the zero-modes given in \eqref{78} and \eqref{81}. Note that even though each matter multiplet in the intermediate region arises from a different ${\bf 16}$ of $Spin(10)$, their component fields have correlated Yukawa and soft supersymmetry breaking parameters due to the $SU(4)_{C}$ symmetry.

What happens when $\chi_{B-L}$ is eventually switched on? By construction, this Wilson line commutes with $ SU(2)_{L} \times U(1)_{T_{3R}} $, but will spontaneously break
\begin{equation}
SU(4)_{C} \longrightarrow SU(3)_{C} \times U(1)_{B-L} \ .
\label{82}
\end{equation}
Decomposing the $SU(4)_{C}$ component of the ${\bf 16}$ representation with respect to $SU(3)_{C} \times U(1)_{B-L}$ gives
\begin{eqnarray}
&&\begin{pmatrix}
 Q \\
 L 
\end{pmatrix} 
=({\bf 4},{\bf 2},0) \rightarrow \chi_{B-L}({\bf 3},{\bf 2},0,1) \oplus 1({\bf 1},{\bf 2},0,-3) \nonumber \\
&&\begin{pmatrix}
 u \\
 \nu 
\end{pmatrix} 
=({\bar{\bf{4}}},{\bf 1}, -1) \rightarrow \chi_{B-L}^{2}({\bar{\bf 3}},{\bf 1},-1,-1) \oplus 1({\bf 1},{\bf 1},-1,3) \label{83} \\
&&\begin{pmatrix}
 d \\
 e 
\end{pmatrix} 
=({\bar{\bf{4}}},{\bf 1}, 1) \rightarrow  \chi_{B-L}^{2}({\bar{\bf 3}},{\bf 1},1,-1) \oplus 1({\bf 1},{\bf 1},1,3)\nonumber
\end{eqnarray}
where we have explicitly displayed the $\chi_{B-L}$ representation. Inserting these expressions into \eqref{77} exactly reproduces the decomposition of the {\bf 16} given in \eqref{28}.
Similarly, the same decomposition of the $SU(4)_{C}$ component of the ${\bf 10}$ representations yields
\begin{eqnarray}
&&({\bf 6},{\bf 1},0) \rightarrow  \chi_{B-L}^{2}({\bf 3}, {\bf 1},0,2) \oplus  \chi_{B-L}({\bar{\bf 3}}, {\bf 1},0,-2) \nonumber \\
&&H=({\bf 1},{\bf 2},1) \rightarrow 1({\bf 1},{\bf 2},1,0) \label{84} \\
&&{\bar{H}}=({\bf 1},{\bf 2},-1) \rightarrow 1({\bf 1},{\bf 2},-1,0) \nonumber
\end{eqnarray}
Inserting these expressions into \eqref{80} reproduces the decomposition of the {\bf 10} given in \eqref{33}. It follows that turning on the $\chi_{B-L}$ Wilson line spontaneously breaks $SU(4)_{C} \times SU(2)_{L} \times U(1)_{T_{3R}} \rightarrow SU(3)_{C} \times SU(2)_{L} \times U(1)_{T_{3R}} \times U(1)_{B-L}$ with exactly the spectrum of the MSSM, as it must. Furthermore, each MSSM field will arise as a ${\mathbb{Z}}_{3}^{B-L}$ invariant from a different intermediate region multiplet. Hence, the correlation between the components of the $SU(4)_{C}$ ${\bf 4}$ matter multiplets is lost. Finally, we note that since $T^{2}=\frac{1}{2\sqrt{6}}Y_{B-L}$ is embedded in the non-Abelian subalgebra ${\mathfrak{su}}(4)_{C}$, the associated field strength cannot mix with the Abelian $T^{1}=\frac{1}{2}Y_{T_{3R}}$ field strength in the intermediate region. It follows that at the scale $M_{\chi_{B-L}}$ the mixing parameter $\alpha$ and, hence, ${\cal{G}}_{M}$ must vanish.\\

Both sequential breaking patterns, specifying the gauge groups and the associated zero-mode spectra, are shown schematically in Figure 3.
\setlength{\unitlength}{.9cm}
\begin{figure}[!ht]
\begin{center}
\scriptsize
\begin{picture}(5,10)(0,1)

\put(1.5,10){\color{blue}$\underline{Spin(10)}$}
\put(-1,9.5){\color{red}\line(1,0){6}}
\put(-2,9.4){\color{red} $M_u$}
\put(2.25,9.5){\vector(2,-1){3}}
\put(1.75,9.5){\vector(-2,-1){3}}
\put(-.5,9){$\chi_{T_{3R}}$}
\put(4.1,9){$\chi_{T_{B-L}}$}

%Pati-Salam type model
\put(-3.5,7.5){\color{blue}$\underline{SU(4)_C\times SU(2)_L\times U(1)_{T_{3R}}}$}
\put(-3,6.5){$\trix{c}Q\\L\notrix=(\textbf{4},\textbf{2},0)$}
\put(-3,5.5){$\trix{c}u\\ \nu\notrix=(\bar {\textbf{4}},\textbf{1},-\frac{1}{2})$}
\put(-3,4.5){$\trix{c}d\\e\notrix=(\bar {\textbf{4}},\textbf{1},\frac{1}{2})$}
\put(-3,3.5){$H=(\textbf{1},\textbf{2},\frac{1}{2})$}
\put(-3,3){$\bar H=(\textbf{1},\textbf{2},-\frac{1}{2})$}

\put(-3,7){\line(-1,0){.2}}
\put(-3,4.1){\line(-1,0){.2}}
\put(-3.2,7){\line(0,-1){2.9}}
\put(-3.2,5.6){\line(-1,0){.2}}
\put(-4,5.5){\textbf{16}}
\put(0,6.8){\oval(.4,.4)[tr]}
\put(.2,6.8){\line(0,-1){1}}
\put(.4,5.8){\oval(.4,.4)[bl]}
\put(.4,5.4){\oval(.4,.4)[tl]}
\put(.2,5.4){\line(0,-1){1.1}}
\put(0,4.3){\oval(.4,.4)[br]}
\put(.6,5.5){$\times 9$}

\put(-3,3.8){\line(-1,0){.2}}
\put(-3,2.9){\line(-1,0){.2}}
\put(-3.2,3.8){\line(0,-1){.9}}
\put(-3.2,3.4){\line(-1,0){.2}}
\put(-4,3.3){\textbf{10}}
\put(.1,3.7){\oval(.2,.2)[tr]}
\put(.2,3.7){\line(0,-1){.3}}
\put(.3,3.4){\oval(.2,.2)[bl]}
\put(.3,3.2){\oval(.2,.2)[tl]}
\put(.2,3.2){\line(0,-1){.3}}
\put(.1,2.9){\oval(.2,.2)[br]}%\qbezier(0,3.8)(0.2,3.4)(0,2.9)
\put(.6,3.3){$\times 2$}

%left right
\put(3.5,7.5){\color{blue}$\underline{SU(3)_C\times SU(2)_L\times SU(2)_R\times U(1)_{B-L}}$}
\put(4,6.5){$L=(\textbf{1},\textbf{2},\textbf{1},-1)$}
\put(4,5.9){$L_R=(\textbf{1},\textbf{1},\textbf{2},1)$}
\put(4,5.3){$Q=(\textbf{3},\textbf{2},\textbf{1},\frac{1}{3})$}
\put(4,4.7){$Q_R=(\bar{\textbf{3}},\textbf{1},\textbf{2},-\frac{1}{3})$}
\put(4,3.9){$\mathcal{H}=(\textbf{1},\textbf{2},\textbf{2},0)$}
\put(4,3.3){$H_C=(\textbf{3},\textbf{1},\textbf{1},\frac{2}{3})$}
\put(4,2.7){${\bar{H}}_C=(\bar{\textbf{3}},\textbf{1},\textbf{1},-\frac{2}{3})$}

\put(4,6.8){\line(-1,0){.2}}
\put(4,4.6){\line(-1,0){.2}}
\put(3.8,6.8){\line(0,-1){2.2}}
\put(3.8,5.7){\line(-1,0){.2}}
\put(3,5.6){\textbf{16}}
\put(7,6.6){\oval(.4,.4)[tr]}
\put(7.2,6.6){\line(0,-1){.7}}
\put(7.4,5.9){\oval(.4,.4)[bl]}
\put(7.4,5.5){\oval(.4,.4)[tl]}
\put(7.2,5.5){\line(0,-1){.7}}
\put(7,4.8){\oval(.4,.4)[br]}
%\qbezier(8,6.8)(8.5,5.9)(8,4.9)
\put(7.6,5.6){$\times 9$}

\put(4,4.2){\line(-1,0){.2}}
\put(4,2.6){\line(-1,0){.2}}
\put(3.8,4.2){\line(0,-1){1.6}}
\put(3.8,3.4){\line(-1,0){.2}}
\put(3,3.3){\textbf{10}}
\put(7.1,4.2){\oval(.2,.2)[tr]}
\put(7.2,4.2){\line(0,-1){.1}}
\put(7.3,4.1){\oval(.2,.2)[bl]}
\put(7.3,3.9){\oval(.2,.2)[tl]}
\put(7.2,3.9){\line(0,-1){.1}}
\put(7.1,3.8){\oval(.2,.2)[br]}
%\put(7,3.9){\big \}}
%\qbezier(8,4.8)(8.2,4.55)(8,4.3)
\put(7.6,3.85){$\times 2$}

\put(-3,2.5){\color{red}\line(1,0){4}}
\put(-5.3,2.4){\color{red}$M_{\chi_{B-L}}=M_I$}
\put(3,2){\color{red}\line(1,0){4}}
\put(7.2,1.9){\color{red}$M_{\chi_{T_{3R}}}=M_I$}
\put(-1.75,2.5){\vector(2,-1){3}}
\put(5.75,2){\vector(-3,-1){3}}
\put(3.5,1.7){$\chi_{T_{3R}}$}
\put(.1,2){$\chi_{T_{B-L}}$}

%SU(3) x SU(2) x U(1)B-L x U(1)T3R
\put(-1,.5){\color{blue}$\underline{SU(3)_C\times SU(2)_L\times U(1)_{T_{3R}}\times U(1)_{B-L}}$}
\put(-.5,-.5){$L=(\textbf{1},\textbf{2},0,-1)$}
\put(-.5,-1){$e=(\textbf{1},\textbf{1},\frac{1}{2},1)$}
\put(-.5,-1.5){$\nu=(\textbf{1},\textbf{1},-\frac{1}{2},1)$}
\put(-.5,-2){$Q=(\bar{\textbf{3}},\textbf{2},0,\frac{1}{3})$}
\put(-.5,-2.5){$u=(\textbf{3},\textbf{1},-\frac{1}{2},-\frac{1}{3})$}
\put(-.5,-3.0){$d=(\textbf{3},\textbf{1},\frac{1}{2},-\frac{1}{3})$}
\put(-.5,-3.5){$H=(\textbf{1},\textbf{2},\frac{1}{2},0)$}
\put(-.5,-4.0){$\bar H=(\textbf{1},\textbf{2},-\frac{1}{2},0)$}

\put(-.5,-.2){\line(-1,0){.2}}
\put(-.5,-3.1){\line(-1,0){.2}}
\put(-.7,-.2){\line(0,-1){2.9}}
\put(-.7,-1.7){\line(-1,0){.2}}
\put(-1.4,-1.8){\textbf{16}}
\put(3,-.4){\oval(.4,.4)[tr]}
\put(3.2,-.4){\line(0,-1){1.1}}
\put(3.4,-1.5){\oval(.4,.4)[bl]}
\put(3.4,-1.9){\oval(.4,.4)[tl]}
\put(3.2,-1.9){\line(0,-1){1}}
\put(3,-2.9){\oval(.4,.4)[br]}
%\qbezier(3,-.2)(3.5,-1.4)(3,-2.6)
\put(3.6,-1.8){$\times 3$}

\put(-.5,-3.2){\line(-1,0){.2}}
\put(-.5,-4.1){\line(-1,0){.2}}
\put(-.7,-3.2){\line(0,-1){.9}}
\put(-.7,-3.7){\line(-1,0){.2}}
\put(-1.4,-3.8){\textbf{10}}

\put(6.6,-1.5){MSSM}
\put(7,-2){+}
\put(5.3,-2.5){3 right-handed neutrino}
\put(6.0,-2.9){supermultiplets}

\end{picture}
\end{center}
\vspace{5cm}
\caption{\small The two sequential Wilson line breaking patterns of $Spin(10)$. The unification and intermediate masses are specified, as well as the particle spectra in the associated scaling regimes.}
\label{fig:3}
\end{figure}
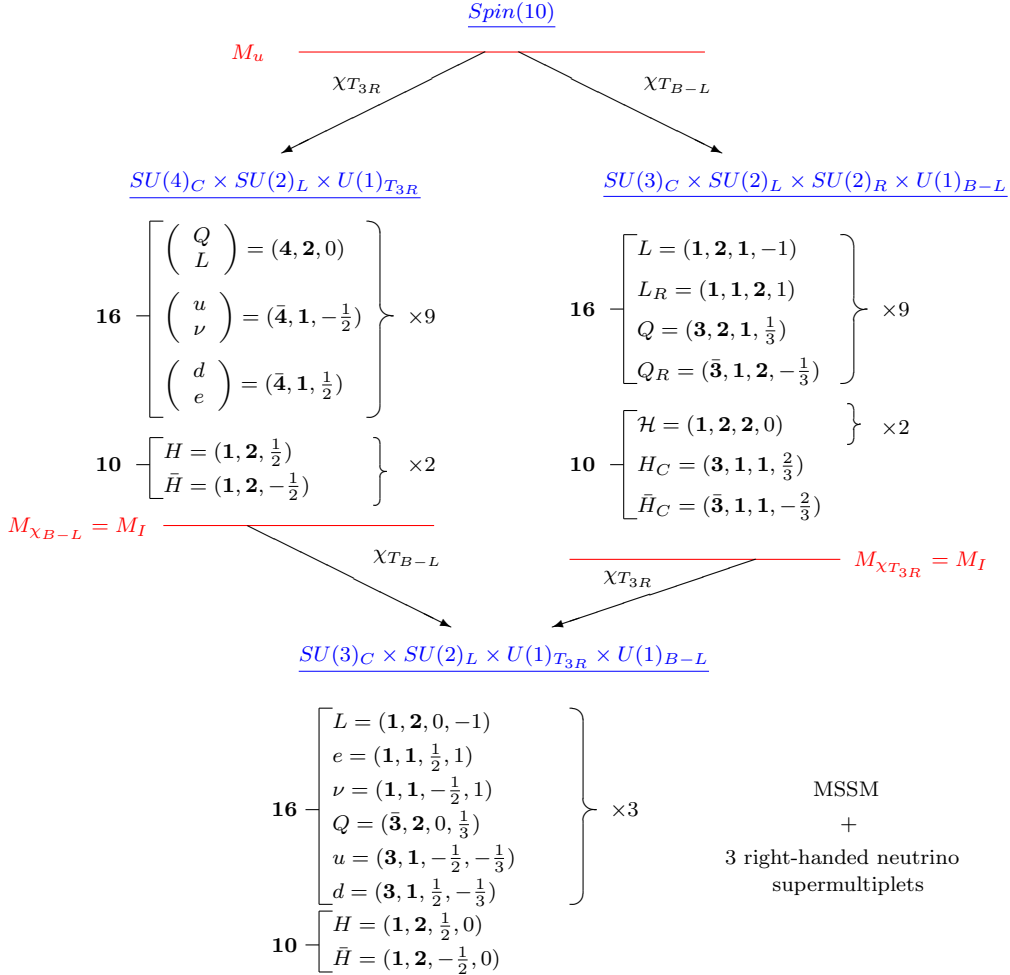
%

%%%%%%%%%%%%%%%%%%%%%%%%%%%%%%%%%%%
\section{Gauge Coupling Unification and the Intermediate Mass Scales}
%%%%%%%%%%%%%%%%%%%%%%%%%%%%%%%%%%%

Within the context of the explicit Wilson line breaking scenarios introduced above, we now present the renormalization group analysis of the gauge coupling parameters. These are chosen for discussion since
%, unlike the gaugino masses, Yukawa couplings and soft supersymmetry breaking parameters,   
their RG running does not depend on introducing initial values for the soft breaking parameters. It is clear from the previous section that the RG flow must be integrated over several distinct regimes, each with a different gauge group and multiplet content and, hence, different gauge couplings and beta functions. We begin, therefore, by carefully elucidating all relevant mass scales and the scaling regimes between them.

%%%%%%%%%%%%%
\subsection{Mass Scales}
%%%%%%%%%%%%%

From the top down, the important mass scales are the following.

\subsubsection*{The Compactification/Unification Scale $(M_{u})$:} The transition from ten-dimensional string theory to the four-dimensional effective field theory is not ``sharp''. Rather, it occurs over a small interval in energy-momentum, roughly centered around the ``average'' inverse radius of the compactification manifold $X/({\mathbb{Z}}_{3} \times {\mathbb{Z}}_{3})$. This compactifiction scale, $M_{c}$, is clearly dependent on the geometric moduli and, hence, difficult to determine directly from string theory. The $E_{8}$ gauge symmetry of the heterotic string is spontaneously broken to $Spin(10)$ by the $SU(4)$ structure group of the vector bundle $V$. Ignoring any further breaking by Wilson lines,  what emerges at the low energy end of the transition region is an effective field theory with unified gauge group $Spin(10)$. We denote this gauge ``unification scale'' by $M_{u}$. If one ignores string ``threshold effects'' due to the transition, then we can identify $M_{c} \simeq M_{u}$. These corrections are expected to be small and, in any case, are not the subject of the present paper. Henceforth, we make this identification.\\

%%%%%%%%%%%%%%%%%%%%%%%%%%%%%%%%
\subsubsection*{The Wilson Line/Intermediate Scale $(M_{I})$:}
%%%%%%%%%%%%%%%%%%%%%%%%%%%%%%%%

If the compactification manifold is relatively ``round'', one expects the inverse radii of the all non-contractible curves in $X/({\mathbb{Z}}_{3} \times {\mathbb{Z}}_{3})$ to be approximately that of the average inverse radius. That is, $M_{\chi_{T_{3R}}} \simeq M_{\chi_{B-L}} \simeq M_{u}$. In this case, both Wilson lines turn on simultaneously and break $Spin(10)$ down to the low energy gauge group $SU(3)_{C} \times SU(2)_{L} \times U(1)_{T_{3R}} \times U(1)_{B-L}$ at the unification scale.

However, if the manifold is in a region of its moduli space for which one of the non-contractable inverse radii is much larger than the other, then the Wilson lines will turn on at different scales. Furthermore, ignoring threshhold effects, it is clear that the larger Wilson line mass should be identified with the unification scale $M_{u}$. In this case, there are two possibilities. First, if $M_{u} \simeq M_{\chi_{B-L}} > M_{\chi_{T_{3R}}}$ then $Spin(10)$ is broken to 
$SU(3)_{C} \times SU(2)_{L} \times SU(2)_{R} \times U(1)_{B-L}$ at $M_{c}$. The second case occurs for $M_{u} \simeq M_{\chi_{T_{3R}}} >  M_{\chi_{B-L}}$. Now $Spin(10)$ is broken at $M_{u}$ to $SU(4)_{C} \times SU(2)_{L} \times U(1)_{T_{3R}}$.

The smaller of the two Wilson line masses introduces a new mass scale, $M_{I}$, into the analysis. Below this ``intermediate'' scale, independently of whether $M_{I} \simeq M_{\chi_{T_{3R}}}$ or
 $M_{I} \simeq M_{\chi_{B-L}}$, the theory is spontaneneously broken to $SU(3)_{C} \times SU(2)_{L} \times U(1)_{T_{3R}} \times U(1)_{B-L}$ with precisely the spectrum of the MSSM with three generations of right-handed neutrino supermultiplets. Of course, in the case where the Wilson line masses are degenerate, the intermediate mass $M_{I}$ is simply identified with the unification scale $M_{u}$.

%%%%%%%%%%%%%%%%%%%%%%%%%%%%
\subsubsection*{The B-L Breaking Scale $(M_{B-L})$:} 
%%%%%%%%%%%%%%%%%%%%%%%%%%%%

At some scale considerably below $M_{I}$, but above the electroweak scale, the $U(1)_{T_{3R}} \times U(1)_{B-L}$ symmetry must be spontaneously broken to $U(1)_{Y}$. Let us briefly analyze how this breaking takes place, the scale $M_{B-L}$ at which it occurs and the boundary conditions it places on the relevant gauge parameters. 

It follows from \eqref{48} and the fact that ${\cal{G}}_{M}=0$ in the canonical basis, that the $U(1)_{T_{3R}} \times U(1)_{B-L}$ part of the covariant derivative is given by
\begin{eqnarray}
\nonumber
D&=&\partial -i(T^{1},T^{2}) 
\begin{pmatrix}
g_{3R} & 0 \\
0 & g_{BL}
\end{pmatrix}
\begin{pmatrix}
W^{0}_{R} \\
B_{B-L} 
\end{pmatrix}\\
&=&\partial -i
\begin{pmatrix}
Y-\frac{1}{2}(B-L), \sqrt{\frac{3}{8}}(B-L)
\end{pmatrix}
\begin{pmatrix}
g_{3R} & 0 \\
0 & g_{BL}
\end{pmatrix}
\begin{pmatrix}
W^{0}_{R} \\
B_{B-L} 
\end{pmatrix}.
\label{85}
\end{eqnarray}
The generators $T^{1},T^{2}$ were defined in \eqref{36}, $g_{3R}$ and $g_{BL}$ are the gauge parameters and we denote the gauge bosons associated with $U(1)_{T_{3R}} \times U(1)_{B-L}$ as $W^{0}_{R}$ and $B_{B-L}$ respectively. To simplify notation, it is useful to define
\begin{equation}
g_{BL}^{\prime}=\sqrt{\frac{3}{2}} g_{BL} \ .
\label{86}
\end{equation}
Recall that the matter spectrum of the $SU(3)_{C} \times SU(2)_{L} \times U(1)_{T_{3R}} \times U(1)_{B-L}$ theory contains three right-handed neutrino chiral supermultiplets, one per family, each of which has charge $(-1/2,1)$ under $U(1)_{T_{3R}} \times U(1)_{B-L}$. 
The potential of one of the associated  sneutrinos, ${\tilde{\nu}}$, can be approximated by 
\begin{equation}
V=m_{{\tilde{\nu}}}^{2}|{{\tilde{\nu}}}|^{2}+\frac{1}{8}(g^{\prime 2}_{BL}+g_{3R}^{2})|{\tilde{\nu}}|^{4} \ ,
\label{87}
\end{equation}
where $m_{{\tilde{\nu}}}^{2}$ is the soft supersymmetry breaking squared mass. As shown in detail in \cite{Ambroso:2009jd,Ambroso:2009sc,Ambroso:2010pe}, $m_{{\tilde{\nu}}}^{2}$, although positive at $M_{I}$, can become negative under the RG at a lower mass scale. It follows that potential \eqref{87} develops a vacuum expectation value given by
\begin{equation}
v_{R}=\sqrt{\frac{-8m_{{\tilde{\nu}}}^{2}}{g^{\prime 2}_{BL}+g_{3R}^{2}}} \ ,
\label{88}
\end{equation}
where $\langle {\tilde{\nu}} \rangle=1/\sqrt{2} v_{R}$. 
This breaks $U(1)_{T_{3R}} \times U(1)_{B-L}$ to a $U(1)$ subgroup. Although growing over a range, $v_{R}$ ``freezes out'', that is, stops significantly evolving, when the energy momentum becomes smaller than the associated sneutrino mass. This mass, which we denote by 
$M_{B-L}$, can be identified with the scale of $U(1)_{T_{3R}} \times U(1)_{B-L}$ breaking.

To analyze this breaking, expand the theory around the sneutrino expectation value. 
In the basis $(W_R^0,B_{B-L})$, the matrix for the squared masses of the gauge bosons is
\begin{equation}
	M_\mathcal Z =
	\begin{pmatrix}
		\frac{1}{4} g_{3R} v_R^2
		&
		-\frac{1}{4} g_{BL}' \, g_{3R} \, v_R^2
	\\
		-\frac{1}{4} g_{BL}' \, g_{3R} \, v_R^2
		&
		\frac{1}{4} g_{BL}'^2 v_R^2
	\end{pmatrix} .
	\label{89}
\end{equation}
This can be diagonalized into the physical states
\begin{align}
	&Z_{3R}  =   \cos \theta_R W_R^0 -\sin \theta_R B_{B-L}, \label{90}
\\
	&B  = \sin \theta_R W_R^0 +\cos \theta_R B_{B-L}   \nonumber
\end{align}
with masses
\begin{eqnarray}
&&	M_{Z_{3R}}^2  = \frac{1}{4} \left( g_{3R}^2 + g_{BL}'^2 \right) v_R^2 = 2 | m_{ {\tilde{\nu}}}|^2, \quad M^{2}_{B}=0 \label{91} 
\\
&&\qquad \qquad 	\cos \theta_R  = \frac{g_{3R}}{\sqrt{g_{3R}^2 + g_{BL}'^2}} \ . \label{92}
\end{eqnarray}
Note that the non-vanishing $Z_{3R}$ mass is closely associated with the soft supersymmetry breaking mass of the sneutrino. $Z_{3R}$ is generically called a $Z'$ boson and $B$ is the massless gauge boson of the unbroken $U(1)$ gauge symmetry. To determine exactly which symmetry has been broken and which persists, rewrite the covariant derivative \eqref{85} in the physical basis. Using \eqref{36}, \eqref{86},\eqref{90} and \eqref{92}, we find that
\begin{equation}
D=\partial  -iYg_{Y} B -i\big(Y\cos^{2}\theta_{R}-\frac{1}{2}(B-L))g_{Z_{3R}}Z_{3R} \ ,
\label{93}
\end{equation}
where
\begin{equation}
g_{Y}= \frac{g^{\prime}_{BL}g_{3R}}{\sqrt{g_{3R}^2 + g_{BL}'^2}}, \quad g_{Z_{3R}}=\sqrt{g_{3R}^{2}+g_{BL}^{'2}} \ .
\label{94}
\end{equation}
Since $B$ is massless, it follows that hypercharge $Y$ is the unbroken gauge charge, that is,
\begin{equation}
U(1)_{T_{3R}} \times U(1)_{B-L} \rightarrow U(1)_{Y} 
\label{95}
\end{equation}
and $B$ is the hypercharge boson. The broken $U(1)$ symmetry is then the combination of $Y$ and $B-L$ specified by the rotated charge generator in the last term of \eqref{93}. This couples to the massive $Z_{3R}$ vector boson.

The expression for the hypercharge parameter in \eqref{94} leads to an important boundary condition at the $M_{B-L}$ scale. Defining $\alpha_{i}=g_{i}^{2}/4\pi$, it follows that 
\begin{equation}
\alpha_{Y}=\frac{1}{\alpha_{3R}^{-1}+\alpha_{BL}^{\prime~-1}} \ .
\label{96}
\end{equation}
However, this is not the most useful form of this relation. First, one should re-express it in terms of $g_{BL}$ using \eqref{86}. Second, note from \eqref{36} that 
\begin{equation}
Y=T^{1}+\sqrt{\frac{2}{3}}T^{2}  \ .
\label{97}
\end{equation}
Using the Killing relations in \eqref{37}, one finds $(Y|Y)=5/6$. This is not the canonical normalization of $1/2$ used for the basis elements of ${\mathfrak{so}(10)}$.  To be consistent with this normalization, define
\begin{equation}
Y^{\prime}=\sqrt{\frac{3}{5}}Y, \quad g_{1}=\sqrt{\frac{5}{3}}g_{Y} \ .
\label{98}
\end{equation}
Then $(Y^{\prime}|Y^{\prime})=1/2$ and \eqref{96} becomes
\begin{equation}
\alpha_{1}=\frac{5}{3\alpha_{3R}^{-1}+2\alpha_{BL}^{-1}} \ .
\label{99}
\end{equation}
When transitioning through the $M_{B-L}$ mass scale, the $T^{1}$,$T^{2}$ and $Y^{\prime}$ gauge parameters will be related using this boundary condition.
Values of $M_{B-L}$ in \cite{Ambroso:2010pe} are found to between $\sim 5 \times10^2~GeV$ and $10^4 ~GeV$, with the former being more ``typical.''

\subsubsection*{The Supersymmetry Breaking Scale $(M_{SUSY})$:}

Below $M_{B-L}$, the massive $Z_{3R}$ vector superfield--both the gauge boson and its 
gaugino--decouple. Furthermore, since the breaking of the extra $U(1)$ gauge factor arises from an expectation value in one of the right-handed sneutrinos, this neutrino supermultiplet--both the right-chiral neutrino and its sneutrino partner--also get a mass of order $M_{B-L}$ and decouple. The resulting theory is exactly the MSSM, that is, gauge group $SU(3)_{C} \times SU(2)_{L} \times U(1)_{Y} $ with three families of quarks/leptons and one Higgs-Higgs conjugate pair, along with the soft breaking interactions. There are also the remaining two right-handed neutrino multiplets but, since they are uncharged under the standard model gauge group, they will not effect the RG running of parameters.

The Higgs contribution to the quadratic potential is given by
\begin{equation}
V_{Higgs}=m^{2}_{H}|H|^{2}+m^{2}_{{\bar{H}}}|{\bar{H}}|^{2}+b(H{\bar{H}}+hc)+\mu^{2}(|H|^{2}
+|{\bar{H}}|^{2}) \ ,
\label{100}
\end{equation}
where $m^{2}_{H}$, $m^{2}_{\bar{H}}$ and  $b$ are the supersymmetry breaking Higgs masses, while $\mu$ is the supersymmetric Higgs mass. The beta function for the $H$ soft mass has a significant contribution from the square of the top quark Yukawa coupling. That is,
\begin{equation}
\frac{dm^{2}_{H}}{dt} \simeq \frac{6}{16\pi^{2}}|\lambda_{t}|^{2}\big(m_{{\tilde{Q}}_{3}}^{2}+m_{{\tilde{t}}}^{2}+m^{2}_{H}  \big) + \dots \ .
\label{101}
\end{equation}
%
%On the other hand, the ${\bar{H}}$ soft mass beta function depends only on the down quark Yukawa couplings and, hence, relative to the $H$ mass scaling
%%
%\begin{equation}
%\frac{dm^{2}_{\bar{H}}}{dt} \simeq 0~\Rightarrow m^{2}_{\bar{H}} \simeq m^{2}_{\bar{H}}(M_{I})>0 \ .
%\label{102}
%\end{equation}
%

Potential \eqref{100} can be diagonalized to 
\begin{equation}
V_{Higgs}=m^{2}_{H^{\prime}}|H^{\prime}|^{2}+m^{2}_{{\bar{H}}^{\prime}}|{\bar{H}}^{\prime}|^{2}
\label{103}
\end{equation}
where, for sufficiently small $b$, 
\begin{equation}
	m^{2}_{H^{\prime}} \simeq m^{2}_{H}-\frac{b^{2}}{m^{2}_{\bar{H}} - m^{2}_{H} },
\quad
	m^{2}_{\bar{H}^{\prime}} \simeq m^{2}_{\bar{H}} + \frac{b^{2}}{m^{2}_{\bar{H}} - m^{2}_{H} }\ .
\label{104}
\end{equation}
Running down from $M_{B-L}$, the left and right stop masses in \eqref{101} drive $m_{H}^{2}$ toward zero, making $m^{2}_{H^{\prime}}$ negative and signaling the radiative breakdown of electroweak symmetry. Note that the Higgs expectation value continues to grow until one reaches the stop mass threshold. At that point, the stops decouple and the Higgs expectation value is relatively fixed. 
A good estimate of the stop decoupling scale is given by\footnote{See~\cite{Gamberini:1989jw} for a more detailed study.}
\begin{equation}
M_{SUSY} \simeq \sqrt{m_{{\tilde{Q}}_{3}}m_{{\tilde{t}}}} \ .
\label{105}
\end{equation}

In addition to this being the scale where radiative Higgs breaking ``freezes out'', it is also a good indicator of the masses of all superpartners. Again due to the large top Yukawa coupling, one finds with approximately universal soft squared masses that all scalar superpartners tend to lie just above \eqref{105}, as shown in \cite{Ambroso:2010pe} for these types of B-L models. Furthermore, the fermionic superpartners lie near or just below the stop threshold. That is, $M_{SUSY}$ is a reasonable estimate of superpartner masses and, therefore, of the scale where supersymmetry is broken. Hence, the subscript $M_{SUSY}$. Finally, we note that the component Higgs and Higgsino fields of the ${\bar{H}}^{\prime}$ supermultiplet also develop masses of order $M_{SUSY}$. 
The value of $M_{SUSY}$ is constrained on the low end by non-observation of SUSY particles, and on the high end by fine tuning arguments. A typical value satisfying all constraints is
\begin{eqnarray}
M_{SUSY}=1~TeV \ .
\label{eq:}
\end{eqnarray}

We conclude that at $M_{SUSY}$ all superpartners and the Higgs conjugate supermultiplet decouple. Hence, at lower energy-momentum the theory has the matter and Higgs spectrum of the non-supersymmetric standard model. Although electroweak symmetry has been broken at $M_{SUSY}$, the vacuum expectation value 
\begin{equation}
\langle H^{\prime} \rangle 
\simeq {\cal{O}}(10^{2}~GeV) << M_{SUSY} \ .
\label{106}
\end{equation}
Therefore, the gauge symmetry below $M_{SUSY}$ remains approximately $SU(3)_{C}\times SU(2)_{L} \times U(1)_{Y}$.

\subsubsection*{The Electroweak Scale $(M_{EW})$:}

At ${\cal{O}}(10^{2}~GeV)$ the Higgs expectation value becomes relevant and spontaneously breaks the gauge group to $SU(3)_{C} \times U(1)_{EM}$. We will identify this scale with the $Z$ boson mass; that is,
\begin{equation}
M_{EW}=M_{Z} \ .
\label{107}
\end{equation}
We will input all experimental data for the gauge parameters, Yukawa couplings and so on at this scale.

\subsection{Scaling Regimes}

From the top down, the scaling regimes are the following.

\subsubsection*{$M_{u} \rightarrow M_{I}$:}

In this regime, the theory has softly broken $N=1$ supersymmetry. The gauge group is either $SU(3)_{C} \times SU(2)_{L} \times SU(2)_{R} \times U(1)_{B-L}$ or $SU(4)_{C} \times SU(2)_{L} \times U(1)_{T_{3R}}$ depending on whether the $\chi_{B-L}$ or $\chi_{T_{3R}}$ Wilson line turns on first. The associated matter and Higgs spectra are the superfields discussed in Subsection \ref{swb} and listed in Figure 3.

\subsubsection*{$M_{I} \rightarrow M_{B-L}$:} The theory remains supersymmetric with soft breaking interactions. Regardless of which breaking pattern occurs above $M_{I}$, in this regime the gauge group is $SU(3)_{C} \times SU(2)_{L} \times U(1)_{T_{3R}} \times U(1)_{B-L}$. The associated spectrum is exactly that of the MSSM with three right-handed neutrino supermultiplets, as discussed in Subsection \ref{swb} and listed in Figure 3.

\subsubsection*{$M_{B-L} \rightarrow M_{SUSY}$:} The theory remains softly broken supersymmetric. However, the gauge group is now reduced to $SU(3)_{C} \times SU(2)_{L} \times U(1)_{Y}$ of the standard model. The spectrum is the same as in the previous regime with the exception that the $Z^{\prime}$ vector supermultiplet as well as one neutrino chiral multiplet have decoupled.

\subsubsection*{$M_{SUSY} \rightarrow M_{Z}$:} In this regime supersymmetry is completely broken with all superpartners and the ${\bar{H}}^{\prime}$ conjugate Higgs integrated out. The gauge group remains $SU(3)_{C} \times SU(2)_{L} \times U(1)_{Y}$ with the spectrum precisely that of the non-supersymmetric standard model.\\

In any of the above regimes, the RGE for each gauge coupling parameter $g_{i}$ is 
\begin{equation}
\frac{d\alpha_{i}}{dt}=\frac{b_{i}}{2\pi} \alpha_{i}^{2} \ ,
\label{108}
\end{equation}
where $\alpha_{i}=\frac{g_{i}^{2}}{4\pi}$. In a regime where the gaugino has been integrated out, the coefficients $b_{i}$ are given by~\cite{Jones:1981we}
\begin{equation}
b_{i}=-\frac{11}{3} C_{2}(G_{i})+\frac{1}{3} \sum_{\rm scalars} I_{i}(R)+\frac{2}{3} \sum_{\rm fermions} I_{i}(R) \ ,
\label{109}
\end{equation}
where $C_{2}(G_{i})$ is the second Casimir invariant for the adjoint representation of $G_{i}$ and $ I_{i}(R)$ is the Dynkin index for the representation $R$ of $G_{i}$ defined in \eqref{18}. For a simplified version relevant for supersymmetric theory see~\cite{Martin:1993zk}.

%%%%%%%%%%%%%%%%%%%%%%%%%%%%%%%%%%%%%%%%%%%%%%%
\subsection{Gauge Unification via the $SU(2)_{L} \times SU(2)_{R}$ ``Left-Right'' Model \label{LR}}
%%%%%%%%%%%%%%%%%%%%%%%%%%%%%%%%%%%%%%%%%%%%%%%

When the $\chi_{B-L}$ Wilson line is turned on first, the scaling region between $M_{u}$ and $M_{I}$ is populated by a ``left-right model''; that is, the theory with $SU(3)_C \times SU(2)_L \times SU(2)_R \times U(1)_{B-L}$ gauge group and the particle content composed of 9 families of quarks/leptons, two Higgs bi-doublets--each one containing a pair of MSSM-like Higgs doublets--and two colored triplets. Proton decay is highly sensitive to the presence of such colored triplets~\cite{Nath:2006ut}, even at scales close to the GUT scale. However, because these fields do not come from the same {\bf 10} as the MSSM Higgs, their couplings are unknown and could be small enough to avoid the current bounds. Alternatively, this can be viewed as a justification for the Pati-Salam type model or the simultaneous Wilson like scenario, discussed in the next two sections, which do not contain these fields below $M_u$.

One of the advantages of seperating $M_u$ and $M_I$ is the extra freedom granted by the latter, which allows the exact unification of gauge couplings, an observation which has been discussed in situations with similar gauge groups but different particle content~\cite{DeRomeri:2011ie,Malinsky:2005bi}. Specifically, $M_I$ can be chosen independently of $M_{B-L}$ so as to allow exact unification of the gauge couplings.  Although not strictly necessary, we will enforce such unification as a way of exactly specifying the low energy theory. Below $M_I$, the left-right model reduces to the gauge group $SU(3)_C \times SU(2)_L \times U(1)_{T_{3R}} \times U(1)_{B-L}$ with the MSSM particle content supplemented by three families of right-handed neutrino chiral multiplets. There are enough boundary conditions at $M_{u}$, $M_{I}$ and $M_{B-L}$ to determine all gauge couplings at $M_{SUSY}$, and, hence, at $M_{Z}$ where they can be compared to the experimental data. The RG analysis requires three things: 1) the boundary conditions for the gauge couplings at $M_{u}$, $M_{I}$, $M_{B-L}$ and $M_{SUSY}$, 2) the beta functions for the gauge couplings between these different scales and 3) the input of the low-energy values of the gauge couplings from experimental data. First consider the boundary conditions. They are given by
\begin{align}
	& \alpha_3(M_{u}) = \alpha_2(M_{u}) = \alpha_R(M_{u}) = \alpha_{BL}(M_{u}) \equiv \alpha_u, \label{110}
\\ 
	& ~~\qquad \qquad \alpha_R(M_I) = \alpha_{3R}(M_I) \label{111}
\end{align}
and
\begin{equation}
\alpha_1(M_{B-L}) = \frac{5}{3 \alpha_{3R}^{-1}(M_{B-L}) + 2 \alpha_{BL}^{-1}(M_{B-L})} \ ,
\label{112}
\end{equation}
where $\alpha_3$ and $\alpha_2$ are gauge couplings for $SU(3)_C$ and $SU(2)_L$ respectively. Gauge couplings without a specific boundary condition at a given scale 
are simply identical above and below that mass.

Below $M_{u}$, the gauge couplings of the $SU(3)_C \times SU(2)_L \times SU(2)_R \times U(1)_{B-L}$ left-right model have the following RG slope factors  in the intermediate regime:
\begin{equation}
b_3 = 10, \quad b_2 = 14, \quad b_R = 14, \quad b_{BL} = 6,
\label{113}
\end{equation}
where \eqref{109} has been used.
The RG for the gauge couplings of the $SU(3)_C \times SU(2)_L \times U(1)_{T_{3R}} \times U(1)_{B-L}$ theory between $M_I$ and $M_{B-L}$ has the slope factors:
\begin{equation}
b_3 = -3, \quad b_2 = 1, \quad b_{3R} = 7, \quad b_{BL} = 6.
\label{114}
\end{equation}

Once $U(1)_{T_{3R}} \times U(1)_{B-L}$ breaks to $U(1)_Y$, the RGE slope factors  are simply the well-known ones of the  MSSM:
\begin{equation}
b_3 = -3, \quad b_2 = 1, \quad b_{1} = \frac{33}{5}.
\label{115}
\end{equation}
Integrating out the superparners and the conjugate Higgs chiral multiplet at $M_{SUSY}$ leaves the familiar standard model result
\begin{equation}
b_3 = -7, \quad b_2 = \frac{19}{16}, \quad b_{1} = \frac{41}{10}.
\label{116}
\end{equation}
Finally, $M_Z$ is the scale where the initial values of the gauge couplings can be inputed from experiment.  They are
\begin{equation}
	\alpha_1 = 0.017, \quad \quad \alpha_2 = 0.034, \quad \quad \alpha_3 = 0.118.
	\label{117}
\end{equation}

Our procedure is to start with the experimental values of $\alpha_1$, $\alpha_2$ and $\alpha_3$ at the $Z$ mass scale and to run them to the SUSY scale using the standard model RGEs. From here, the MSSM RGEs are used up to $M_{B-L}$. Above this scale, gauged hypercharge is replaced by $U(1)_{T_{3R}} \times U(1)_{B-L}$ whose coupling parameters $\alpha_{3R}$ and $\alpha_{BL}$, as yet not determined, are related to $\alpha_1$ by the boundary condition \eqref{99}. However, $\alpha_2$ and $\alpha_3$ remain valid gauge couplings and can be evolved up to the point where they become equal to each other. We will, for specificity, equate the scale of $\alpha_2$ and $\alpha_3$ unification with the unification mass $M_{u}$. Note that this scale is independent of $M_I$ since the additional particle content in the intermediate region fits into complete multiplets of $Spin(10)$. Hence, the value of  the $M_{u}$ is completely determined. We denote the unified coupling parameter by $\alpha_{u}$. This quantity, however, is dependent on both the value of $M_{I}$ as well as the boundary conditions at $M_{B-L}$.

The unification of $\alpha_{3}$ and $\alpha_{2}$ supplies a necessary boundary condition for $\alpha_{BL}$ and $\alpha_R$; namely, that they be equal to $\alpha_u$ at $M_{u}$. Now $\alpha_{BL}$ and $\alpha_R$  can be evolved back down to the B-L scale, remembering that $g_{3R}(M_I) = g_R(M_I)$.  The values of $\alpha_{BL}$ and $\alpha_R$ at the B-L scale are not independent because of gauge coupling unification and will furthermore depend on $M_I$. However, \eqref{112} can be used to solve for $M_I$. 

To procede, we must choose values of $M_{SUSY}$ and $M_{B-L}$. For the former, the typical value already discussed is $1~TeV$. For the latter, \cite{Ambroso:2010pe} shows a range of possible values. In order to distinguish between the SUSY and B-L scales, both for the plots and to avoid any appearance of conflating them, we will first give our results for a rather high $M_{B-L}=10~TeV$. Since such a large $M_{B-L}$ could be invisible to the LHC, and is disfavored by \cite{Ambroso:2010pe}, we will also give results for $M_{B-L}=1~TeV$. An exhaustive analysis of these mass thresholds will be presented in future publications \cite{Preparation}.

\begin{figure}[h!]
	\centering
	\includegraphics[scale=.9]{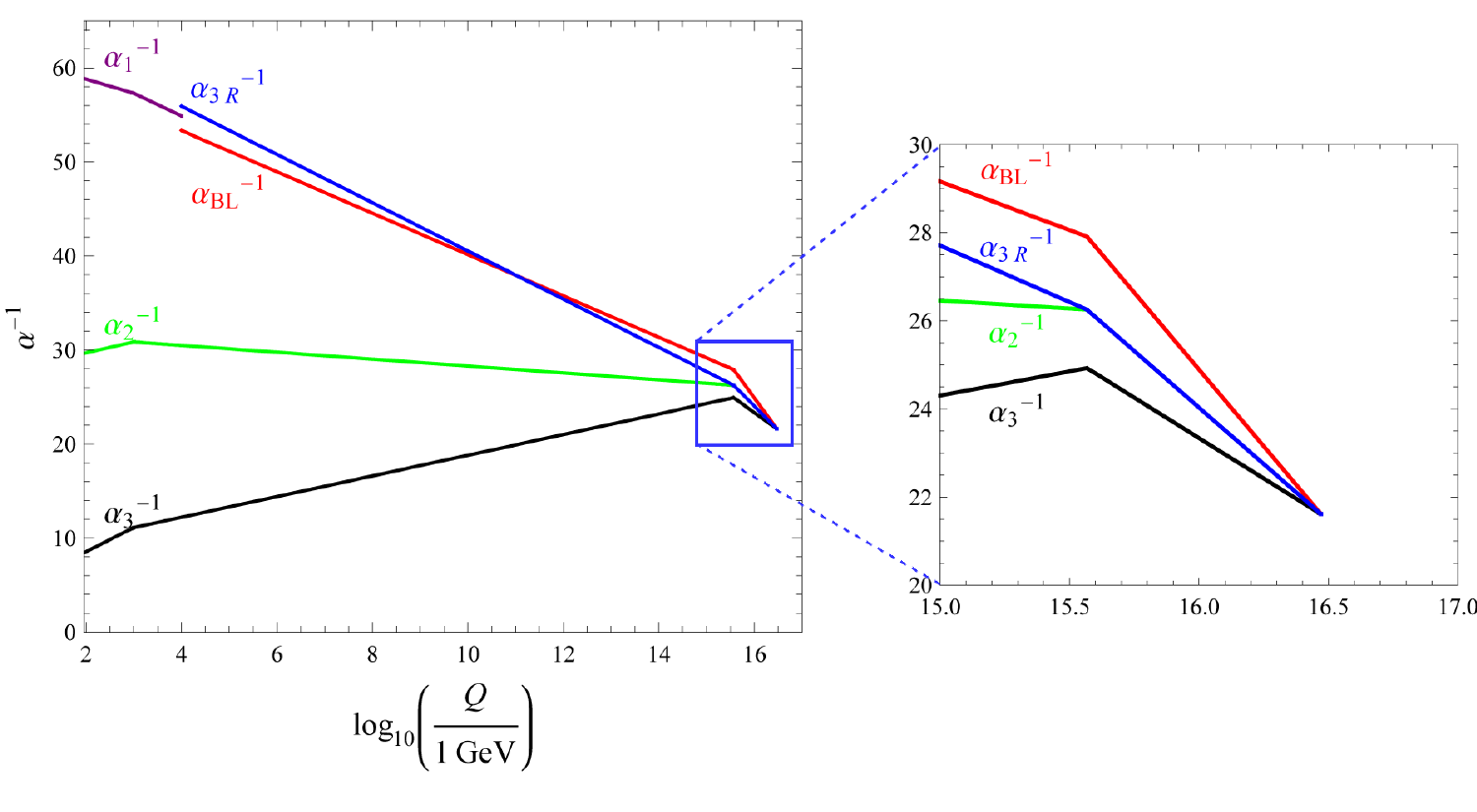}
	\caption{\small One-loop RGE running of the inverse gauge couplings, $\alpha_i^{-1}$ in the case of the left-right model with $M_{B-L}=10~TeV$ with an enlarged image of the intermediate region.}
	\label{LR.GUT}
\end{figure}

Taking the values
\begin{equation}
M_{SUSY} = 1~TeV, \quad  M_{B-L} = 10~TeV \ ,
\label{burt1}
\end{equation}
we find that
\begin{eqnarray}
\label{LR.values}
&&\quad \quad M_{u} = 3.0 \times 10^{16}~GeV, \quad M_I = 3.7 \times 10^{15}~GeV \label{burt2} \\
&&\alpha_{u}=0.046, \quad \alpha_{3R}(M_{B-L}) = 0.0179, \quad \alpha_{BL}(M_{B-L})=0.0187 \ . \nonumber
\end{eqnarray}
The associated running coupling parameters are plotted in Figure~\ref{LR.GUT}.

\begin{figure}[h]
	\centering
	\includegraphics[scale=.9]{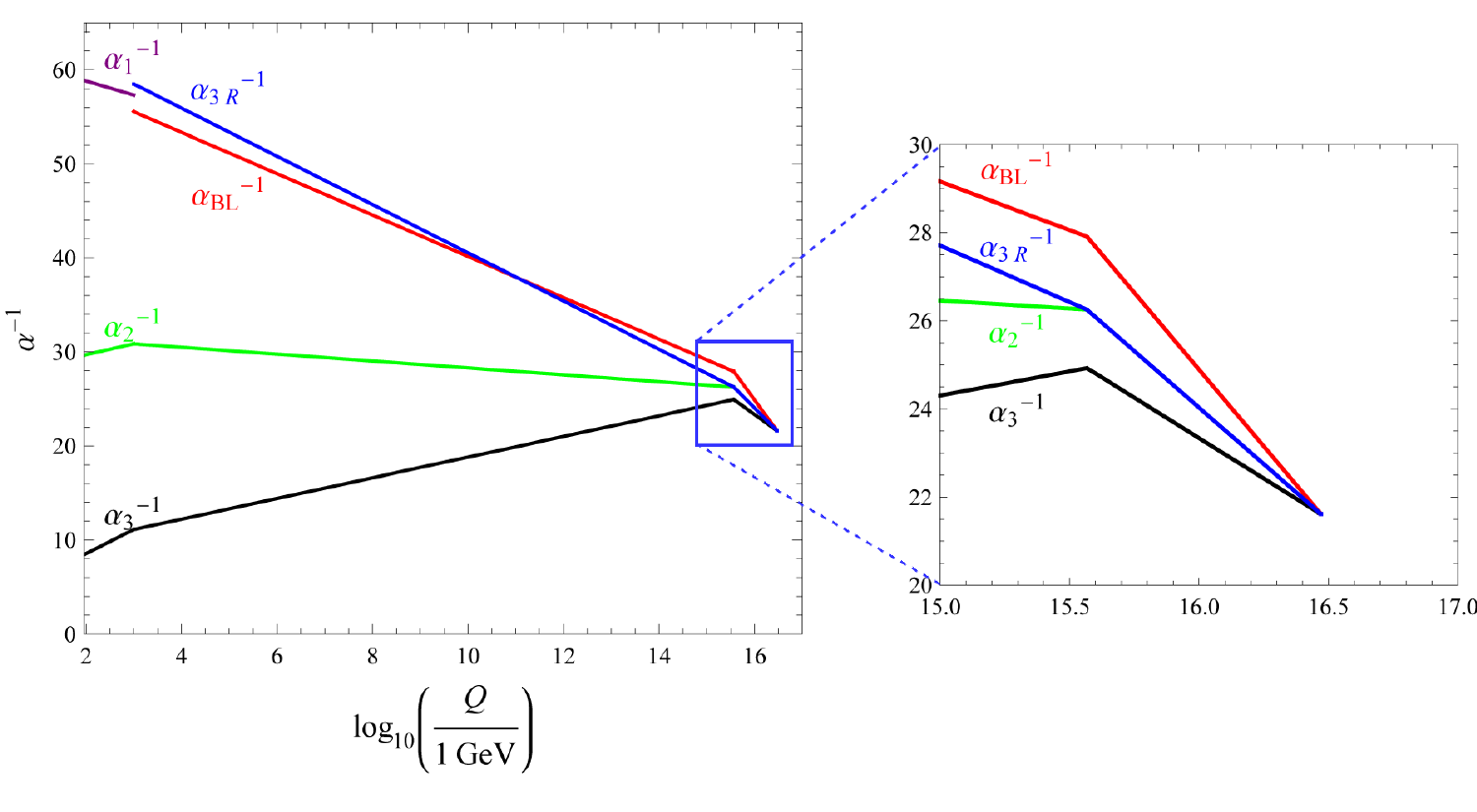}
	\caption{\small One-loop RGE running of the inverse gauge couplings, $\alpha_i^{-1}$ in the case of the left-right model with $M_{B-L}=1~TeV$ with an enlarged image of the intermediate region.}
	\label{LR.GUT2}
\end{figure}

Similarly, for
\begin{equation}
M_{SUSY} = 1~TeV, \quad  M_{B-L} = 1~TeV \ ,
\label{burt1a}
\end{equation}
we find 
\begin{eqnarray}
\label{LR.1values}
&&\quad \quad M_{u} = 3.0 \times 10^{16}~GeV, \quad M_I = 3.7 \times 10^{15}~GeV \label{burt2a} \\
&&\alpha_{u}=0.046, \quad \alpha_{3R}(M_{B-L}) = 0.0171, \quad \alpha_{BL}(M_{B-L})=0.0180 \ . \nonumber
\end{eqnarray}
The running coupling parameters are plotted in Figure~\ref{LR.GUT2}. It is worth noting that changing the value of $M_{B-L}$ did not affect any of the running above $M_{B-L}$. That is, $M_u$, $M_I$, and $\alpha_u$ are unchanged.
%

%%%%%%%%%%%%%%%%%%%%%%%%%%%%%%%%%%%%%%%%%%%
\subsection{Gauge Unification via the $SU(4)_{C} \times SU(2)_{L}$ Pati-Salam Type Model}
%%%%%%%%%%%%%%%%%%%%%%%%%%%%%%%%%%%%%%%%%%%

When the $\chi_{T{3R}}$ Wilson line is turned on first, the scaling region between $M_{u}$ and $M_{I}$ is populated by the Pati-Salam type model; that is, a theory with an $SU(4)_{C} \times SU(2)_L \times U(1)_{T_{3R}}$ gauge group whose particle content again has nine quark/lepton families and  two pairs of $H$-${\bar{H}}$ doublets. However, unlike the left-right case, there are no color triplets.

Below $M_I$, the Pati-Salam-like model again reduces to the gauge group $SU(3)_C \times SU(2)_L \times U(1)_{T_{3R}} \times U(1)_{B-L}$ with the MSSM particle content supplemented by three right-handed chiral neutrino supermultiplets. The scenario that follows is much like that in Subsection~\ref{LR}. First, one inputs the boundary conditions at the various transition masses, as well as giving the beta function coefficients for the gauge couplings between these scales. As in the left-right scenario, we will {\it enforce gauge coupling unification} as a way of specifying, and simplifying, the low energy analysis. The unification condition then fixes the value of $M_I$. There remain enough conditions to completely determine all gauge couplings at the low scale. 

Specifically, the boundary conditions are
\begin{align}
	\alpha_4&(M_u) = \alpha_2(M_u) = \alpha_{3R}(M_u) \equiv \alpha_u, \label{118}
\\
	& \alpha_4(M_I) = \alpha_{BL}(M_I) = \alpha_3(M_I) \label{119}
\end{align}
and
\begin{equation}
 \alpha_1(M_{B-L}) = \frac{5}{3 \alpha_{3R}^{-1}(M_{B-L}) + 2 \alpha_{BL}^{-1}(M_{B-L})},
\label{120}
\end{equation}
where $\alpha_4$ is the $SU(4)_C$ gauge coupling. Gauge parameters without a specific boundary condition at a given scale 
are simply identical above and below that mass.
Using \eqref{109}, the gauge coupling beta function coefficients can be calculated in the scaling regime $M_u \rightarrow M_{I}$. The result is
\begin{equation}
b_4 = 6, \quad b_2 = 14, \quad b_{3R} = 20 \ .
\label{121}
\end{equation}
Since below $M_I$ the theory is identical to the left-right case, all subsequent beta function coefficients are given in \eqref{114}, \eqref{115} and \eqref{116} respectively. 

Again, the analysis begins with the experimental values of $\alpha_1$, $\alpha_2$ and $\alpha_3$ at $M_{Z}$ given in \eqref{117}. We then follow the procedure discussed in Subsection~\ref{LR}. One notable difference is that the particle content in the $M_{u} \rightarrow M_{I}$ intermediate region does not contain full multiplets of $Spin(10)$. This shifts the unification scale $M_{u}$ from its value in the previous example, as well as the value of $M_{I}$.
The $\alpha_2$ and $\alpha_3$ gauge couplings continue to be used to fix the unification scale, although, at this scale, $\alpha_3$ is replaced by $\alpha_4$. The coupling $\alpha_{3R}$ can be evolved down from the unification scale and $\alpha_{BL}$ can be scaled from the intermediate mass.

\begin{figure}[H]
	\centering
	\includegraphics[scale=.9]{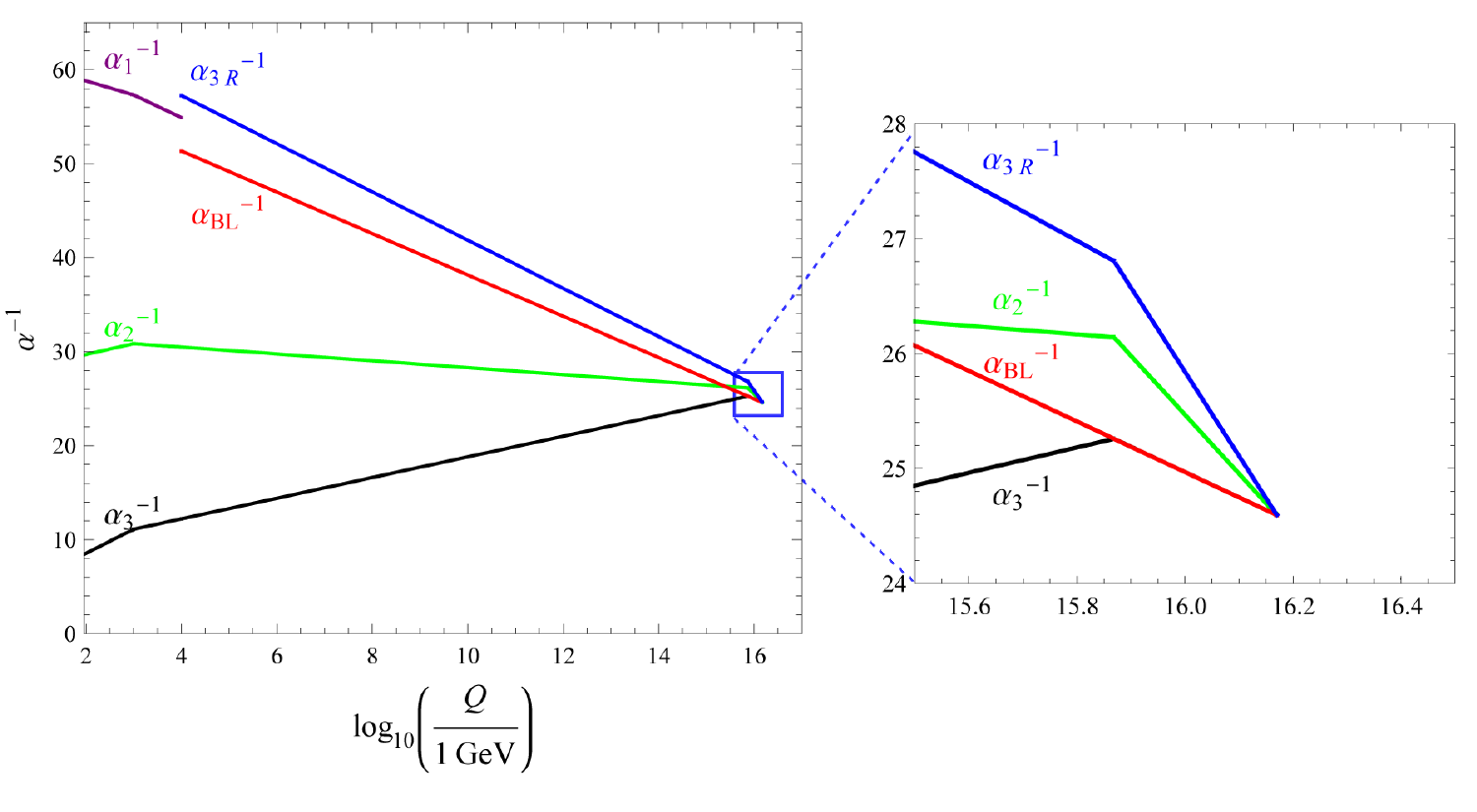}
	\caption{\small One-loop RGE running of the inverse gauge couplings, $\alpha_i^{-1}$ in the case of the Pati-Salam type model with $M_{B-L}=10~TeV$ with an enlarged image of the intermediate region.}
	\label{PS.GUT}
\end{figure}

Taking the values
\begin{equation}
M_{SUSY} = 1~TeV, \quad  M_{B-L} = 10~TeV \ ,
\label{burt1b}
\end{equation}
we find that
\begin{eqnarray}
&&\quad \quad M_{u} = 1.5 \times 10^{16}~GeV, \quad M_I = 7.4 \times 10^{15}~GeV \label{burt3} \\
&&\alpha_{u}=0.041, \quad \alpha_{3R}(M_{B-L}) = 0.0175, \quad \alpha_{BL}(M_{B-L})=0.0195 \ . \nonumber
\end{eqnarray}
The associated running coupling parameters are plotted in Figure \ref{PS.GUT}.

\begin{figure}[h]
	\centering
	\includegraphics[scale=.9]{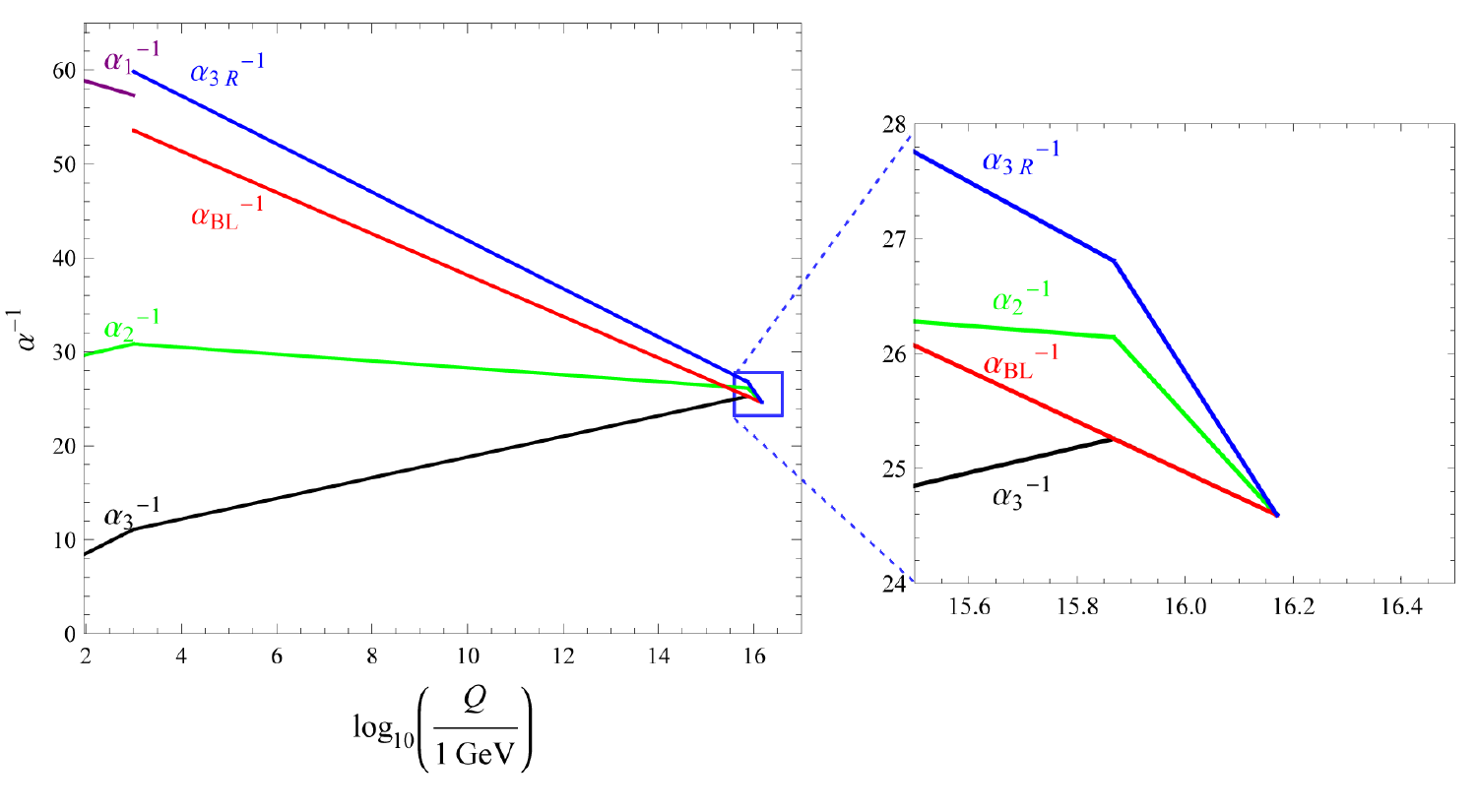}
	\caption{\small One-loop RGE running of the inverse gauge couplings, $\alpha_i^{-1}$ in the case of the Pati-Salam type model with $M_{B-L}=1~TeV$ with an enlarged image of the intermediate region.}
	\label{PS.GUT2}
\end{figure}

Similarly, for
\begin{equation}
M_{SUSY} = 1~TeV, \quad  M_{B-L} = 1~TeV \ ,
\label{burt1c}
\end{equation}
we find that
\begin{eqnarray}
&&\quad \quad M_{u} = 1.5 \times 10^{16}~GeV, \quad M_I = 7.4 \times 10^{15}~GeV \label{burt3c} \\
&&\alpha_{u}=0.041, \quad \alpha_{3R}(M_{SUSY}) = 0.0167, \quad \alpha_{BL}(M_{SUSY})=0.0187 \ . \nonumber
\end{eqnarray}
The running coupling parameters are plotted in Figure \ref{PS.GUT2}. Again, the running above $M_{B-L}$ is not affected by the change in $M_{B-L}$.
%

%%%%%%%%%%%%%%%%%%%%%%%%%%%%%%%
\subsection{Gauge Unification with Simultaneous Wilson Lines}
%%%%%%%%%%%%%%%%%%%%%%%%%%%%%%%

When both Wilson lines turn on simultaneously, so that $M_I=M_u$, the intermediate region is absent and $Spin(10)$ is immediately broken to $SU(3)_C\times SU(2)_L\times U(1)_{T_{3R}}\times U(1)_{B-L}$ with the MSSM particle content supplemented by three families of right-handed neutrino chiral mulitplets.

Naively, one might try to impose the boundary condition
\begin{align}
	& \alpha_3(M_u) = \alpha_2(M_u) = \alpha_{3R}(M_u) = \alpha_{BL}(M_u) \ . \label{122} 
\end{align}
However, as we will see below, unlike in the left-right and Pati-Salam cases, this unification condition is inconsistent with the experimental values of $\alpha_{3}$, $\alpha_{2}$ and $\alpha_{1}$ at $M_{Z}$ within the assumptions we have made about the mass thresholds. Hence, we will not input this condition. Rather, we will scale up to $M_{u}$ from the experimental input at $M_{Z}$ and examine to what extent unification is violated. Of course, the boundary condition 
\begin{equation}
 \alpha_1(M_{B-L}) = \frac{5}{3 \alpha_{3R}^{-1}(M_{B-L}) + 2 \alpha_{BL}^{-1}(M_{B-L})},
\label{123}
\end{equation}
at $M_{B-L}$ continues to hold.
Since the theory is identical to the left-right and Pati-Salam cases below $M_I$, the beta functions in all subsequent scaling regimes are given in \eqref{114}, \eqref{115} and \eqref{116}.

In the previous two sections, the final step of the RG procedure was to solve for $M_I$. In both cases, there was a unique solution for $M_I$ that satisfied the boundary conditions--including gauge coupling unification at $M_{u}$. In the simultaneous Wilson lines case, however, we are fixing $M_I=M_u$ in advance. Hence, if we continue to use the full set of boundary conditions mandated in the previous sections, the system will be overdetermined. Specifically, we find that one cannot simultaneously impose \eqref{122} and \eqref{123} while also matching the low energy experimental input \eqref{117}. To proceed, some boundary condition must be relaxed. Constraint \eqref{122} has the greatest uncertainty due to string threshold effects. Hence, we will no longer impose it. There is no flexibility in the running of $\alpha_3$ and $\alpha_2$, their running and unification being completely determined by the experimental input. However, the low energy value of $\alpha_Y$ along with \eqref{123} can be used to write a relationship between $\alpha_{3R}$ and $\alpha_{BL}$ at $M_{B-L}$, but not fix them. Most choices for these two couplings will lead to neither of them unifying with $\alpha_{3}$, $\alpha_{2}$ at $M_{u}$. However, it is possible to choose one of them so that it indeed unifies at $M_{u}$. In this case, however, the other coupling, calculated from the first using \eqref{123}, will not unify.
And vice, versa.

Let us first demand that $\alpha_{BL}$ unify with $\alpha_3$, $\alpha_2$ at $M_{u}$. Using \eqref{123} to solve for $\alpha_{3R}$ at $M_{B-L}$, we find that $\alpha_{3R}(M_{u})$ will miss unification by $\sim 8\%$. To be precise,
\begin{eqnarray}
\left|\frac{\alpha_{BL}(M_u)-\alpha_{3R}(M_u)}{\alpha_{3R}(M_u)}\right|\approx 8\%.
\label{124}
\end{eqnarray}
As one might expect from the previous sections, this result is unchanged for the range of $M_{B-L}$ from 1 to 10 $TeV$.

Another, potentially instructive, way to think of this procedure is to start with the left-right model of Subsection 2.3 and move $M_I$ continuously up to $M_u$, without changing any of the RG running below $M_I$. Recall that the unification scale of $\alpha_3$ and $\alpha_2$ is independent of $M_I$, since the additional particle content in the intermediate region fits into complete multiplets of $Spin(10)$. Note that $\alpha_{BL}$ is affected in the same way, which means that all three of these couplings will continue to unify at the same scale as we move $M_I$ up toward $M_{u}$. However, $\alpha_{3R}$ will be affected differently because, at the intermediate scale, it changes from a $U(1)_{T_{3R}}$ coupling to an $SU(2)_R$ coupling. Hence, it will not continue to unify with the others as $M_I$ approaches $M_{u}$.

\begin{figure}[h]%
\centering
\subfloat[]{\includegraphics[scale=.8]{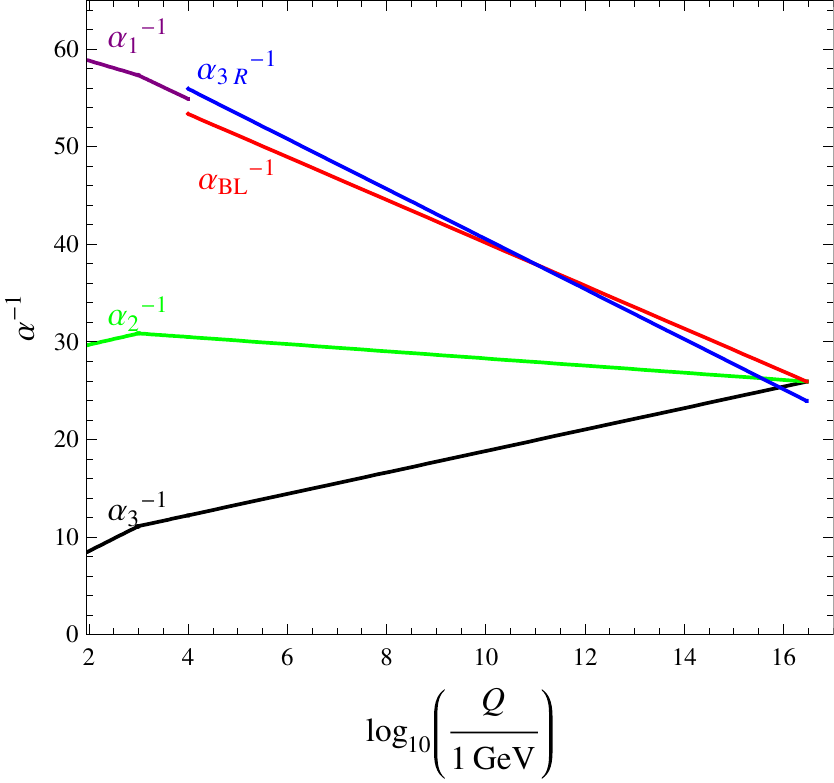}}%
\subfloat[]{\includegraphics[scale=.8]{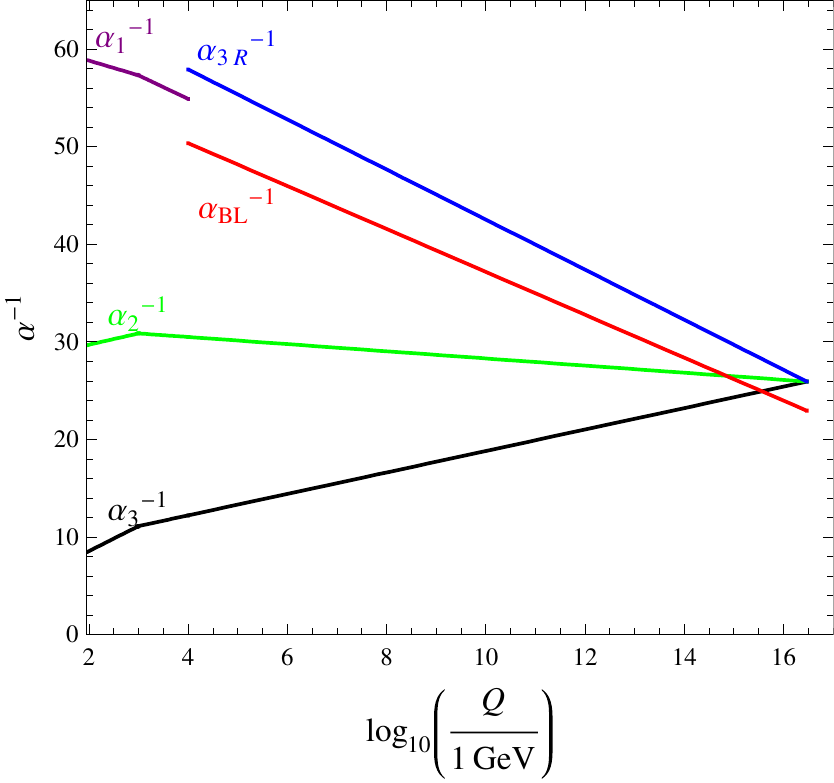}}%
\caption{\small The gauge couplings do not unify exactly if the two Wilson lines turn on simultaneously. In (a), $\alpha_{BL}$ is chosen to unify exactly. In (b) $\alpha_{3R}$ is chosen to unify exactly. $M_{B-L} = 10~TeV$ in both plots.}%
\label{fig:133}%
\end{figure}

If we demand that $\alpha_{3R}$ unify with $\alpha_3$, $\alpha_2$, and use \eqref{123} to solve for $\alpha_{BL}$ at $M_{B-L}$, We find that $\alpha_{3R}(M_{u})$ will miss unification by $\sim 13\%$. The RG running of the gauge coupling in each of these scenarios is shown in Figure \ref{fig:133}.

\begin{figure}[t!]%
\centering
{\includegraphics[scale=.8]{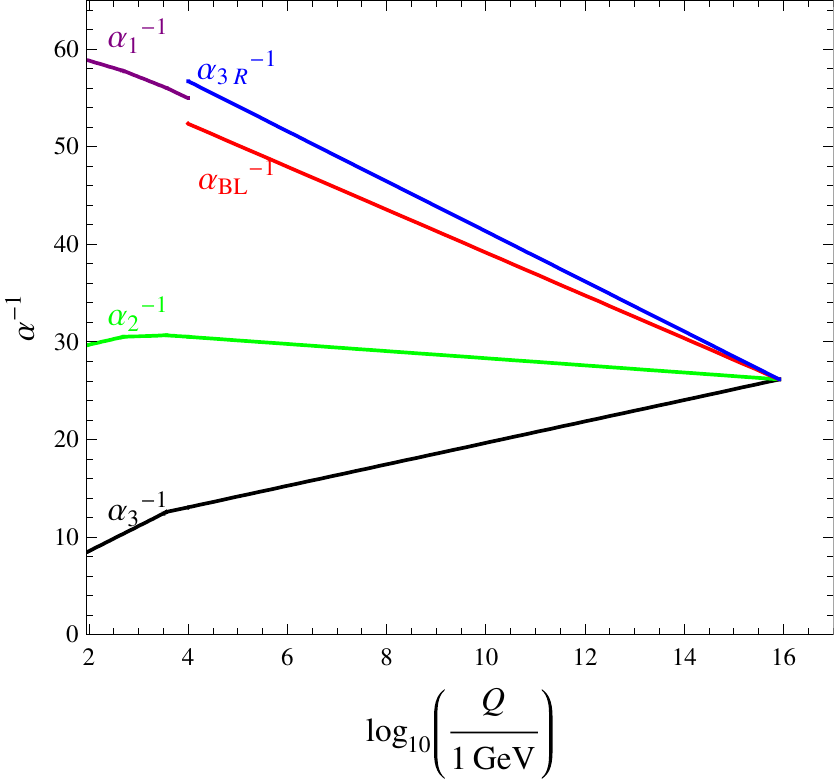}}%
\caption{\small Exact unification for simultaneous Wilson lines, requiring a splitting between the colored and non-colored superpartners.}%
\label{fig:sim.uni}%
\end{figure}

It is interesting to note that exact unification with simultaneous Wilson lines can be achieved by accounting for the fact that all superpartners will not have exactly the same mass, and, therefore, will not all decouple at exactly the same scale, $M_{SUSY}$. It is sufficient to assume that the scale at which the colored superpartners decouple, $M_{SUSY_c}$, is higher than the scale at which the non-colored superpartners decouple, $M_{SUSY_n}$. The beta functions below $M_{SUSY_n}$ and above $M_{SUSY_c}$ are unchanged. In between these two scales, the theory is the MSSM without the colored superpartners. Using \eqref{109}, the beta function coefficients in the regime $M_{SUSY_c} \to M_{SUSY_n}$ are calculated to be
\begin{equation}
	b_3 = -7, \quad b_2 = -\frac{1}{2}, \quad b_1 = \frac{11}{2}.
\end{equation}
Choosing $M_{SUSY_n}$ and $M_{B-L}$ and demanding unification of gauge couplings specifies the value of $M_{SUSY_c}$. A specific example is shown in Figure~\ref{fig:sim.uni}, where 
\begin{equation}
\label{sim.mass.scales}
	M_{SUSY_n} = 500 ~GeV, \quad M_{B-L} = 10~TeV
\end{equation}
is chosen. This yields
\begin{eqnarray}
\label{SWL.values}
&&\quad \quad M_{u} = 8.3 \times 10^{15}~GeV, \quad M_{SUSY_c} = 3.7~TeV \\
&&\alpha_{u}=0.038, \quad \alpha_{3R}(M_{B-L}) = 0.0176, \quad \alpha_{BL}(M_{B-L})=0.0191 \ . \nonumber
\end{eqnarray}

The wino and gluino play a critical role in allowing unification here. Note that the ratio of the non-colored to the colored SUSY scale (and therefore the masses of the wino and gluino) is approximately 
\begin{eqnarray}
M_2:M_3\sim M_{SUSY_n}:M_{SUSY_c}\sim1:7
\label{eq:wpe}
\end{eqnarray}
in this case. In the next section we will examine whether or not this can happen in our model.

The output parameters of interests for the three unification models (left-right, Pati-Salam, and simultaneous Wilson lines) are displayed in Table~\ref{tab:param}. The low energy observables, such as $\sin^2 \theta_R$, are too similar between the different cases to be experimentally distinguishable at the LHC.
\begin{table}[h!]
\begin{center}
\begin{tabular}{|c|c|c|c|}
\hline
Value & $\ $ LR $\ $ & $\ $ PS $\ $ & SWL
\\
\hline
$M_u$ & $3.0 \times 10^{16}$  & $1.5 \times 10^{16}$ & $8.3\times10^{15}$
\\
$M_I$ & $3.7 \times 10^{15}$  & $7.4 \times 10^{15}$  & $8.3\times10^{15}$
\\
$\alpha_u$ & 0.046 & 0.043 & 0.038
\\
$ \ \sin^2 \theta_R \ $ & 0.61 & 0.63 & 0.61
\\
$g_{Z_{3R}}$ &0.76 & 0.77 & 0.76
\\
\hline
\end{tabular}
\end{center}
\caption{Values of interest specified by gauge coupling unification for the left-right model in the intermediate regime (LR), the Pati-Salam type model in the intermediate regime (PS) and for the simultaneous Wilson lines (SWL). The results do not depend significantly on $M_{B-L}$, but these are evaluated for $M_{B-L} = 10~TeV$.}
\label{tab:param}
\end{table}%
%%%%%%%%%%%%%%%%%%%%%%%%%%%%%%%%%%%%%%
%
	\subsection{Running Gaugino Masses}
%
%%%%%%%%%%%%%%%%%%%%%%%%%%%%%%%%%%%%%%
One of the predictive consequences of assuming the unification of the gauge couplings at $M_u$ is the unification of the gauginos into a single gaugino field of $Spin(10)$. These individual components therefore have the same mass, $ {M}_{1/2}$, at $M_u$. The RGE for the gaugino mass is \cite{Martin:1993zk}
\begin{equation}
\frac{d}{dt} {M}_{a} = \frac{b_a \alpha_a {M}_a}{2 \pi},
\end{equation}
where the $b_a$ are the same coefficients as for the gauge couplings in the different regimes. The boundary conditions at the scales of interest follow in a similar way from those of the gauge couplings. At $M_u$ in the left-right model,
\begin{eqnarray}
{M}_3(M_u) = {M}_2(M_u) = {M}_R(M_u) = {M}_{BL}(M_u) \equiv {M}_{1/2},
\label{eq:}
\end{eqnarray}
and in the Pati-Salam type model,
\begin{eqnarray}
{M}_4(M_u) = {M}_2(M_u) = {M}_{3R}(M_u) \equiv {M}_{1/2}.
\label{eq:}
\end{eqnarray}
At $M_I$ in the left-right model,
\begin{eqnarray}
\quad & {M}_{3R}(M_I) = {M}_R(M_I),
\label{eq:}
\end{eqnarray}
and in the Pati-Salam type model,
\begin{eqnarray}
\quad & {M}_4(M_I) = {M}_{BL}(M_I) = {M}_{3}(M_I).
\label{eq:}
\end{eqnarray}
Solving the RGE using unification and the boundary conditions at $M_I$ yields that $M_a(\mu)$ is proportional to $\alpha_a(\mu)$.
%, where the proportionality constant is $M_{1/2}$, the mass of the $Spin(10)$ gaugino at unification.
That is,
\begin{equation}
{M}_a(\mu) = \frac{{M}_{1/2}}{\alpha_u} \alpha_a(\mu) \ .
\label{eq:M.gen}
\end{equation}
Note that we have not discussed the boundary condition at $M_{B-L}$, so this solution is only valid above that scale.
Therefore, 
%in the case of gauge coupling unification, 
the ratio of the gaugino mass parameters at the $B-L$ scale depends only on the ratio of the gauge couplings.  We now turn to the discussion of the hypercharge gaugino mass once $B-L$ is broken.

The relevant mass matrix 
%for the $U(1)$ gauginos in the $B-L$ sector 
in the basis $(\nu, \tilde B_{B-L}, \tilde W_R^0)$, neglecting electroweak effects,  is 
\begin{equation}
{M}_{\tilde \chi} = 
\begin{pmatrix}
0
&
\cos \theta_R M_{Z_{3R}}
&
\sin \theta_R M_{Z_{3R}}
\\
\cos \theta_R M_{Z_{3R}}
&
{M}_{3R}
&
0
\\
\sin \theta_R M_{Z_{3R}}
&
0
&
{M}_{BL},
\end{pmatrix}
\label{gaugino.mass}
\end{equation}
where we remind the reader that $M_{Z_{3R}}$ is the mass of the $Z'$ gauge boson.
%Before calculating the mass of the hypercharge gaugino, we will present a physical argument in the limit $M_{B-L}\gg M_{SUSY}$. 
It will be useful, motivated by  \eqref{99},  to define, 
\begin{eqnarray}
\label{eq:HatDef}
\hat M_1 = \left.\frac{M_{1/2}}{\alpha_u}\alpha_1\right|_{M_{B-L}}= \left.\frac{{M}_{1/2}}{\alpha_u} \frac{5}{2\alpha_{BL}^{-1} + 3 \alpha_{3R}^{-1}} \right|_{M_{B-L}}\ .
\end{eqnarray}
%In the limit $M_{B-L} \gg M_{SUSY}$, the effects of supersymmetry breaking can be ignored and the rotation in the gaugino sector will match that of the gauge sector. Specifically, there will be two states with mass $M_{Z_{3R}}$: one is the superpartner of $Z_{3R}$ and the other the superpartner of the VEV-acquiring right-handed sneutrino, which also has a mass of $M_{Z_{3R}}$. The third state is massless and is the superpartner of the hypercharge gauge boson: the hypercharge gaugino. As SUSY breaking effects are introduced, it is possible that these states will drift from these identities. If this is a small pertubation, the hypercharge-like gaugino will still be approximately the superpartner of the hypercharge gauge boson. Following \eqref{eq:M.gen}, its mass would then be close to
%\begin{equation}
%\label{M1.mass}
%{M}_1 \simeq \frac{{M}_{1/2}}{\alpha_u} \alpha_1 = \hat M_1 \ .
%\end{equation}
%This argument depends on $M_{B-L}\gg M_{SUSY}$. We will now show, by 
Let us now diagonalize \eqref{gaugino.mass}. Using \eqref{eq:M.gen} and \eqref{eq:HatDef}, the gaugino masses ${M}_{3R}$ and ${M}_{BL}$ can be parameterized in terms of $\hat{M}_1$:
\begin{equation}
{M}_{3R} = \hat {M}_1(1 + \epsilon_{3R}),
\quad
{M}_{BL} = \hat {M}_1(1 + \epsilon_{BL}),
\end{equation}
where
\begin{equation}
\label{epsilon}
\epsilon_i = \frac{\alpha_i}{\alpha_1} - 1 \ .
\end{equation}
Note that in any of the unification scenarios, the $\epsilon$ values at the $B-L$ scale  will be small since $\alpha_1$ is close to $\alpha_{3R}$ and $\alpha_{BL}$. (For example, in the left-right case, with $M_{B-L}=1~TeV$, $\alpha_1(M_{B-L}) = 0.017$. Using values for $\alpha_{3R}(M_{B-L})$ and $\alpha_{BL}(M_{B-L})$ from \eqref{LR.1values}, we find $\epsilon_{BL}\approx.06$.) Therefore, we can diagonalize \eqref{gaugino.mass} as a perturbative expansion in 
$\epsilon$.
%yielding the physical state which closely matches the hypercharge gaugino (the effective hypercharge gaugino). Its 
The hypercharge gaugino mass is found to be
\begin{equation}
{M}_1 = \hat{M}_1(1 + \cos^2 \theta_R \epsilon_{BL} + \sin^2 \theta_R \epsilon_{3R}+\mc O(\epsilon^2)) \ .
\end{equation}
Using \eqref{epsilon} along with \eqref{92}, \eqref{94}, and \eqref{98}, the two terms proportional to $\epsilon$ cancel, leaving,
\begin{equation}
M_1=\hat M_1(1+\mc O(\epsilon^2))\simeq \hat M_1
\end{equation}
Therefore, the hypercharge gaugino mass is, to first order in $\epsilon$, equal to ${\hat{M}}_{1}$ at $M_{B-L}$. The consequences are that, when calculated at $M_{SUSY}$, the ratios between the MSSM gauginos are 
%still only the ratios of the MSSM gauge couplings:
\begin{equation}
{M}_1:{M}_2:{M}_3 = \alpha_1:\alpha_2:\alpha_3 \sim 1:2:5 \ .
\label{gaugino.ratio}
\end{equation}

This is a prediction that should be visible to the LHC. However, the same prediction would come from other supersymmetric models in which the standard model gauge group unifies into a single gauge group with a single gaugino, so it is not unique to our theory. An important aspect of this conclusion for our model is that it is not consistent with the scenario of exact unification with simultaneous Wilson lines presented in the previous section. According to \eqref{eq:wpe} the ratio ${M}_2:{M}_3$ would need to be approximately $1:7$ for unification in the case of simultaneous Wilson lines, whereas \eqref{gaugino.ratio} predicts it to be $2:5$. We find that if $\alpha_{BL}$ is chosen to unify, $\alpha_{3R}$ will miss uniification by $\sim4\%$. This leads to the following important point:\\

\noindent ${\bullet}$ {\it The predictions for the gaugino mass ratios from gauge unification are inconsistent with the ratio required for exact unification with simultaneous Wilson lines. Therefore unification cannot be achieved naturally with simultaneous Wilson lines in our model.}

%%%%%%%%%%%%%%%%%%%%%%%%%%%%%%%%%%%%%%
%
	\section{Conclusion}
%
%%%%%%%%%%%%%%%%%%%%%%%%%%%%%%%%%%%%%%
MSSM extensions by an Abelian gauge group with the MSSM particle content plus three right-handed neutrinos can be derived from $E_8 \times E_8$ heterotic string theory and have also been proposed by the model building community as attractive minimal TeV scale options. In the former case, such models arise from ${\mathbb{Z}}_{3} \times {\mathbb{Z}}_{3}$ Wilson lines breaking $Spin(10)$. Here, a detailed search for the most general $U(1) \times U(1)$ subgroup of $Spin(10)$ consistent with this framework was conducted and a canonical model was found: $U(1)_{B-L} \times U(1)_{T_{3R}}$, that is, baryon minus lepton number and the third component of right-handed isospin. It has the following four appealing and important features:
\begin{enumerate}
\item 
	As mentioned above, the particle content is simply that of the MSSM supplemented by three right-handed neutrinos while the gauge group has one rank more than the SM gauge group.
\item
	Each quark/lepton and Higgs superfield of the low energy Lagrangian descends from a different ${\bf{16}}$ and ${\bf{10}}$ representation of $Spin(10)$ respectively, indicating that Yukawa couplings and soft SUSY breaking masses are uncorrelated at low energy by $Spin(10)$ relations .
\item
	At the scale of $Spin(10)$ breaking, there is no kinetic mixing between the two $U(1)$ symmetries since the generators of the canonical basis are Killing orthogonal.
\item
	Furthermore, no kinetic mixing is generated through renormalization group effects because the trace of the product of the two $U(1)$ charges over the entire low energy particle content is zero. The physics of kinetic mixing was discussed in detail.
\end{enumerate}
Lastly, there are several predictions associated with the breaking of B-L, among them the spontaneous breaking of R-parity. These features hold at the low scale regardless of whether the Wilson lines turn on simultaneously or at different scales. The latter situation is interesting, however, since the extra freedom associated with the size of the second Wilson line, the intermediate scale, allows for the unification of the gauge couplings. The gauge symmetry in the intermediate regime depends on the order of the Wilson line turn on-- with $SU(3)_C \times SU(2)_L \times SU(2)_R \times U(1)_{B-L}$ symmetry (left-right model)  in the case of the $B-L$  Wilson line turning on first and $SU(4)_C \times SU(2)_L \times U(1)_{T_{3R}}$ symmetry (Pati-Salam type) for the $T_{3R}$ line turning on first. The features of both spectra are summarized in Figure~3.

The threshold scales and beta functions necessary for a one-loop analysis of the gauge coupling running were carefully outlined for the three possible scenarios: $B-L$ Wilson line first, $T_{3R}$ Wilson line first and simultaneous Wilson lines, with the results displayed in Figures~\ref{LR.GUT},\ref{LR.GUT2}, in Figures~\ref{PS.GUT},\ref{PS.GUT2} and Figures~\ref{fig:133},\ref{fig:sim.uni} respectively. Unification in the simultaneous Wilson line case is possible, but the necessary gaugino masses are not consistent with gaugino unification. Low energy observables are summarized in Table~\ref{tab:param}, although their values are too similar to be used to differentiate between the scenarios. The results of this paper also help to set up a full analysis of the soft mass RGEs in order to investigate boundary conditions consistent with $B-L$ and electroweak symmetry breaking: an analysis which will be conducted in a future publication.

%%%%%%%%%%%%%%%%%%%%%%%%%%%%%%%%%%%%%%
%
\section*{Appendix A: Non-Canonical Bases}
%
%%%%%%%%%%%%%%%%%%%%%%%%%%%%%%%%%%%%%%

In this Appendix, we briefly analyze  ``non-canonical'' bases and their relevant properties. For specificity, we consider one such basis, before going on to prove an important  general theorem. 

\subsection*{A.1-- A Non-Canonical Basis}

Define
\begin{equation}
Y_{1}=4Y_{T_{3R}}-2Y_{B-L}, \quad Y_{2}=3Y_{T_{3R}}+Y_{B-L} \ . 
\label{84a}
\end{equation}
Note from \eqref{8}, \eqref{10} and \eqref{13} that
\begin{equation}
Y_{1}=8\big (Y-\frac{5}{4}(B-L) \big), \quad Y_{2}=6Y \ . 
\label{84b}
\end{equation}
Using \eqref{14}, it follows that $Y_{1}$,$Y_{2}$ satisfy the Killing brackets
\begin{equation}
(Y_{1}|Y_{1})=80, \quad (Y_{2}|Y_{2})=30, \quad (Y_{1}|Y_{2})=0 \ .
\label{84c}
\end{equation}
Hence, they form an orthogonal--but ``non-canonical''--basis of the two-dimensional subspace ${\mathfrak{h}}_{3 \oplus 2}$ of the Cartan subalgebra  that commutes with ${\mathfrak{su}}(3)_C \oplus {\mathfrak{su}}(2)_L$. 

The explicit form of $Y_{1}$ and $Y_{2}$ in the $\bf{16}$ representation of ${\mathfrak{so}}(10)$ is easily constructed from \eqref{15},\eqref{16}. 
We find that 
\begin{eqnarray}
&[Y_{1}]_{\bf 16} = (-2){\bf 1}_{3} \oplus (6){\bf 1}_{3}  \oplus (-2){\bf 1}_{6} \oplus (6){\bf 1}_{2} \oplus (-10){\bf 1}_{1} \oplus (-2){\bf 1}_{1} \ ,\label{84d} \\ \nonumber \\
&[Y_{2}]_{\bf 16} = (-4){\bf 1}_{3} \oplus (2){\bf 1}_{3}  \oplus (1){\bf 1}_{6} \oplus (-3){\bf 1}_{2} \oplus (0){\bf 1}_{1} \oplus (6){\bf 1}_{1} \ . \label{84e}
\end{eqnarray}
Similarly, using  \eqref{21},\eqref{22} the explicit form of $Y_{1}$ and $Y_{2}$ in the $\bf{10}$ representation is given by
\begin{eqnarray}
&[Y_{1}]_{\bf 10}=(-4){\bf 1}_{3} \oplus (4){\bf 1}_{3} \oplus (4){\bf 1}_{2}\oplus (-4){\bf 1}_{2} \ ,\label{84f} \\
\nonumber \\
&[Y_{2}]_{\bf 10}=(2){\bf 1}_{3} \oplus (-2){\bf 1}_{3} \oplus (3){\bf 1}_{2}\oplus (-3){\bf 1}_{2} \ .\label{84g}
\end{eqnarray}

\subsection*{A.2-- Properties of the Non-Canonical Basis}

The non-canonical basis \eqref{84a} shares three of the four fundamental properties of the canonical basis. We analyze this as follows.

\subsubsection*{Wilson Lines and the MSSM:}

Consider the two Wilson lines associated with the non-canonical basis. As abstract ${\mathfrak{so}}(10)$ group elements, these are
\begin{equation}
\chi_{1}=e^{iY_{1}\frac{2\pi}{3}}, \quad \chi_{2}=e^{iY_{2}\frac{2\pi}{3}} \ . 
\label{84h}
\end{equation}
Note that $\chi_{1}^{3}=\chi_{2}^{3}=1$ and, hence, each generates a finite ${\mathbb{Z}}_{3}$ subgroup of $Spin(10)$. When turned on simultaneously, these Wilson lines spontaneously break
\begin{equation}
Spin(10) \rightarrow SU(3)_{C} \times SU(2)_{L} \times U(1)_{Y_{1}} \times U(1)_{Y_{2}} \ .
\label{84i}
\end{equation}
The ${\mathbb{Z}}_{3} \times {\mathbb{Z}}_{3}$ isometry acts equivariantly on the chosen vector bundle $V$ and, hence, the associated sheaf cohomology groups of tensor products of $V$ carry a representation of ${\mathbb{Z}}_{3} \times {\mathbb{Z}}_{3}$. To determine the zero modes of the Dirac operator twisted by $V$ and, hence, the low energy spectrum, one takes each $H^{1}(X,U_{R}(V))$, tensors it with the associated representation $R$, and then chooses the invariant subspace 
$(H^{1}(X,U_{R}(V)) \otimes R)^{{\mathbb{Z}}_{3} \times {\mathbb{Z}}_{3}}$. 
%Let us carry this out for each of the relevant representations of $Spin(10)$.

For $R={\bf{16}}$, the associated sheaf cohomology $H^{1} \big(X,V \big)$ was given in \eqref{25}. The explicit representation of ${\mathbb{Z}}_{3} \times {\mathbb{Z}}_{3}$ on this 
linear space was presented in \eqref{26}. Choosing the Wilson line generators in the non-canonical basis \eqref{84a}, it follows from \eqref{84d} and \eqref{84e} that the action of the Wilson lines \eqref{84h} on each $\bf{16}$ is given by
\begin{eqnarray}
{\bf 16}&=& \chi_{1} \cdot  \chi_{2}^{2} ({\bf{\bar{3}},\bf {1}},-2,-4)\oplus   1 \cdot \chi_{2}^{2} ({\bf{\bar{3}},\bf {1}},6, 2) \label{84j} \\
&& \oplus \chi_{1} \cdot  \chi_{2} ({\bf{3}},{\bf {2}},-2,1) \oplus 1 \cdot 1({\bf{1}},{\bf {2}},6,-3) \oplus  \chi_{1}^{2} \cdot 1 ({\bf{1}},{\bf {1}}, -10, 0) \nonumber\\
&& \oplus  \chi_{1} \cdot 1({\bf{1}},{\bf {1}},-2,6)~. \nonumber
\end{eqnarray}
Using this and \eqref{26}, we find that $(H^{1}(X,V) \otimes {\bf 16})^{{\mathbb{Z}}_{3} \times {\mathbb{Z}}_{3}}$ consists of {\it three families} of quarks and leptons, each family transforming as
\begin{equation} 
Q= ({\bf{3}},{\bf {2}},-2,1), \quad u=({\bf{\bar{3}},\bf {1}},-2,-4), \quad d=({\bf{\bar{3}},\bf {1}},6, 2)
\label{84k}
\end{equation}
and 
\begin{equation}
L=({\bf{1}},{\bf {2}},6,-3), \quad \nu=({\bf{1}},{\bf {1}}, -10,0), \quad e=({\bf{1}},{\bf {1}},-2,6)
\label{84l}
\end{equation}
under $SU(3)_{C} \times SU(2)_{L} \times U(1)_{Y_{1}} \times U(1)_{Y_{2}}$.

For $R={\bf 10}$, the associated sheaf cohomology is $H^{1}(X,\wedge^{2}V)$. The explicit representation of ${\mathbb{Z}}_{3} \times {\mathbb{Z}}_{3}$ on this 
linear space was presented in \eqref{31}. Choosing the Wilson line generators in the non-canonical basis \eqref{84a}, it follows from \eqref{84f} and \eqref{84g} that the action of the Wilson lines \eqref{84h} on each $\bf{10}$ is given by
\begin{eqnarray}
{\bf 10}&= &\chi_{1}^{2} \cdot \chi_{2}^{2} ({\bf 3},{\bf 1},-4,2) \oplus \chi_{1} \cdot \chi_{2} ({\bar{\bf 3}},{\bf 1},4,-2) \nonumber \\
&&\oplus \chi_{1} \cdot 1({\bf 1},{\bf 2},4,3) \oplus \chi_{1}^{2} \cdot 1 ({\bf 1},{\bf 2},-4,-3)~. 
\label{84m}
\end{eqnarray}
Using this and \eqref{31}, we find that $(H^{1}(X,\wedge^{2}V) \otimes {\bf 10})^{{\mathbb{Z}}_{3} \times {\mathbb{Z}}_{3}}$ consists of a {\it single pair} of Higgs doublets transforming as
\begin{equation}
H=({\bf 1},{\bf 2},4,3), \quad \bar{H}= ({\bf 1},{\bf 2},-4,-3) 
\label{84n}
\end{equation}
under $SU(3)_{C} \times SU(2)_{L} \times U(1)_{Y_{1}} \times U(1)_{Y_{2}}$. These results lead to the following property of non-canonical basis \eqref{84a}.\\

\noindent $\bullet$ {\it When the two Wilson lines of the non-canonical basis are turned on simultaneously, the resulting low energy spectrum is precisely that of the MSSM--that is, three families of quark/lepton chiral superfields, each family with a right-handed neutrino supermultiplet, and one pair of Higgs-Higgs conjugate chiral multiplets}. \\

This non-canonical basis exhibits a second, related, property.
%that has important consequences for the the low energy effective Lagrangian. 
Consider, once again, the $R={\bf 16}$ case and the ${\mathbb{Z}}_{3} \times {\mathbb{Z}}_{3}$ invariant tensor product of $H^{1}(X,V)$ in \eqref{25},\eqref{26} with the ${\bf 16}$ decomposition in \eqref{84j}. 
%Note that, with the exception of $\chi_{B-L}$, $\chi_{T_{3R}}\chi_{B-L}^{2}$ and $\chi_{T_{3R}}^{2}\chi_{B-L}^{2}$ in \eqref{26} which project out all terms in \eqref{28}, each of the remaining six entries in each RG form a ${\mathbb{Z}}_{3} \times {\mathbb{Z}}_{3}$ invariant with only one term in a ${\bf 16}$. That is, 
Note that each quark and lepton chiral multiplet in the low energy theory arises from a different ${\bf 16}$ representation of $Spin(10)$. Now consider the $R={\bf 10}$ case. It is easily seen from \eqref{31} and \eqref{84m} that
%, with the exception of $\chi_{T_{3R}}\chi_{B-L}^{2}$ and $\chi_{T_{3R}}^{2}\chi_{B-L}$ in $\eqref{31}$ which project out all terms in \eqref{33}, the remaining two entries $\chi_{T_{3R}}$ and $\chi_{T_{3R}}$ each form a ${\mathbb{Z}}_{3} \times {\mathbb{Z}}_{3}$ invariant  with only a single component of \eqref{33}, 
the Higgs and Higgs conjugate chiral multiplets each arise from a different {\bf 10} representation of $Spin(10)$. This leads to the second property of the non-canonical basis. \\

\noindent ${\bullet}$ {\it Since each quark/lepton and Higgs superfield  of the low energy Lagrangian arises from a different ${\bf 16}$ and ${\bf 10}$ representation of $Spin(10)$ respectively, the parameters of the effective theory, and specifically the Yukawa couplings and the soft supersymmetry breaking parameters, are uncorrelated by the $Spin(10)$ unification.} \\
%For example, the soft mass squared parameters of the right-handed sneutrinos need not be universal with the remaining slepton supersymmetry breaking parameters.}

Thus, the non-canonical basis \eqref{84a} shares these two important properties with the canonical basis. Now consider the following.

\subsubsection*{The Kinetic Mixing Parameter:}

Prior to turning on the ${\mathbb{Z}}_{3} \times {\mathbb{Z}}_{3}$ Wilson lines, the conventionally normalized kinetic energy part of the gauge field Lagrangian is $Spin(10)$ invariant and given in
\eqref{35}, where $\{T^{a}_{R}, a=1,\dots.45\}$ is an orthogonal basis of ${\mathfrak{so}}(10)$ in any representation $R$ , each basis element Killing normalized to $\frac{1}{\sqrt{2}}$. Defining
\begin{equation}
T^{1}= \sqrt{\frac{2}{5}}\big(Y-\frac{5}{4}(B-L)\big)=\frac{1}{4\sqrt{10}} Y_{1}, \quad T^{2}=\sqrt{\frac{3}{5}}Y=\frac{1}{2\sqrt{15}}Y_{2}
\label{84o}
\end{equation}
we see from \eqref{84c} that 
\begin{equation}
(T^{1}|T^{1})=(T^{2}|T^{2})=\frac{1}{2}, \quad (T^{1}|T^{2})=0
\label{84p}
\end{equation}
and, hence,
\begin{equation}
{\cal{L}}_{kinetic}=-\frac{1}{4}(F_{\mu\nu}^{1})^{2}-\frac{1}{4}(F_{\mu\nu}^{2})^{2} +\dots ~.
\label{84q}
\end{equation}
That is, there is no kinetic mixing term of the form $F_{\mu\nu}^{1} F^{2\mu\nu}$. This is a consequence of the fact that the non-canonical basis elements $Y_{1}$ and $Y_{2}$ are Killing orthogonal, and is of little importance while $Spin(10)$ remains unbroken. However, if both Wilson lines are turned on simultaneously, the gauge group is spontaneously broken to $SU(3)_{C} \times SU(2)_{L} \times U(1)_{Y_{1}} \times U(1)_{Y_{2}}$. For general $U(1) \times U(1)$, the two Abelian field strengths can exhibit  kinetic mixing; that is,
\begin{equation}
{\cal{L}}_{kinetic}=-\frac{1}{4}((F_{\mu\nu}^{1})^{2}+2\alpha F_{\mu\nu}^{1} F^{2\mu\nu}+(F_{\mu\nu}^{2})^{2} +\dots) ~.
\label{84r}
\end{equation}
for some real parameter $\alpha$. However, for $U(1)_{Y_{1}} \times U(1)_{Y_{2}}$ the normalized canonical generators satisfy \eqref{84p} and, specifically, are orthogonal in ${\mathfrak{so}}(10)$. It follows that the initial value of $\alpha$ at the unification scale, $M_{u}$, must vanish. Hence, there is a third property that the non-canonical basis \eqref{84a} shares with the canonical basis.\\

\noindent $\bullet$ {\it Since the generators of the non-canonical basis are Killing orthogonal in
${\mathfrak{so}}(10)$, the value of the kinetic field strength mixing parameter $\alpha$ must vanish at the unification scale. That is, $\alpha(M_{u})=0$.}\\

Once the $Spin(10)$ symmetry is broken by both Wilson lines, either by turning them on at the same scale or sequentially, one expects the mixing parameter $\alpha$ to regrow due to radiative corrections. In this case, the Abelian field strengths develop a non-vanishing mixing term which greatly complicates the renormalization group analysis of the low energy effective theory. As discussed in detail in Subsection \ref{PCB}, at an arbitrary scale the covariant derivative can be written in an ``upper triangular'' realization given by
\begin{equation}
D=\partial -i(T^{1},T^{2}) 
\begin{pmatrix}
{\cal{G}}_{1} & {\cal{G}}_{M} \\
0 & {\cal{G}}_{2}
\end{pmatrix}
\begin{pmatrix}
{\cal{A}}^{1} \\
{\cal{A}}^{2} 
\end{pmatrix} \ ,
\label{84s}
\end{equation}
with
\begin{equation}
{\cal{G}}_{1}=g_{1}, \quad {\cal{G}}_{2}=\frac{g_{2}}{\sqrt{1-\alpha^{2}}}, \quad {\cal{G}}_{M}=\frac{-g_{1}\alpha}{\sqrt{1-\alpha^{2}}} \ .
\label{84t}
\end{equation}
%
%Note that in the limit that $\alpha \rightarrow 0$, ${\cal{G}}_{2}=g_{2}$ and ${\cal{G}}_{M}=0$.
The RGE for the off-diagonal coupling ${\cal{G}}_{M}$ was presented in \eqref{50},\eqref{51} and \eqref{52}. Recall that if the mixing parameter $\alpha$ and, hence, the off-diagonal coupling ${\cal{G}}_{M}$ vanish at some initial scale, as they will for the non-canonical basis \eqref{84a}, then a non-vanishing ${\cal{G}}_{M}$ will be generated at a lower scale if and only if the charges $T^{1}$ and $T^{2}$ are such that
\begin{equation}
B_{12}=Tr(T^{1}T^{2})\neq0 \ .
\label{84u}
\end{equation}
The trace in \eqref{84u} is over the entire matter and Higgs spectrum of the MSSM.
Let us break $Spin(10)$ to $U(1)_{Y_{1}}\times U(1)_{Y_{2}}$ with both Wilson lines of the non-canonical basis \eqref{84a}. The associated normalized charges $T^{1}$ and $T^{2}$ were presented in \eqref{84o} and satisfy
\begin{equation}
(T^{1}|T^{2})=0 \ .
\label{84v}
\end{equation}
It then follows from \eqref{17} that 
\begin{equation}
Tr([T^{1}]_{R}[T^{2}]_{R})=0
\label{84w}
\end{equation}
for any complete ${\mathfrak{so}}(10)$ representation $R$. Recalling that each quark/lepton family with a right-handed neutrino fills out a complete ${\bf 16}$ multiplet, one can conclude that
\begin{equation}
Tr([T^{1}]_{quarks/leptons}[T^{2}]_{quarks/leptons})=0 \ .
\label{84x}
\end{equation}
However, in the reduction to the zero-mode spectrum the color triplet Higgs $H_{C}$ and ${\bar{H}}_{C}$ are projected out. Hence, the electroweak Higgs doublets $H$ and ${\bar{H}}$ do not make up a complete ${\bf 10}$ of ${\mathfrak{so}}(10)$. Therefore, the trace of $T^{1}T^{2}$ over the Higgs fields of the MSSM is not guaranteed to vanish. It is straightforward to compute this trace using \eqref{84f}, \eqref{84g} and \eqref{84o}. If we ignore the color triplet components, then
\begin{eqnarray}
&&[Y_{1}]_{H,{\bar{H}}}= (4){\bf 1}_{2}\oplus (-4){\bf 1}_{2} \label{84y} \\ \nonumber \\
&&[Y_{2}]_{H,{\bar{H}}}=(3){\bf 1}_{2}\oplus (-3){\bf 1}_{2} \ . \label{84z}
\end{eqnarray}
It then follows from \eqref{84o} and \eqref{84y}, \eqref{84z} that
\begin{equation}
Tr([T^{1}]_{H,{\bar{H}}}[T^{2}]_{H,{\bar{H}}})= \frac{1}{8\sqrt{150}}Tr([Y_{1}]_{H,{\bar{H}}}[Y_{2}]_{H,{\bar{H}}})=\frac{\sqrt 6}{5} \ .
\label{84aa}
\end{equation}
We conclude from \eqref{84x} and \eqref{84aa} that
\begin{equation}
B_{12}=\frac{\sqrt 6}{5} \neq 0 \ .
\label{84bb}
\end{equation}
Therefore, for the non-canonical basis \eqref{84a}, although the initial value of $\alpha$ and, hence, ${\cal{G}}_{M}$ vanish, these off-diagonal parameters will re-grow at any lower scale. That is, unlike the case for the canonical basis, here kinetic mixing will re-emerge due to radiative corrections. We conclude that the non-canonical basis \eqref{84a} does not share the fourth property possessed by the canonical basis.\\

\noindent $\bullet$ {\it The generators of the non-canonical basis \eqref{84a} are such that $Tr(T^{1}T^{2}) \neq 0$ when the trace is performed over the matter and Higgs spectrum of the MSSM. Thus, unlike the canonical basis, even though the original kinetic mixing parameter vanishes, 
$\alpha$ and, hence, ${\cal{G}}_{M}$ will regrow under the RG at any scale. This property of  kinetic mixing greatly complicates the renormalization group analysis of the $SU(3)_{C} \times SU(2)_{L} \times U(1)_{Y_{1}}\times U(1)_{Y_{2}}$ low energy theory}.\\

\subsection*{A.3-- General Theorem About Kinetic Mixing}

In this final subsection, we will prove that this property of basis \eqref{84a} is shared by all orthogonal non-canonical bases. That is, the only basis with the property that kinetic mixing vanishes at all scales is the canonical basis, or any appropriate multiples of this basis. Consider a generic basis of ${\mathfrak{h}}_{3 \oplus 2}$ given by
\begin{equation}
Y_{1}=mY_{T_{3R}}+nY_{B-L}, \quad Y_{2}=pY_{T_{3R}}+qY_{B-L} \ .
\label{84cc}
\end{equation}
Using \eqref{14}, we find that
\begin{eqnarray}
&&(Y_{1}|Y_{1})=2(m^{2}+6n^{2}), \quad (Y_{2}|Y_{2})=2(p^{2}+6q^{2}) \label{84dd} \\
&& \quad \qquad \qquad (Y_{1}|Y_{2})=2(mp+6nq) \ .\label{84ee}
\end{eqnarray}
In order for the initial value of the mixing parameter $\alpha(M_{u})$ to vanish, we want to consider bases for which $(Y_{1}|Y_{2})=0$. Hence, one must choose
\begin{equation}
mp=-6nq \ .
\label{84ff}
\end{equation}
As previously, define the normalized generators 
\begin{equation}
T^{1}=\frac{1}{2\sqrt{m^{2}+6n^{2}}}Y_{1}, \quad T^{2}=\frac{1}{2\sqrt{p^{2}+6q^{2}}}Y_{2} 
\label{84gg}
\end{equation}
so that
\begin{equation}
(T^{1}|T^{1})=(T^{2}|T^{2})=\frac{1}{2}, \quad (T^{1}|T^{2})=0 \ .
\label{84hh}
\end{equation}
We will, henceforth, make the assumption that the associated Wilson lines lead to exactly the MSSM spectrum with three families of right-handed neutrino supermultiplets with no vector-like pairs or exotic fields. The non-canonical basis \eqref{84a} serves as a proof that, in addition to the canonical basis, there is at least one other basis with this property.

Having guaranteed that the initial mixing parameter $\alpha$ vanishes, we want to decide when this parameter can remain zero at any lower scale under the RG. As discussed above, this will be the case if and only if  the charges $T^{1}$ and $T^{2}$ are such that
\begin{equation}
B_{12}=Tr(T^{1}T^{2})=0 \ ,
\label{84ii}
\end{equation}
where the trace is over the entire matter and Higgs spectrum of the MSSM.
The normalized charges $T^{1}$ and $T^{2}$ in \eqref{84gg} satisfy
\begin{equation}
(T^{1}|T^{2})=0 
\label{84jj}
\end{equation}
and, hence, 
\begin{equation}
Tr([T^{1}]_{R}[T^{2}]_{R})=0
\label{84kk}
\end{equation}
for any complete ${\mathfrak{so}}(10)$ representation $R$. Recalling that each quark/lepton family with a right-handed neutrino fills out a complete ${\bf 16}$ multiplet, one can conclude that
\begin{equation}
Tr([T^{1}]_{quarks/leptons}[T^{2}]_{quarks/leptons})=0 \ .
\label{84ll}
\end{equation}
However, in the reduction to the zero-mode spectrum the color triplet Higgs $H_{C}$ and ${\bar{H}}_{C}$ are projected out. Hence, the electroweak Higgs doublets $H$ and ${\bar{H}}$ do not make up a complete ${\bf 10}$ of ${\mathfrak{so}}(10)$. Therefore, the trace of $T^{1}T^{2}$ over the Higgs fields of the MSSM is not guaranteed to vanish. It is straightforward to compute this trace using \eqref{21}, \eqref{22} and \eqref{84cc}. If we ignore the color triplet components, then
\begin{eqnarray}
&&[Y_{1}]_{H,{\bar{H}}}= (m){\bf 1}_{2}\oplus (-m){\bf 1}_{2} \label{84mm} \\ \nonumber \\
&&[Y_{2}]_{H,{\bar{H}}}=(p){\bf 1}_{2}\oplus (-p){\bf 1}_{2} \ . \label{84nn}
\end{eqnarray}
It follows from \eqref{84gg} and \eqref{84mm}, \eqref{84nn} that
\begin{eqnarray}
&Tr([T^{1}]_{H,{\bar{H}}}[T^{2}]_{H,{\bar{H}}})= \frac{1}{4\sqrt{(m^{2}+6n^{2})(p^{2}+6q^{2})}}Tr([Y_{1}]_{H,{\bar{H}}}[Y_{2}]_{H,{\bar{H}}}) \\
& \quad = \frac{mp}{\sqrt{(m^{2}+6n^{2})(p^{2}+6q^{2})}} \ . 
\label{84oo}
\end{eqnarray}
We conclude from \eqref{84ll} and \eqref{84oo} that
\begin{equation}
B_{12}= \frac{mp}{\sqrt{(m^{2}+6n^{2})(p^{2}+6q^{2})}} \ .
\label{84pp}
\end{equation}
Therefore,  for $B_{12}$ to satisfy \eqref{84ii} one must choose
\begin{equation}
mp=0 \ .
\label{84qq}
\end{equation}

We conclude that to have kinetic mixing vanish at any energy-momentum scale, a generic basis \eqref{84cc} must satisfy both \eqref{84ff} and \eqref{84qq}. It is straightforward to show that the only bases that satisfy these two constraints is the canonical basis, and appropriate multiples of the canonical basis.\\

\noindent $\bullet$ {\it The only basis of ${\mathfrak{h}}_{3 \oplus 2} \subset {\mathfrak{h}}$ for which $U(1)_{Y_{1}} \times U(1)_{Y_{2}}$ kinetic mixing vanishes at all values of energy-momentum is the canonical basis $Y_{T_{3R}}$, $Y_{B-L}$ and appropriate multiples of this basis.}

%%%%%%%%%%%%%%%%%%%%%%%%%%%%%%%%%%%

\section*{Acknowledgments}
The work of Burt Ovrut  and Austin Purves is supported in part by the DOE under contract No. DE-AC02-76-ER-03071.  B.A.O. acknowledges partial support from the NSF RTG grant DMS-0636606 and from NSF Grant 555913/14 for International Collaboration.
%%%%%%%%%%%%%%%%%%%%%%%%%%%%%%%%%%%


\begin{thebibliography}{99}

\bibitem{Nakamura:2010zzi}
  K.~Nakamura {\it et al.}  [Particle Data Group Collaboration],
  ``Review of particle physics,''
  J.\ Phys.\ G G {\bf 37} (2010) 075021.
  %%CITATION = JPHGB,G37,075021;%%

\bibitem{MSSM1}
  P.~Fayet,
  ``Supersymmetry and Weak, Electromagnetic and Strong Interactions,''
  Phys.\ Lett.\ B {\bf 64} (1976) 159.
  %%CITATION = PHLTA,B64,159;%%
  
\bibitem{MSSM2}
  P.~Fayet,
  ``Spontaneously Broken Supersymmetric Theories of Weak, Electromagnetic and Strong Interactions,''
  Phys.\ Lett.\ B {\bf 69} (1977) 489.
  %%CITATION = PHLTA,B69,489;%%  

\bibitem{MSSM3}
  S.~Dimopoulos and H.~Georgi,
  ``Softly Broken Supersymmetry and SU(5),''
  Nucl.\ Phys.\ B {\bf 193} (1981) 150.
  %%CITATION = NUPHA,B193,150;%%

\bibitem{e1} 
 A.~Lukas, B.~A.~Ovrut and D.~Waldram,
 ``On the Four-Dimensional Effective Action of Strongly Coupled Heterotic String Theory'',
 Nucl.\ Phys.\  B {\bf 532}, 43 (1998),
 [arXiv:9710208 [hep-th]]

\bibitem{Lukas:1998yy} 
  A.~Lukas, B.~A.~Ovrut, K.~S.~Stelle and D.~Waldram,
  ``The Universe as a domain wall,''
  Phys.\ Rev.\ D {\bf 59}, 086001 (1999)
  [hep-th/9803235]
  
  \bibitem{Lukas:1998tt} 
  A.~Lukas, B.~A.~Ovrut, K.~S.~Stelle and D.~Waldram,
  ``Heterotic M theory in five-dimensions,''
  Nucl.\ Phys.\ B {\bf 552}, 246 (1999)
  [hep-th/9806051]
  
  \bibitem{Lukas:1999kt} 
  A.~Lukas, B.~A.~Ovrut and D.~Waldram,
  ``Five-branes and supersymmetry breaking in M theory,''
  JHEP {\bf 9904}, 009 (1999)
  [hep-th/9901017]
  
  \bibitem{Donagi:1999jp} 
  R.~Donagi, B.~A.~Ovrut and D.~Waldram,
  ``Moduli spaces of five-branes on elliptic Calabi-Yau threefolds,''
  JHEP {\bf 9911}, 030 (1999)
  [hep-th/9904054]
  
  \bibitem{Donagi:1999ez} 
  R.~Donagi, B.~A.~Ovrut, T.~Pantev and D.~Waldram,
  ``Standard models from heterotic M theory,''
  Adv.\ Theor.\ Math.\ Phys.\  {\bf 5}, 93 (2002)
  [hep-th/9912208]
  
  \bibitem{e2} 
  L.~B.~Anderson, Y.~H.~He and A.~Lukas,
  ``Heterotic compactification, an algorithmic approach'',
  JHEP {\bf 0707}, 049 (2007),
  [hep-th/0702210]

\bibitem{e3} 
 L.~B.~Anderson, Y.~H.~He and A.~Lukas,
 ``Monad Bundles in Heterotic String Compactifications'',
 JHEP {\bf 0807}, 104 (2008),
 [arXiv:0805.2875 [hep-th]]


  %%%%%%%%%%%%%%%%%%%%%%%%%%%%%%%%%%

\bibitem{Donagi:2003tb} 
  R.~Donagi, B.~A.~Ovrut, T.~Pantev and R.~Reinbacher,
  ``SU(4) instantons on Calabi-Yau threefolds with Z(2) x Z(2) fundamental group,''
  JHEP {\bf 0401}, 022 (2004)
  [hep-th/0307273]
  
  \bibitem{Braun:2004xv} 
  V.~Braun, B.~A.~Ovrut, T.~Pantev and R.~Reinbacher,
  ``Elliptic Calabi-Yau threefolds with Z(3) x Z(3) Wilson lines,''
  JHEP {\bf 0412}, 062 (2004)
  [hep-th/0410055]
  
  %%%%%%%%%%%%%%%%%%%%%%%%%%%%%%%%%
  
  \bibitem{Braun:2005ux} 
  V.~Braun, Y.~-H.~He, B.~A.~Ovrut and T.~Pantev,
  ``A Heterotic standard model,''
  Phys.\ Lett.\ B {\bf 618}, 252 (2005)
  [hep-th/0501070]
  
  \bibitem{Braun:2005bw} 
  V.~Braun, Y.~-H.~He, B.~A.~Ovrut and T.~Pantev,
  ``A Standard model from the E(8) x E(8) heterotic superstring,''
  JHEP {\bf 0506}, 039 (2005)
  [hep-th/0502155]
  
  \bibitem{Braun:2005zv} 
  V.~Braun, Y.~-H.~He, B.~A.~Ovrut and T.~Pantev,
  ``Vector bundle extensions, sheaf cohomology, and the heterotic standard model,''
  Adv.\ Theor.\ Math.\ Phys.\  {\bf 10}, 4 (2006)
  [hep-th/0505041]
  
  \bibitem{Braun:2005fk} 
  V.~Braun, Y.~-H.~He, B.~A.~Ovrut and T.~Pantev,
  ``Heterotic standard model moduli,''
  JHEP {\bf 0601}, 025 (2006)
  [hep-th/0509051]
  
  \bibitem{Ambroso:2008kb} 
  M.~Ambroso, V.~Braun and B.~A.~Ovrut,
  ``Two Higgs Pair Heterotic Vacua and Flavor-Changing Neutral Currents,''
  JHEP {\bf 0810}, 046 (2008)
  [arXiv:0807.3319 [hep-th]] 

\bibitem{Buchbinder:2002ji} 
  E.~Buchbinder, R.~Donagi and B.~A.~Ovrut,
  ``Vector bundle moduli and small instanton transitions,''
  JHEP {\bf 0206}, 054 (2002)
  [hep-th/0202084]
  
  \bibitem{Buchbinder:2002pr} 
  E.~I.~Buchbinder, R.~Donagi and B.~A.~Ovrut,
  ``Vector bundle moduli superpotentials in heterotic superstrings and M theory,''
  JHEP {\bf 0207}, 066 (2002)
  [hep-th/0206203] 
  
  
  %%%%%%%%%%%%%%%%%%%%%%%%%%%%%%%%
  
  \bibitem{Braun:2005nv} 
  V.~Braun, Y.~-H.~He, B.~A.~Ovrut and T.~Pantev,
  ``The Exact MSSM spectrum from string theory,''
  JHEP {\bf 0605}, 043 (2006)
  [hep-th/0512177]
  
  \bibitem{lukas}  
L.~B.~Anderson, J.~Gray, Y.~H.~He and A.~Lukas,
``Exploring Positive Monad Bundles And A New Heterotic Standard Model'',     
 [arXiv:0911.1569 [hep-th]]

  
%%%%%%%%%%%%%%%%%%%%%%%%%%%%%%%%% 

 \bibitem{Ambroso:2009jd} 
  M.~Ambroso and B.~Ovrut,
  ``The B-L/Electroweak Hierarchy in Heterotic String and M-Theory,''
  JHEP {\bf 0910}, 011 (2009)
  [arXiv:0904.4509 [hep-th]] 
  
  \bibitem{Ambroso:2009sc} 
  M.~Ambroso and B.~A.~Ovrut,
  ``The B-L/Electroweak Hierarchy in Smooth Heterotic Compactifications,''
  Int.\ J.\ Mod.\ Phys.\ A {\bf 25}, 2631 (2010)
  [arXiv:0910.1129 [hep-th]]
  
  \bibitem{Ambroso:2010pe} 
  M.~Ambroso and B.~A.~Ovrut,
  ``The Mass Spectra, Hierarchy and Cosmology of B-L MSSM Heterotic Compactifications,''
  arXiv:1005.5392 [hep-th]
  
\bibitem{Brelidze:2010hf} 
  T.~Brelidze and B.~A.~Ovrut,
  ``B-L Cosmic Strings in Heterotic Standard Models,''
  JHEP {\bf 1007}, 077 (2010)
  [arXiv:1003.0234 [hep-th]]

%%%%%%%%%%%%%%%%%%%%%%%%%%%%%%%%%%%%%%%%%%%%%%%%%%%%%%%%%
\bibitem{Mohapatra:1986aw}
  R.~N.~Mohapatra,
  ``Mechanism For Understanding Small Neutrino Mass In Superstring Theories,''
  Phys.\ Rev.\ Lett.\  {\bf 56} (1986) 561.
  %%CITATION = PRLTA,56,561;%%

\bibitem{Ghosh:2010hy}
  D.~K.~Ghosh, G.~Senjanovic and Y.~Zhang,
  ``Naturally Light Sterile Neutrinos from Theory of R-parity,''
  Phys.\ Lett.\ B {\bf 698} (2011) 420
  [arXiv:1010.3968 [hep-ph]].
  %%CITATION = ARXIV:1010.3968;%%

\bibitem{Barger:2010iv}
  V.~Barger, P.~Fileviez Perez and S.~Spinner,
  ``Three Layers of Neutrinos,''
  Phys.\ Lett.\ B {\bf 696} (2011) 509
  [arXiv:1010.4023 [hep-ph]].
  %%CITATION = ARXIV:1010.4023;%%
%%%%%%%%%%%%%%%%%%%%%%%%%%%%%%%%%%%%%%%%%%%%%%%%%%%%%%%%%  

%\cite{Salam:1974xa}
\bibitem{R1}
  A.~Salam and J.~A.~Strathdee,
  ``Supersymmetry and Fermion Number Conservation,''
  Nucl.\ Phys.\ B {\bf 87} (1975) 85.
  %%CITATION = NUPHA,B87,85;%%  
  
%\cite{Fayet:1974pd}
\bibitem{R2}
  P.~Fayet,
  ``Supergauge Invariant Extension of the Higgs Mechanism and a Model for the electron and Its Neutrino,''
  Nucl.\ Phys.\ B {\bf 90} (1975) 104.
  %%CITATION = NUPHA,B90,104;%%

\bibitem{AM}
  C.~S.~Aulakh and R.~N.~Mohapatra,
  ``Neutrino as the Supersymmetric Partner of the Majoron,''
  Phys.\ Lett.\ B {\bf 119} (1982) 136.
  %%CITATION = PHLTA,B119,136;%%
  
\bibitem{Hayashi}
  M.~J.~Hayashi and A.~Murayama,
  ``Radiative Breaking of SU(2)-R x U(1)-(B-L) gauge symmetry induced by broken N=1 supersymmetry in a left-right symmetry model,''
  Phys.\ Lett.\ B {\bf 153} (1985) 251.
  %%CITATION = PHLTA,B153,251;%%  

\bibitem{Masiero:1990uj}
  A.~Masiero and J.~W.~F.~Valle,
  ``A Model For Spontaneous R Parity Breaking,''
  Phys.\ Lett.\ B {\bf 251} (1990) 273.
  %%CITATION = PHLTA,B251,273;%%

\bibitem{Kuchimanchi:1993jg}
  R.~Kuchimanchi and R.~N.~Mohapatra,
  ``No parity violation without R-parity violation,''
  Phys.\ Rev.\ D {\bf 48} (1993) 4352
  [hep-ph/9306290].
  %%CITATION = HEP-PH/9306290;%%

\bibitem{Martin:1996kn}
  S.~P.~Martin,
  ``Implications of supersymmetric models with natural R-parity conservation,''
  Phys.\ Rev.\ D {\bf 54} (1996) 2340
  [hep-ph/9602349].
  %%CITATION = HEP-PH/9602349;%%
  
\bibitem{Aulakh:1999cd}
  C.~S.~Aulakh, A.~Melfo, A.~Rasin and G.~Senjanovic,
  ``Seesaw and supersymmetry or exact R-parity,''
  Phys.\ Lett.\ B {\bf 459} (1999) 557
  [hep-ph/9902409].
  %%CITATION = HEP-PH/9902409;%%
 
\bibitem{Aulakh:2000sn}
  C.~S.~Aulakh, B.~Bajc, A.~Melfo, A.~Rasin and G.~Senjanovic,
  ``SO(10) theory of R-parity and neutrino mass,''
  Nucl.\ Phys.\ B {\bf 597} (2001) 89
  [hep-ph/0004031].
  %%CITATION = HEP-PH/0004031;%% 

\bibitem{Feldman:2011ms}
  D.~Feldman, P.~Fileviez Perez and P.~Nath,
  ``R-parity Conservation via the Stueckelberg Mechanism: LHC and Dark Matter Signals,''
  arXiv:1109.2901 [hep-ph].
  %%CITATION = ARXIV:1109.2901;%%

\bibitem{FileviezPerez:2008sx}
  P.~Fileviez Perez and S.~Spinner,
  ``Spontaneous R-Parity Breaking and Left-Right Symmetry,''
  Phys.\ Lett.\ B {\bf 673} (2009) 251
  [arXiv:0811.3424 [hep-ph]].
  %%CITATION = ARXIV:0811.3424;%%

\bibitem{Barger:2008wn}
  V.~Barger, P.~Fileviez Perez and S.~Spinner,
  ``Minimal gauged U(1)(B-L) model with spontaneous R-parity violation,''
  Phys.\ Rev.\ Lett.\  {\bf 102} (2009) 181802
  [arXiv:0812.3661 [hep-ph]].
  %%CITATION = ARXIV:0812.3661;%%  

\bibitem{Everett:2009vy}
  L.~L.~Everett, P.~Fileviez Perez and S.~Spinner,
  ``The Right Side of Tev Scale Spontaneous R-Parity Violation,''
  Phys.\ Rev.\ D {\bf 80} (2009) 055007
  [arXiv:0906.4095 [hep-ph]].
  %%CITATION = ARXIV:0906.4095;%%

\bibitem{FileviezPerez:2012mj}
  P.~Fileviez Perez and S.~Spinner,
  ``The Minimal Theory for R-parity Violation at the LHC,''
  arXiv:1201.5923 [hep-ph].
  %%CITATION = ARXIV:1201.5923;%%

\bibitem{Preparation} In Preparation.

\bibitem{Babu:1996vt}
  K.~S.~Babu, C.~F.~Kolda and J.~March-Russell,
  ``Leptophobic U(1)s and the R(b) - R(c) crisis,''
  Phys.\ Rev.\  D {\bf 54}, 4635 (1996)
  [arXiv:hep-ph/9603212].
  %%CITATION = PHRVA,D54,4635;%%

\bibitem{Georgi} 
  Howard Georgi,
  ``Lie Algebras in Particle Physics''
  The Benjamin/Cummings Publishing Company (1982)
  
\bibitem{Slansky:1981yr}
  R.~Slansky,
  ``Group Theory For Unified Model Building,''
  Phys.\ Rept.\  {\bf 79}, 1 (1981).
  %%CITATION = PRPLC,79,1;%%
    
\bibitem{delAguila:1988jz}
  F.~del Aguila, G.~D.~Coughlan and M.~Quiros,
  ``Gauge Coupling Renormalization With Several U(1) Factors,''
  Nucl.\ Phys.\  B {\bf 307}, 633 (1988)
  [Erratum-ibid.\  B {\bf 312}, 751 (1989)].
  %%CITATION = NUPHA,B307,633;%%
  
\bibitem{Holdom:1985ag}
  B.~Holdom,
  ``Two U(1)'S And Epsilon Charge Shifts,''
  Phys.\ Lett.\  B {\bf 166}, 196 (1986).
  %%CITATION = PHLTA,B166,196;%%
  
\bibitem{Dienes:1996zr}
  K.~R.~Dienes, C.~F.~Kolda and J.~March-Russell,
  ``Kinetic mixing and the supersymmetric gauge hierarchy,''
  Nucl.\ Phys.\  B {\bf 492}, 104 (1997)
  [arXiv:hep-ph/9610479].
  %%CITATION = NUPHA,B492,104;%%

\bibitem{Foot:1991kb}
  R.~Foot and X.~G.~He,
  ``Comment On Z Z-Prime Mixing In Extended Gauge Theories,''
  Phys.\ Lett.\  B {\bf 267}, 509 (1991).
  %%CITATION = PHLTA,B267,509;%%
  
\bibitem{Fonseca:2011vn}
  R.~M.~Fonseca, M.~Malinsky, W.~Porod and F.~Staub,
  ``Running soft parameters in SUSY models with multiple U(1) gauge factors,''
  Nucl.\ Phys.\ B {\bf 854} (2012) 28
  [arXiv:1107.2670 [hep-ph]].
  %%CITATION = ARXIV:1107.2670;%%

\bibitem{Mohapatra:1974gc}
  R.~N.~Mohapatra and J.~C.~Pati,
  ``A Natural Left-Right Symmetry,''
  Phys.\ Rev.\ D {\bf 11} (1975) 2558.
  %%CITATION = PHRVA,D11,2558;%%

\bibitem{Senjanovic:1975rk}
  G.~Senjanovic and R.~N.~Mohapatra,
  ``Exact Left-Right Symmetry and Spontaneous Violation of Parity,''
  Phys.\ Rev.\ D {\bf 12} (1975) 1502.
  %%CITATION = PHRVA,D12,1502;%%

\bibitem{Senjanovic:1978ev}
  G.~Senjanovic,
  ``Spontaneous Breakdown of Parity in a Class of Gauge Theories,''
  Nucl.\ Phys.\ B {\bf 153} (1979) 334.

\bibitem{Pati:1974yy}
  J.~C.~Pati and A.~Salam,
  ``Lepton Number as the Fourth Color,''
  Phys.\ Rev.\ D {\bf 10} (1974) 275
   [Erratum-ibid.\ D {\bf 11} (1975) 703].
  %%CITATION = PHRVA,D10,275;%%
  %%CITATION = NUPHA,B153,334;%%
  
\bibitem{Carena:2002es} 
  M.~S.~Carena and H.~E.~Haber,
  ``Higgs boson theory and phenomenology,''
  Prog.\ Part.\ Nucl.\ Phys.\  {\bf 50}, 63 (2003)
  [hep-ph/0208209].
  
%\cite{Gamberini:1989jw}
\bibitem{Gamberini:1989jw}
  G.~Gamberini, G.~Ridolfi and F.~Zwirner,
  ``On Radiative Gauge Symmetry Breaking in the Minimal Supersymmetric Model,''
  Nucl.\ Phys.\  B {\bf 331}, 331 (1990).
  %%CITATION = NUPHA,B331,331;%%

\bibitem{Jones:1981we}
  D.~R.~T.~Jones,
  ``The Two Loop beta Function for a G(1) x G(2) Gauge Theory,''
  Phys.\ Rev.\ D {\bf 25}, 581 (1982).
  %%CITATION = PHRVA,D25,581;%%
  
\bibitem{Martin:1993zk} 
  S.~P.~Martin and M.~T.~Vaughn,
  ``Two loop renormalization group equations for soft supersymmetry breaking couplings,''
  Phys.\ Rev.\ D {\bf 50}, 2282 (1994)
  [Erratum-ibid.\ D {\bf 78}, 039903 (2008)]
  [hep-ph/9311340].

\bibitem{Nath:2006ut}
  P.~Nath and P.~Fileviez Perez,
  ``Proton stability in grand unified theories, in strings and in branes,''
  Phys.\ Rept.\  {\bf 441} (2007) 191
  [hep-ph/0601023].
  %%CITATION = HEP-PH/0601023;%%

\bibitem{Malinsky:2005bi}
  M.~Malinsky, J.~C.~Romao and J.~W.~F.~Valle,
  ``Novel supersymmetric SO(10) seesaw mechanism,''
  Phys.\ Rev.\ Lett.\  {\bf 95} (2005) 161801
  [hep-ph/0506296].
  %%CITATION = HEP-PH/0506296;%%

\bibitem{DeRomeri:2011ie}
  V.~De Romeri, M.~Hirsch and M.~Malinsky,
  ``Soft masses in SUSY SO(10) GUTs with low intermediate scales,''
  Phys.\ Rev.\ D {\bf 84} (2011) 053012
  [arXiv:1107.3412 [hep-ph]].
  %%CITATION = ARXIV:1107.3412;%%
%%%%%%%%%%%%%%%%%%%%%%%%%%%%%%%%%%%%%%%%%%%%%%%%%%%%%%%%%  
  
\end{thebibliography}
\end{document}